\documentclass{article}

\usepackage{arxiv}

\usepackage{amsthm}
\usepackage{amssymb}
\usepackage{mathtools}
\usepackage{mathpartir}
\usepackage{stmaryrd}
\usepackage{leftindex}
\usepackage{wasysym}
\usepackage{esvect}
\usepackage{wasysym}

\usepackage[utf8]{inputenc} 
\usepackage[T1]{fontenc}    
\usepackage{hyperref}       
\usepackage{url}            
\usepackage{booktabs}       
\usepackage{amsfonts}       
\usepackage{nicefrac}       
\usepackage{microtype}      
\usepackage{cleveref}       
\usepackage{graphicx}
\usepackage{natbib}
\usepackage{doi}
\usepackage{subcaption}

\theoremstyle{definition}
\newtheorem{definition}{Definition}[section]
\newtheorem{example}{Example}[section]
\theoremstyle{plain}
\newtheorem{proposition}{Proposition}[section]
\newtheorem{theorem}{Theorem}[section]

\newtheorem{lemma}[theorem]{Lemma}

\theoremstyle{remark}

\usepackage{proof}

\usepackage{algorithm}
\usepackage{algpseudocode}

\algnewcommand\algorithmicforeach{\textbf{for each}}
\algdef{S}[FOR]{ForEach}[1]{\algorithmicforeach\ #1\ \algorithmicdo}

\newcommand{\braces}[1]{{\{}#1{\}}}
\newcommand{\anglebrackets}[1]{{\langle}#1{\rangle}}

\usepackage{fancyvrb}
\usepackage{listings}
\usepackage{xcolor}

\usepackage{CJKutf8}

\lstdefinelanguage{Seni}{
  keywords={default, type, system, context, async, sync, refine, prior, var, init, expose, as, with, under, prop, not, and, or, AF, G, F, True, False, before, after, ltl, ctl, el, admitted, print, read, write},
  ndkeywords={Int, String, Bool, Graph},
  ndkeywordstyle=\color{teal}\bfseries,
  identifierstyle=\color{black},
  sensitive=true,
  comment=[l]{//},
  morecomment=[s]{/*}{*/},
  commentstyle=\color{gray}\ttfamily,
  stringstyle=\color{darkgray}\ttfamily,
  morestring=[b]',
  morestring=[b]"
}
\lstdefinelanguage{Solidity}{
  keywords={function, public, void, for, in, return},
  ndkeywords={uint8, uint256},
  ndkeywordstyle=\color{teal}\bfseries,
  identifierstyle=\color{black},
  sensitive=true,
  comment=[l]{//},
  morecomment=[s]{/*}{*/},
  commentstyle=\color{gray}\ttfamily,
  stringstyle=\color{darkgray}\ttfamily,
  morestring=[b]',
  morestring=[b]"
}
\title{Transition-Oriented Programming: \linebreak{}Developing Provably Correct Systems}

\date{}

\newif\ifuniqueAffiliation
\uniqueAffiliationtrue

\ifuniqueAffiliation 
\author{ \href{https://orcid.org/0000-0002-6996-9333}{\includegraphics[scale=0.06]{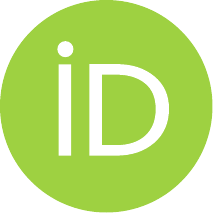}\hspace{1mm}Yepeng~Ding}\thanks{\url{https://yepengding.github.io/}} \\
	The University of Tokyo \\
	\texttt{yepengding@g.ecc.u-tokyo.ac.jp} \\
}
\else
\usepackage{authblk}

\newbox{\orcid}\sbox{\orcid}{\includegraphics[scale=0.06]{orcid.pdf}} 
\author[1]{%
	\href{https://orcid.org/0000-0000-0000-0000}{\usebox{\orcid}\hspace{1mm}David S.~Hippocampus\thanks{\texttt{hippo@cs.cranberry-lemon.edu}}}%
}
\author[1,2]{%
	\href{https://orcid.org/0000-0000-0000-0000}{\usebox{\orcid}\hspace{1mm}Elias D.~Striatum\thanks{\texttt{stariate@ee.mount-sheikh.edu}}}%
}
\affil[1]{Department of Computer Science, Cranberry-Lemon University, Pittsburgh, PA 15213}
\affil[2]{Department of Electrical Engineering, Mount-Sheikh University, Santa Narimana, Levand}
\fi

\renewcommand{\shorttitle}{Transition-Oriented Programming}

\hypersetup{
pdftitle={Transition-Oriented Programming: Developing Provably Correct Systems},
pdfsubject={Transition-oriented programming},
pdfauthor={Yepeng~Ding},
pdfkeywords={Transition-oriented programming, Formal methods, Software engineering, Information security, Distributed computing},
}

\begin{document}
\maketitle

\begin{abstract}
Correctness is a necessary condition for systems to be effective in meeting human demands, thus playing a critical role in system development. However, correctness often manifests as a nebulous concept in practice, leading to challenges in accurately creating specifications, effectively proving correctness satisfiability, and efficiently implementing correct systems. Motivated by tackling these challenges, this paper introduces Transition-Oriented Programming (TOP), a programming paradigm to facilitate the development of provably correct systems by intertwining correctness specification, verification, and implementation within a unified theoretical framework.

This paper establishes a formalized foundation for correctness with formal models and formal logics. Based on this foundation, this paper introduces Graph Transition Systems (GTSs), a formalism proficient in capturing stateful dynamics and structural transformations of systems. The core features of TOP in practice are centered around GTSs, including progressive specification, modularization, context perception, and proof automation.

These features are exemplified through practical examples programmed in Seni, a fully-fledged TOP language. The Seni language excels at intuitively constructing GTSs and articulating temporal and functional properties. It combines bottom-up (modularization) and top-down (progressive specification) approaches to improve development efficiency via model refinement, parameterization, integration, and composition. Seni also incorporates context perception that introduces type refinement and transformation prioritization to improve the precision of specifications and efficiency of verification processes. Besides, Seni leverages bounded model checking and iterative deepening theorem proving techniques in correctness verification to automate proofs at compile time.

Furthermore, this paper demonstrates the applicability of TOP in diverse practical scenarios. It introduces transition-oriented development, a development process aligned with the principles of TOP to develop provably correct distributed protocols. It also elaborates on transition-oriented security analysis for security infrastructures against threat models, exemplified by a security framework for self-sovereign identity systems. Additionally, this paper presents transition-oriented program analysis, exemplified by context-sensitive concolic verification, for uncovering vulnerabilities in decentralized applications, particularly in decentralized finance protocols. These applications underscore the potential of TOP in addressing the real-world challenges of developing provably correct systems.

This paper concludes with a summary of its contributions, followed by a detailed discussion on broadening the applicability of extending TOP in compiling tool development, explainable artificial intelligence, and hardware design, alongside the future directions of TOP and the Seni language.
\end{abstract}

\keywords{Transition-oriented programming \and Formal methods \and Software engineering \and Information security \and Distributed computing}

\section{Introduction}
\label{sec:introduction}

Systems are developed to meet human demands. A system is a group of interconnected elements acting as a unified whole to satisfy its specification that formalizes human demands. Consequently, a system is effective if 

\begin{itemize}
    \item \textbf{Expressiveness} its specification accurately describe a human demand, and
    \item \textbf{Correctness} it satisfies its specification.
\end{itemize}

However, human demands are subjective and not always precisely describable in languages that humans and machines can understand, making the expressiveness proof an open philosophical challenge. Nevertheless, the correctness proof concerning the objective satisfaction relation is an approximately solvable problem. Therefore, a system is correct if it satisfies its specification, regardless of whether it meets human demands or not.

In software development, demand analysis is indispensable yet remains a formidable challenge regarding soundness and completeness. This difficulty stems primarily from the diverse and dynamic nature of user demands, coupled with the constraints of limited analysis resources \cite{leffingwell_managing_2000}. Despite diligent efforts in demand analysis, the resultant requirement specifications, though meticulously documented and formalized, inevitably fall short of meeting all user demands perpetually. Consequently, the efficacy of software is inherently time-bound and user-specific, even when it adheres to the correctness condition.

This paper delves into the concept of correctness within the context of computer science and engineering, while acknowledging the challenge of the expressiveness proof in this specific academic field. The focus is predominantly on programs as exemplars of computer systems, although the discourse is applicable to other system typologies. At its core, the paper contends that a program’s correctness is contingent upon the satisfaction of its requirement specification. However, this premise gives rise to two pivotal questions as follows.

\begin{itemize}
    \item[\textbf{Q1}] How to define languages for specification?
    \item[\textbf{Q2}] How to prove correctness satisfiability?
\end{itemize}

\textbf{Q1} and \textbf{Q2} are intertwined, as the selection of a specification language is crucial for proving correctness satisfiability. While natural languages can be used for program specification, their inherent ambiguity and lack of verifiability pose significant challenges. To illustrate, consider the program presented in Listing~\ref{lst:vulnerable_smart_contract}, which is a distilled version of real-world incidents, notably the 2018 BeautyChain hack. A specification about this program, stating "The execution of function \textit{process} yields a result where $x$ is greater than $0$", may seem clear but harbors ambiguity. It could be interpreted in multiple ways, such as $\forall t \in [0, 2^{256}): \textit{process}(t) \implies x > 0$, $\exists t \in [0, 2^{256}): \textit{process}(t) \implies x > 0$, and $\forall t \in [0, 2^8): \textit{process}(t) \implies x > 0$. Each interpretation significantly alters the perceived behavior of the program, thereby accentuating the critical need for precise and rigorous specification languages.

\begin{lstlisting}[caption={Pseudocode of a vulnerable smart contract.}, language=Solidity,mathescape=true, numbers=left, label={lst:vulnerable_smart_contract},basicstyle=\footnotesize]
public uint8 x := 1;
function process(uint256 t) -> uint8 {
    for i in [0..t):
        x := x + 1;
    return x;
}

process(Random.UINT256)

\end{lstlisting}

Moreover, the verifiability of specification languages, particularly conducive to automated verification, is paramount. The reliance on manual proofs for correctness satisfiability is fraught with challenges, especially when dealing with extensive inputs and outputs, due to the high risk of human error and the impracticality of mental computation. Taking the example of the simple program outlined in Listing~\ref{lst:vulnerable_smart_contract}, even a seemingly straightforward manual proof for the specification $\forall t \in [0, 2^{256}): \textit{process}(t) \implies x > 0$ can be problematic. A plausible proof constructed on the premises that $x = 1 + t$ and $t \geq 0$ to deduce that $x > 0$, however, is flawed. For instance, when $t = 2^8 - 1$, it results in $x = 0$ due to integer overflow, a typical issue overlooked in manual proofs. Besides, the variable $x$ is publicly accessible and subject to modification by other functions, which adds another layer of complexity to the proof. The limitations of manual proofs in such scenarios underscore the necessity for automated verification. In conjunction with precise and verifiable specifications, automated verification has the potential to navigate through the complexities of efficiently constructing reliable proof.

Formal languages, exemplified by representing programs as either transition systems or rewriting systems, can effectively mitigate ambiguity and enhance verifiability. However, these models of computation are incomparable, as each of them has unique strengths in representing different aspects of programs. Generally, program specifications can be categorized into temporal and functional properties. Transition systems are adept at capturing stateful dynamics, representing changes in system states with respect to time, exemplified by properties like "There exists an execution path of function \textit{process} where $x \geq 0$ eventually holds", i.e., $\exists \Diamond (x \geq 0)$. These properties are commonly verified by model checking \cite{baier_principles_2008,clarke_jr_model_2018} (e.g., SPIN \cite{holzmann_model_1997}, NuSMV \cite{cimatti_nusmv_2002}, CBMC \cite{kroening_cbmcc_2014}) and concolic execution \cite{baldoni_survey_2018} (e.g., PathFinder \cite{visser_test_2004}, KLEE \cite{cadar_klee_2008}, PyExZ3 \cite{irlbeck_deconstructing_2015}) techniques.

Conversely, rewriting systems are tailored to the representation of structural transformations with respect to rules, suitable for specifying functional properties such as $\textit{process}(v+1) = \textit{process}(v) + 1$. These properties are predominantly verified by theorem proving \cite{bibel_automated_2013} (Isabelle \cite{paulson_isabelle_1994}, F* \cite{swamy_dependent_2016}, Lean \cite{moura_lean_2021}) and term rewriting \cite{baader_term_1998} techniques (AProVE \cite{giesl_automated_2004}, TNT \cite{gupta_proving_2008}, TTT \cite{korp_tyrolean_2009}). The incomparable expressiveness requires considerable effort, such as selecting and mastering appropriate languages and tools, for both specification and verification.

Therefore, an intuitive answer to \textbf{Q1} and \textbf{Q2} is to define a specification language that is comprehensible to humans and processable by machines. Such a language should be precise and concise to reflect human demands, while also being computable and verifiable to automate correctness satisfiability proving. Nevertheless, in practical scenarios, repairing erroneous programs to preserve specific properties often involves arduous efforts and introduces new complexities and risks. For the vulnerable smart contract presented in Listing~\ref{lst:vulnerable_smart_contract}, modifying the variable type of $x$ to \textit{uint256} seems a viable solution to the integer overflow issue and fulfilling the specification $\forall t \in [0, 2^8): \textit{process}(t) \implies x \geq 0$. However, this modification is fraught with risks and incurs significant costs. For instance, $x$ may only be adequate for other functions with type \textit{uint8}. Besides, altering its type not only increases gas consumption but may also disrupt the storage layout of the deployed smart contract, potentially leading to malfunctions. Consequently, our focus may shift towards the following question.

\begin{itemize}
    \item[\textbf{Q3}] How to implement correct systems?
\end{itemize}

\textbf{Q3} presents a harder question on account of the fact that an effective and efficient solution to \textbf{Q3} implies an effective solution to \textbf{Q2}. In practice, system complexity is escalating rapidly to accommodate a variety of demands, propelled by advancements in Information and Communications Technologies (ICTs), which, in turn, catalyze significant societal shifts, as evidenced in Industry 4.0 and Web 3.0. The increasing complexity carries the promise of enriching user experience, streamlining functionalities, and spawning innovative solutions, thereby bolstering productivity across various domains. Industry 4.0, characterized by the integration of cyber-physical systems \cite{baheti_cyber-physical_2011,lee_past_2015}, the Internet of Things (IoT) \cite{da_xu_internet_2014,li_internet_2015}, artificial intelligence \cite{jordan_machine_2015,mahesh_machine_2020}, and cloud computing \cite{dillon_cloud_2010,mao_survey_2017}, is transforming traditional manufacturing processes into smart, interoperable systems. Simultaneously, Web 3.0 is reshaping the landscape of the Internet by advocating decentralized, personalized, and intelligent systems \cite{ding_derepo_2020,ding_dagbase_2020,ding_bloccess_2023} through blockchain \cite{wood_ethereum_2014,nakamoto_bitcoin_2019,zheng_blockchain_2018} and extended reality \cite{ratcliffe_extended_2021} technologies.

Therefore, for \textbf{Q3}, intuitively, an implementation of a correct system should not only adhere to its specified requirements but also be cost-effective. This consideration becomes particularly crucial in large-scale and complex systems, where the complexity of individual components and their interactions necessitates a systematic method to balance correctness and cost. Additionally, correctness satisfiability may not be theoretically provable for implementations in cases like those involving undecidable problems. Therefore, an implementation is provably correct if and only if its correctness satisfiability is provable.

\begin{figure}[ht]
\centering
\includegraphics[width=0.4\textwidth]{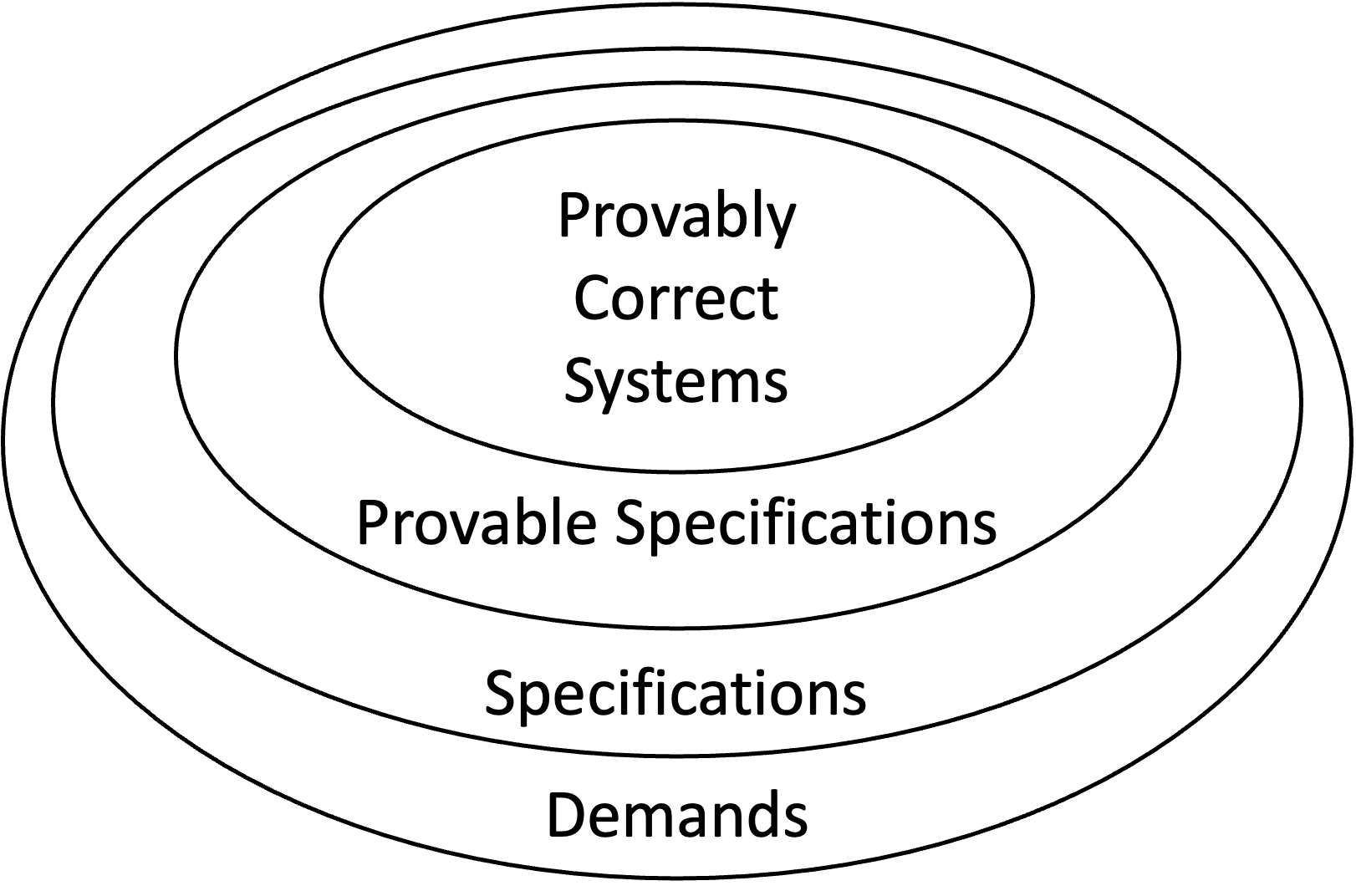}
\caption[Inclusion of Provably Correct Systems]{Inclusion of demands, specifications, provable specifications, and provably correct systems.}
\label{fig:inclusion_relation}
\end{figure}

As illustrated in Figure~\ref{fig:inclusion_relation}, only a subset of human demands is specifiable by a specification language. However, an effective response to \textbf{Q1} holds the potential to expand this subset to approximate the human demands, thereby amplifying correctness specifiability. Besides, only a subset of these specifications is technically provable. Nonetheless, an effective and efficient solution to \textbf{Q2} can streamline the establishment of correctness, thereby enhancing correctness verifiability. Furthermore, among the specifications that are technically provable, only a subset proving to be correct is practically implementable in a manner that produces provably correct systems. Addressing \textbf{Q3} effectively and efficiently can extend the set of provably correct systems, consequently advancing correctness implementability. This hierarchical interplay highlights the nuanced and interconnected nature of human demands, specifications, provable specifications, and provably correct systems.

This paper is primarily driven by the objective of addressing \textbf{Q1}, \textbf{Q2}, and \textbf{Q3}. It commences with a formalization of correctness by a systematic illustration of formal models and logic. The formalization streamlines the representations of transition systems and rewriting systems, along with their typical variants, thereby laying a foundation for the subsequent discussions.

Central to this paper is the introduction of \textbf{Transition-Oriented Programming} (TOP), a novel programming paradigm designed to facilitate the development of provably correct systems. The essence of TOP lies in intertwining specification, verification, and implementation within a singular theoretical framework, which ensures that a provably correct specification is also a provably correct implementation. The theoretical underpinnings of TOP are grounded in the concept of Graph Transition Systems (GTSs) that are adept at capturing both stateful dynamics and structural transformations inherent in complex systems. Additionally, TOP is distinct in its intrinsic support for correctness verification to prove correctness satisfiability regarding the formulated GTSs associated with temporal and functional properties. The correctness verification follows the principles of the established model-checking and theorem-proving techniques with tailored adaptations for application to GTSs. Moreover, TOP bridges the gap between formal models and practical implementations. Formal models are not only immediately executable but are also capable of handling side effects with practical refinements effectively, thereby marking a stride toward applicability and usability in practice.

This paper also introduces the Seni language, a TOP language with fully-fledged TOP features. Seni is designed to facilitate the intuitive construction of GTSs and the expressive formulation of both temporal and functional properties. A distinctive feature of Seni is its integration of both bottom-up (modularization) and top-down (progressive specification) approaches to optimize the specification process. Modularization is geared toward amplifying scalability and reusability by enabling model parameterization and composition. In parallel, progressive specification is augmented by model refinement techniques, fostering an iterative and incremental methodology for both model and property specification and enabling verification at multiple abstract levels. Besides, Seni incorporates context perception, a feature that improves specification precision and verification efficiency via type refinement and transformation prioritization. Furthermore, Seni stands out by offering intrinsic support for correctness verification at compile time, leveraging fully automated bounded model checking and iterative deepening theorem proving techniques.

Additionally, this paper discusses the applicability of TOP through three application scenarios, including transition-oriented development of distributed protocols, transition-oriented security analysis of security infrastructures, and transition-oriented program analysis of decentralized applications. It further envisions the potential of TOP in compiling tool development, explainable artificial intelligence, and hardware design.
\section{Correctness}
\label{sec:correctness}

A system is correct if it satisfies its specification. A specification uses a language to describe a set of properties that a system needs to hold. Intriguingly, specifications themselves can be represented through systems. Consequently, system correctness is a type of satisfaction relation between two systems. 

However, practical systems, such as computer programs, usually manifest a significant degree of complexity and encompass numerous details that are extraneous to their correctness. In contrast, succinct and expressive specifications are paramount for articulating the necessary system properties. This dichotomy underscores the indispensable role of formal models and formal logic in formalizing and practicalizing correctness.

Intuitively, system correctness revolves around two fundamental operations: verification and implementation. Verification entails the rigorous process of proving correctness satisfaction, whereas implementation refers to the development of a system according to its specifications.

This section formalizes correctness to elucidate the essence of \textbf{Q1}, \textbf{Q2}, and \textbf{Q3} as introduced in Section~\ref{sec:introduction} with formal models and formal logic as preliminaries.

\subsection[Formal Model]{Formal Model}
\label{sec:formal_model}

Formal models are essential in abstracting complex systems. Given that computer programs represent computations, models of computation serve as appropriate abstractions for these programs.

A model of computation is a formal system that specifies the mechanism of a function producing an output for an input. Such formal systems play a pivotal role in program analysis, as they provide a foundational basis to which all computer programs can be theoretically reduced. This section presents two representative models of computation: transition system and rewriting system.

\subsubsection{Transition System}

\paragraph{Basic Transition System}

A transition system, formally defined in Definition~\ref{def:ts}, is a model of computation to specify states and their dynamic transitions.

\begin{definition}[Transition System]
\label{def:ts}
    A \textbf{Transition System} (TS) is a pair
    \[
        \anglebrackets{S, {\to}}
    \]
    \begin{itemize}
        \item $S$ is a set of states, and
        \item ${\to} \subseteq S \times S$ is a transition relation.
    \end{itemize}
\end{definition}

\begin{figure}[ht]
     \centering
     \begin{subfigure}[b]{0.45\textwidth}
         \centering
         \includegraphics[width=0.5\textwidth]{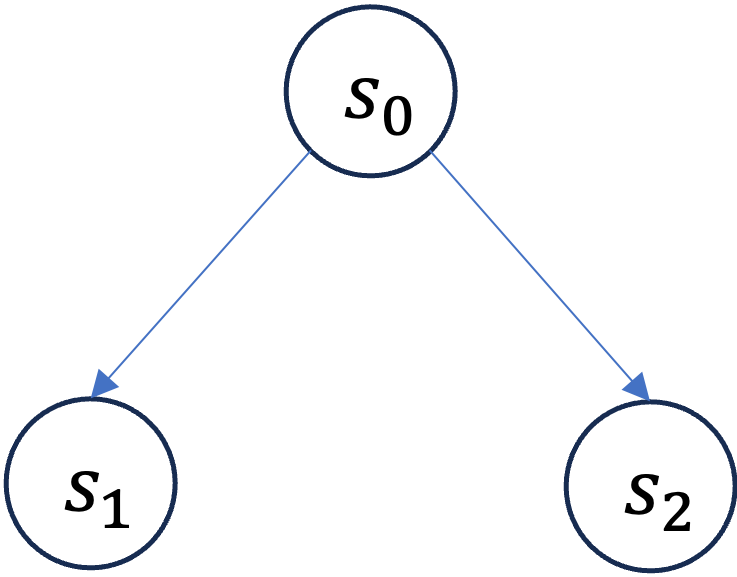}
         \caption[Transition System]{Transition system example.}
         \label{fig:ts}
     \end{subfigure}
     \hfill
     \begin{subfigure}[b]{0.45\textwidth}
         \centering
         \includegraphics[width=0.5\textwidth]{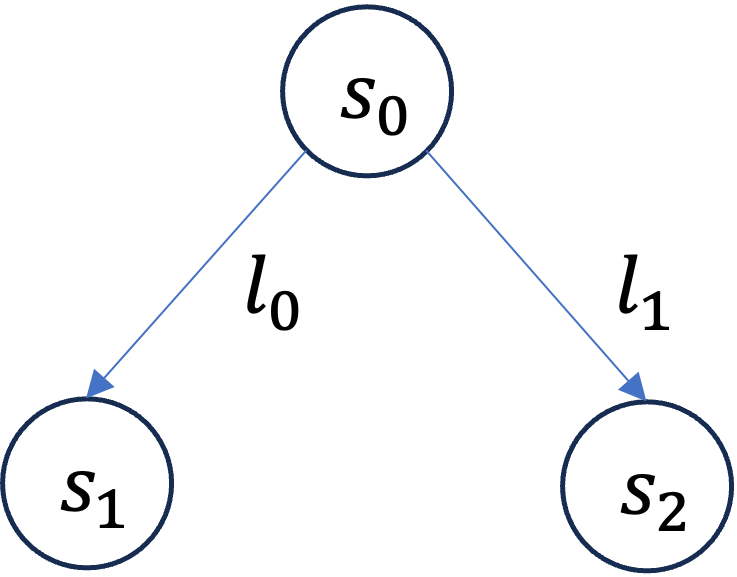}
         \caption[Labeled Transition System]{Labeled transition system example.}
         \label{fig:lts}
     \end{subfigure}
    \caption[Transition System Examples]{Examples of transition systems.}
    \label{fig:ts_examples}
\end{figure}

\begin{example}[Transition System]
\label{eg:ts}
A transition system example is visualized in Figure~\ref{fig:ts}, where $S = \braces{s_0, s_1, s_2}$ and ${\to} = \braces{(s_0, s_1), (s_0, s_2)}$.
\end{example}

Many variants of transition systems have emerged from Definition~\ref{def:ts} to accommodate a wide range of scenarios, such as labeled transition systems in Definition~\ref{def:lts}, Kripke structure in Definition~\ref{def:ks}, and labeled Kripke structure in Definition~\ref{def:lks}.

Labeled transition systems, formally defined in Definition~\ref{def:lts}, introduce the label set to represent elements varying with contexts, such as inputs to consume, conditions to trigger transitions, actions performed during transitions, and effects emitted after the transition.

\begin{definition}[Labeled Transition System]
\label{def:lts}
    A \textbf{Labeled Transition System} (LTS) is a triple
    \[
        \anglebrackets{S, L, {\to}}
    \]
    \begin{itemize}
        \item $S$ is a set of states,
        \item $L$ is a set of labels, and
        \item ${\to} \subseteq S \times L \times S$ is a transition relation.
    \end{itemize}
\end{definition}

\begin{example}[Labeled Transition System]
\label{eg:lts}
    An example of a labeled transition system is depicted in Figure~\ref{fig:lts}, where $S = \braces{s_0, s_1, s_2}$, $L = \braces{l_0, l_1}$, and ${\to} = \braces{(s_0, l_0, s_1), (s_0, l_1, s_2)}$.
\end{example}

While LTSs focus on state changes via introducing labels over transitions, Kripke structures, formally defined in Definition~\ref{def:ks}, abstract away specific transitions and focus on state properties by assigning atomic propositions to states.

\paragraph{Kripke Structure}

\begin{definition}[Kripke Structure]
\label{def:ks}
    A \textbf{Kripke Structure} (KS) over set $A$ of atomic propositions is a quadruple
    \[
        \anglebrackets{S, \leftindex^0{S}, {\to}, \mathcal{L}}
    \]
    \begin{itemize}
        \item $S$ is a set of states,
        \item $\leftindex^0{S} \subseteq S$ is a set of initial states,
        \item ${\to} \subseteq S \times S$ is a transition relation, and
        \item $\mathcal{L}: S \mapsto \wp(A)$ is a total labeling function.
    \end{itemize}
\end{definition}

\begin{figure}
     \centering
    \begin{subfigure}[b]{0.45\textwidth}
        \centering
        \includegraphics[width=0.6\textwidth]{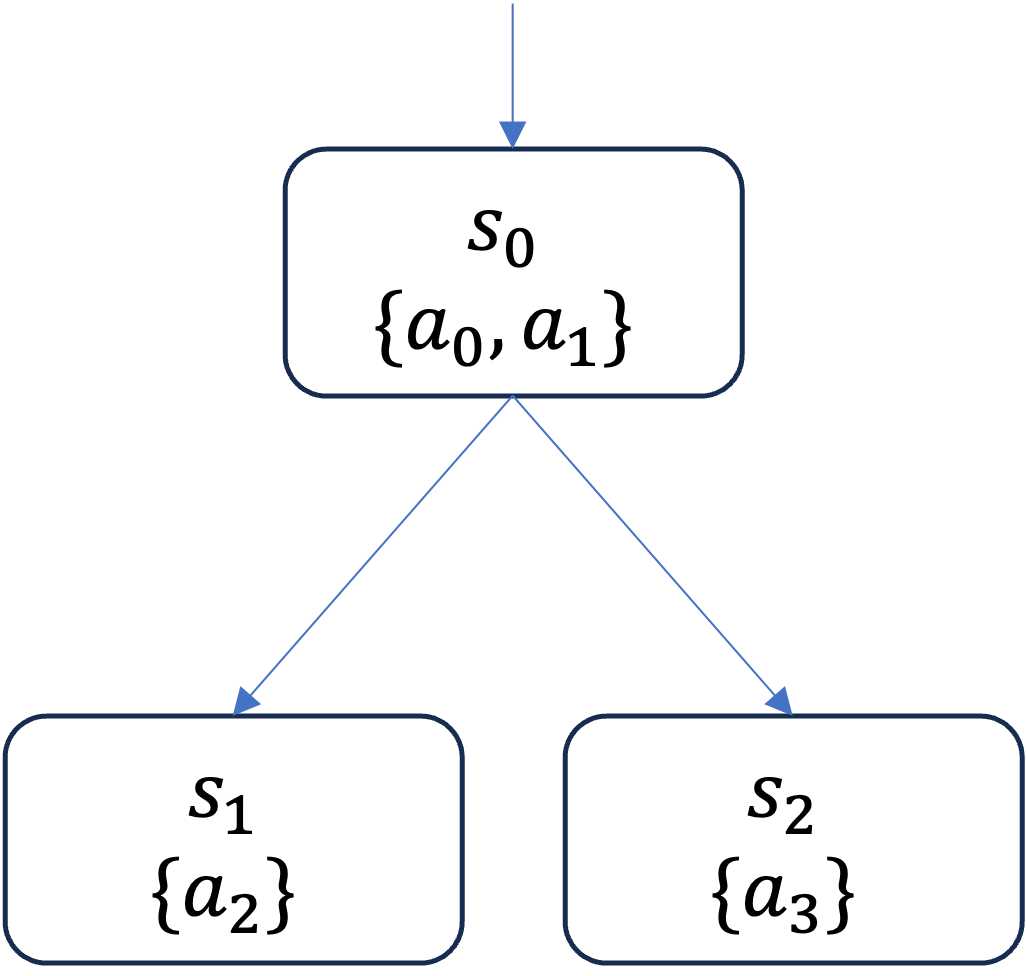}
        \caption[Kripke Structure]{Kripke structure example.}
        \label{fig:ks}
    \end{subfigure}
    \hfill
    \begin{subfigure}[b]{0.45\textwidth}
        \centering
        \includegraphics[width=0.6\textwidth]{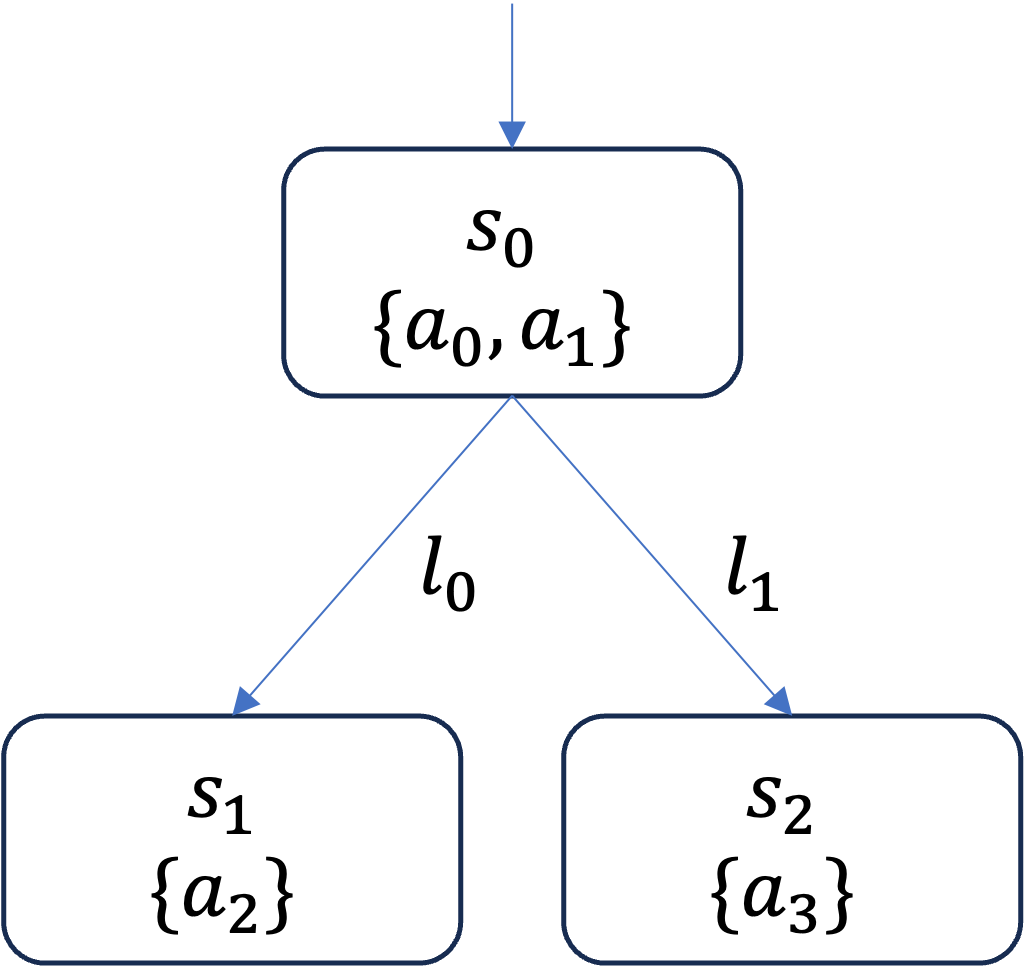}
        \caption[Labeled Kripke Structure]{Labeled Kripke structure example.}
        \label{fig:lks}
    \end{subfigure}
     
    \caption[Kripke Structure Examples]{Examples of Kripke structures.}
    \label{fig:ks_examples}
\end{figure}

\begin{example}[Kripke Structure]
    Figure~\ref{fig:ks} illustrates a Kripke structure over $A = \braces{a_0, a_1, a_2, a_3}$, where $S = \braces{s_0, s_1, s_2}$, $\leftindex^0{S} = \braces{s_0}$, ${\to} = \braces{(s_0, s_1), (s_0, s_2)}$, and $\mathcal{L}$ relates a set atomic propositions in curly brackets to a corresponding state, such as $\mathcal{L}(s_0) = \braces{a_0,a_1}$.
\end{example}

Combining the characteristics of LTSs and KSs, labeled Kripke structures, formally defined in Definition~\ref{def:lks}, have both labels over transitions and atomic propositions assigned to states.

\begin{definition}[Labeled Kripke Structure]
\label{def:lks}
    A \textbf{Labeled Kripke Structure} (LKS) over set $A$ of atomic propositions is a quintuple
    \[
        \anglebrackets{S, \leftindex^0{S}, L, {\to}, \mathcal{L}}
    \]
    \begin{itemize}
        \item $S$ is a set of states,
        \item $\leftindex^0{S} \subseteq S$ is a set of initial states,
        \item $L$ is a set of labels,
        \item ${\to} \subseteq S \times L \times S$ is a transition relation,
        \item $\mathcal{L}: S \mapsto \wp(A)$ is a total labeling function.
    \end{itemize}

    A LKS is terminal if it contains a \textbf{terminal state}\index{Terminal State}, i.e., $\exists s \in S: \mathcal{S}(s) = \emptyset$.
\end{definition}

\begin{example}[Labeled Kripke Structure]
    An example of a LKS over $A = \braces{a_0, a_1, a_2, a_3}$ is visualized in Figure~\ref{fig:lks}. It retain the same sets $S$, $\leftindex^0{S}$, and function $\mathcal{L}$ as defined in Example~\ref{def:ks}, while adopting the $L$ and ${\to}$ from Example~\ref{def:lts}.

    This LKS models interactive state changes governed by $L$ in a system. When $l_0$ is activated, the system transitions from the initial state $s_0$ to $s_1$. Similarly, the activation of $l_1$ results in a transition from $s_0$ to $s_2$. 
\end{example}

\subsubsection{Rewriting System}

\paragraph{Basic Rewriting System}

An abstract rewriting system, formally defined in Definition~\ref{def:ars}, is a model of computation to specify structures and their transformation rules.

\begin{definition}[Abstract Rewriting System]
\label{def:ars}
An \textbf{Abstract Rewriting System} (ARS) is a pair
\[
\anglebrackets{S, {\to}}
\]
\begin{itemize}
    \item $S$ is a set of structures, and
    \item ${\to} \subseteq S \times S$ is a transformation relation containing a set of rules.
\end{itemize}
\end{definition}

We can observe that Definition~\ref{def:ars} of ARSs aligns with Definition~\ref{def:ts} of TSs, despite differing terminologies. While TSs emphasize specific state transitions, ARSs focus on the transformation rules governing transitions.

\begin{figure}[ht]
     \centering
     \begin{subfigure}[b]{0.45\textwidth}
         \centering
         \includegraphics[width=0.5\textwidth]{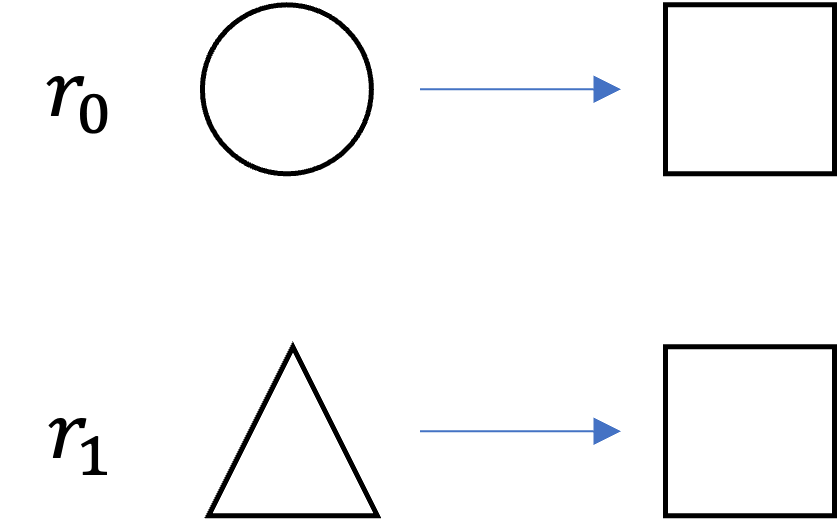}
         \caption[Abstract Rewriting System]{Abstract rewriting system example.}
         \label{fig:ars}
     \end{subfigure}
     \hfill
    \begin{subfigure}[b]{0.45\textwidth}
        \centering
        \includegraphics[width=0.8\textwidth]{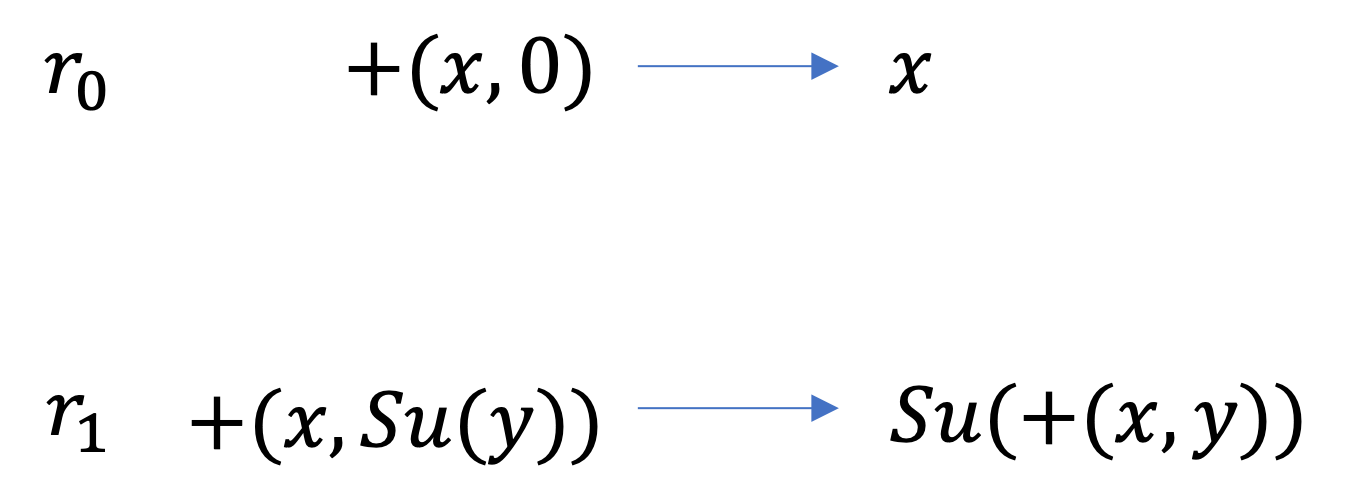}
        \caption[Term Rewriting System]{Term rewriting system example.}
        \label{fig:trs}
    \end{subfigure}

    \caption[Rewriting System Examples]{Examples of rewriting systems.}
    \label{fig:rs_examples}
\end{figure}

\begin{example}[Abstract Rewriting System]
    An example of an ARS is shown in Figure~\ref{fig:ars}, where $S$ is a set of objects $\braces{\circ, \square, \triangle}$ and ${\to} = \braces{r_0, r_1}$. Rule $r_0$ transforms $\circ$ into $\square$, while rule $r_1$ transforms $\triangle$ to $\square$.
\end{example}

Theoretically, ARS structures can represent various mathematical objects, including terms and graphs. A term rewriting system and a graph rewriting system are defined in Definition~\ref{def:trs} and Definition~\ref{def:grs} respectively.

\begin{definition}[Term Rewriting System]
\label{def:trs}
    A \textbf{Term Rewriting System} (TRS) over set $F$ of function symbols and set $V$ of variables, where $F \cap V = \emptyset$, is a pair
    \[
    \anglebrackets{S, {\to}}
    \]
    \begin{itemize}
        \item $S \subseteq \textit{Term}(F, V)$ is a set of terms defined inductively:
        \begin{itemize}
            \item $V \subseteq \textit{Term}(F, V)$,
            \item $\forall f \in F^i: f(t_1, \dots, t_i) \in \textit{Term}(F, V)$, where $F^i, i \in \mathbb{N}$ is the set of $i$-ary function symbols, and $t_1, \dots, t_i \in \textit{Term}(F, V)$.
        \end{itemize}
        \item ${\to} \subseteq S \times S$ is a transformation relation.
    \end{itemize}
\end{definition}

\begin{example}[Term Rewriting System]
\label{eg:trs}
    Figure~\ref{def:trs} depicts an example of a TRS over $F=\braces{+, 0, \textit{Su}}$ and $V = \braces{x, y}$, where $S = \braces{+(x,0), x, +(x, \textit{Su}(y)}$, and ${\to} = \braces{r_0, r_1} = \braces{(+(x,0), x), (+(x, \textit{Su}(y)), \textit{Su}(+(x, y)))}$.

    This TRS models the binary addition function of natural numbers. For instance, the initial term $+(\textit{Su}(0), \textit{Su}(0))$ is utilized to compute $1+1$, where $\textit{Su}(0)$ represents the natural number $1$ by the unary successor function $\textit{Su}$. The computational (rewriting) process involves sequentially applying the transformation rules defined in ${\to}$ to the initial term, as exemplified in the computation of $1+1$ as follows.
    \[
        +(\textit{Su}(0), \textit{Su}(0)) \xrightarrow[r_1]{} \textit{Su}(+(\textit{Su}(0), 0))
        \xrightarrow[r_0]{} \textit{Su}(\textit{Su}(0))
    \]
    The process terminates when no further rules in ${\to}$ can be applied to the term. In this case, the process terminates with the term $\textit{Su}(\textit{Su}(0))$ representing the natural number $2$.
\end{example}

\paragraph{Graph Rewriting System}

Terms can be represented by node-labeled graphs defined in Definition~\ref{def:nlg}.

\begin{definition}[Node-Labeled Graph]
\label{def:nlg}
    A \textbf{Node-Labeled Graph} (NLG) over set $F$ of function symbols and set $V$ of variables, where $F \cap V = \emptyset$ is a quadruple
    \[
    \anglebrackets{N, \leftindex^0{n}, \mathcal{L}, \mathcal{S}}
    \]
    \begin{itemize}
        \item $N$ is a set of nodes,
        \item $\leftindex^0{n} \in N$ is the root,
        \item $\mathcal{L}: N \mapsto F \cup V$ is a total labeling function, and
        \item $\mathcal{S}: N \mapsto N^*$ is a partial successor function well-defined for all $n \in N$ such that $\mathcal{L}(n) \in F$, where $N^*$ is the Kleene closure of $N$.
    \end{itemize}

    A NLG is \textbf{closed}\index{Closed Node-Labeled Graph} if $V = \emptyset$ or $\forall n \in N: \mathcal{L}(n) \in F$. Thus, a \textbf{Closed NLG} (CNLG) is a quadruple
    \[
    \anglebrackets{N, \leftindex^0{n}, \mathcal{L}, \mathcal{S}}
    \]
    \begin{itemize}
        \item $N$ is a set of nodes,
        \item $\leftindex^0{n}$ is the root,
        \item $\mathcal{L}: N \mapsto F$ is a total labeling function, and
        \item $\mathcal{S}: N \mapsto N^*$ is a total successor function.
    \end{itemize}

    A \textbf{subgraph} $g/\leftindex^0{n}'$ of a NLG $g$ rooted at $\leftindex^0{n}'$ is a NLG 
    \[
    \anglebrackets{N', \leftindex^0{n}', \mathcal{L}', \mathcal{S}'}
    \]
    \begin{itemize}
        \item $N' \subseteq N$ is a set of nodes reachable by paths from $\leftindex^0{n}'$,
        \item $\leftindex^0{n}'$ is the root,
        \item $\mathcal{L}: N' \mapsto F \cup V$ is a total labeling function, and
        \item $\mathcal{S}: N' \mapsto N'^*$ is a partial successor function well-defined for all $n \in N'$ such that $\mathcal{L}(n) \in F$.
    \end{itemize}

    A node $n$ is a \textbf{function node}\index{Function Node} if $\mathcal{L}(n) \in F$ and a \textbf{variable node}\index{Variable Node} if $\mathcal{L}(n) \in V$. For any NLG $g$, $\textit{Function}(g) = \braces{\mathcal{L}(n) \in F \mid n \in N}$, $\textit{Variable}(g) = \braces{\mathcal{L}(n) \in V \mid n \in N}$.
\end{definition}

This paper uses a simple language to describe NLGs, as detailed in Definition~\ref{def:language_nlg}.

\begin{definition}[Language for Node-Labeled Graphs]
\label{def:language_nlg}
    The \textbf{Extended Backus-Naur Form} (EBNF)\footnote{\url{https://www.iso.org/standard/26153.html}} standardized in ISO/IEC 14977 of the language for NLGs is defined as follows.
\begin{align*}
    \textit{NLG} &= [\text{"\textasciicircum"}], \textit{Node} \mid \textit{Node}, \text{"$\oplus$"}, \textit{NLG} \\
    \textit{Node} &= \textit{Function}, \text{"("}, \braces{\textit{Node}}, \text{")"} \mid \textit{ID} \mid \textit{ID}, \text{"["}, \textit{Function}, \text{"("}, \braces{\textit{Node}}, \text{")"}, \text{"]"}
\end{align*}
Here, \text{\textasciicircum} identifies a node as the root and must occur exactly once. $\oplus$ denotes a graph concatenation operation. $\textit{Function}$ ranges over $F$, and $\textit{ID}$ ranges over a set of identifier symbols such that $F \cap \textit{ID} = \emptyset$ and $V \subseteq \textit{ID}$. Besides, for any identifier symbol appearing in a graph, this symbol must occur exactly once in the form $\textit{ID}, \text{"["}, \textit{Function}, \text{"("}, \braces{\textit{Node}}, \text{")"}, \text{"]"}$. For brevity, a nullary ($0$-ary) function $\textit{Function}()$ is abbreviated as $\textit{Function}$.
\end{definition}

\begin{example}[Terms as Graphs]
    The term $+(\textit{Su}(0), \textit{Su}(0))$ in Example~\ref{eg:trs} can be represented as a NLG defined in Definition~\ref{def:nlg} over $F$ and $V$, where $N = \braces{\leftindex^0{n}, n_1, n_2, n_3, n_4}$, $n^0 = \leftindex^0{n}$, $\mathcal{L} = \braces{(\leftindex^0{n}, +), (n_1, \textit{Su}), (n_2, \textit{Su}), (n_3, 0), (n_4, 0)}$, and $\mathcal{S} = \braces{(\leftindex^0{n}, \braces{n_1, n_2}), (n_1, \braces{n_3}), (n_2, \braces{n_4})}$. By Definition~\ref{def:language_nlg}, we can describe the NLG by $\hat{+}(\textit{Su}(0()), \textit{Su}(0()))$, abbreviated to $\hat{+}(\textit{Su}(0), \textit{Su}(0))$ and visualized in Figure~\ref{fig:nlg_0}. Notably, this is to be distinguished from $\hat{+}(1[\textit{Su}(0)], 1)$, as visualized in Figure~\ref{fig:nlg_1}.

    \begin{figure}[ht]
     \centering
     \begin{subfigure}[b]{0.45\textwidth}
         \centering
         \includegraphics[width=0.5\textwidth]{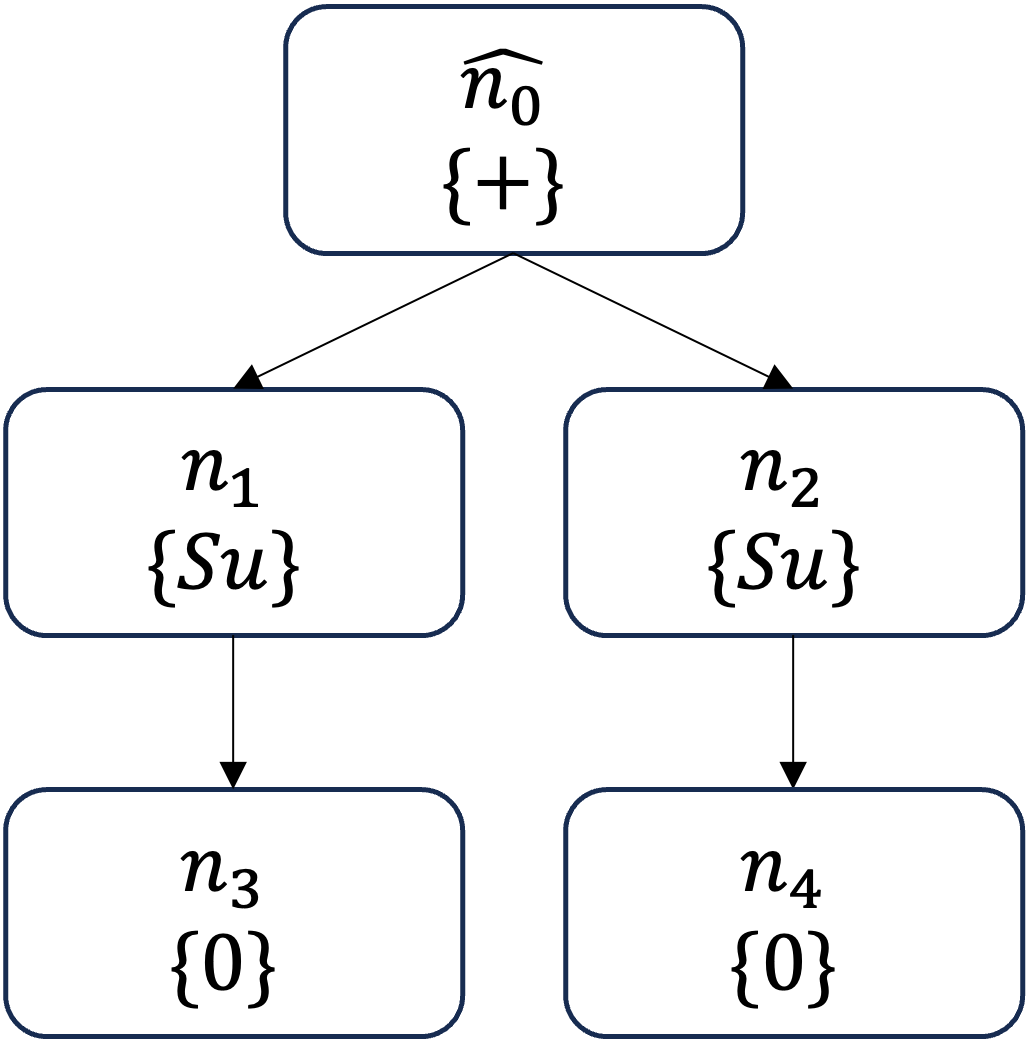}
         \caption{Node-labeled graph $\hat{+}(\textit{Su}(0), \textit{Su}(0))$.}
         \label{fig:nlg_0}
     \end{subfigure}
     \hfill
    \begin{subfigure}[b]{0.45\textwidth}
        \centering
        \includegraphics[width=0.3\textwidth]{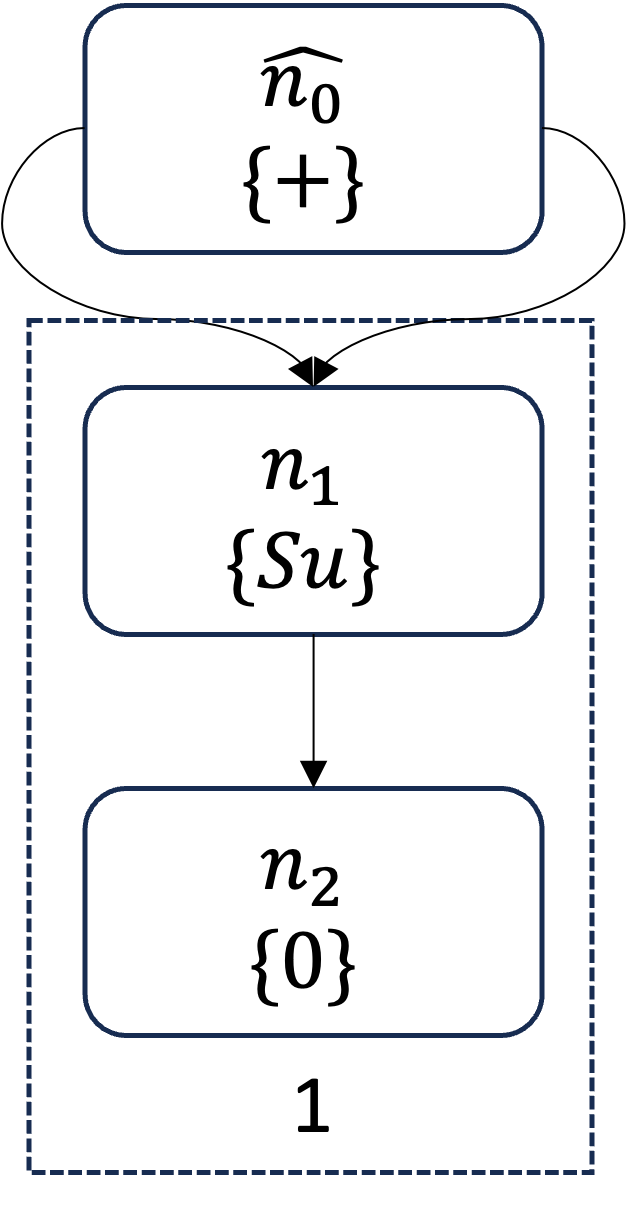}
        \caption{Node-labeled graph $\hat{+}(1[\textit{Su}(0)], 1)$.}
        \label{fig:nlg_1}
    \end{subfigure}
    
    \caption[Node-Labeled Graph Examples]{Examples of node-labeled graphs.}
    \label{fig:ngl_examples}
\end{figure}

\end{example}

\begin{definition}[Node-Labeled Graph Homomorphism]
\label{def:nlg_homomorphism}
    A \textbf{node-labeled graph homomorphism} $g_0 \leadsto g_1$ from a NLG $g_0 = \anglebrackets{N_0, \leftindex^0{n}_0, \mathcal{L}_0, \mathcal{S}_0}$ to $g_1 = \anglebrackets{N_1, \leftindex^0{n}_1, \mathcal{L}_1, \mathcal{S}_1}$ over $F$ and $V$, where $F \cap V = \emptyset$, is a function $\mathcal{H}: N_0 \mapsto N_1$, such that $\mathcal{H}(\leftindex^0{n}_0) = \leftindex^0{n}_1$, and for all $n \in N_0, \mathcal{L}_0(n) \in F$,
    \begin{itemize}
        \item $\mathcal{L}_1(\mathcal{H}(n)) = \mathcal{L}_0(n)$, and
        \item For each $i$-th element $n_i \triangleq \mathcal{S}_0(n)[i]$ in $\mathcal{S}_0(n)$, $\mathcal{H}(n_i) = \mathcal{S}_1(\mathcal{H}(n))[i]$, where $1 \leq i \leq |\mathcal{S}_0(n)|$.
    \end{itemize}
\end{definition}

\begin{definition}[Graph Rewriting System]
\label{def:grs}
    A \textbf{Graph Rewriting System} (GRS) over set $F$ of function symbols and set $V$ of variables, where $F \cap V = \emptyset$, is a pair
    \[
    \anglebrackets{S, {\to}}
    \]
    \begin{itemize}
        \item $S \subseteq \textit{NLG}(F, V)$ is a set of NLGs over $F$ and $V$ as per Definition~\ref{def:nlg}, and
        \item ${\to} \subseteq S \times N$ is a transformation relation, such that $\forall s \in S: (s, n) \in {\to} \implies n \in N \land n \neq \leftindex^0{n}$, where $s = \anglebrackets{N, \leftindex^0{n}, \mathcal{L}, \mathcal{S}}$.
    \end{itemize}

    A transformation rule $(s, n) \in {\to}$ is applicable to a graph $g$ if $s/\leftindex^0{n} \leadsto g$, as per Definition~\ref{def:nlg_homomorphism}, with $n$ the \textbf{post-transformation root}\index{Post-Transformation Root}.
\end{definition}

\begin{definition}[NLG Rewriting]
\label{def:nlg_rewriting}
$\textit{NLGRewrite}: G \times \tilde{G} \times \tilde{N} \mapsto G$ is a rewriting function, where $G, \tilde{G} \subseteq \textit{NLG}(F, V)$, $\forall \tilde{g} \in \tilde{G}, \tilde{g} = \anglebrackets{\tilde{N}, \tilde{\mathcal{L}}, \tilde{\leftindex^0{n}}, \tilde{\mathcal{S}}}: (\tilde{g}, \tilde{n}) \in \tilde{G} \times \tilde{N} \implies \tilde{n} \in \tilde{N} \land \tilde{n} \neq \tilde{\leftindex^0{n}}$, and $\forall g \in G: \exists (\tilde{g}, \tilde{n}) \in \tilde{G} \times \tilde{N}: \tilde{g}/\tilde{\leftindex^0{n}} \leadsto g$.

Function $\textit{NLGRewrite}(\anglebrackets{N, \mathcal{L}, \leftindex^0{n}, \mathcal{S}}, \anglebrackets{\tilde{N}, \tilde{\mathcal{L}}, \tilde{\leftindex^0{n}}, \tilde{\mathcal{S}}}, \tilde{n})$ contains three steps as follows.
\begin{enumerate}
    \item Construct $g_1 \triangleq \anglebrackets{N_1, \leftindex^0{n}_1, \mathcal{L}_1, \mathcal{S}_1}$ by adding an isomorphic copy of $\tilde{N}_{\tilde{g}/\tilde{n}} \setminus \tilde{N}_{\tilde{g}/\tilde{\leftindex^0{n}}}$ to $g$.
    \begin{itemize}
        \item $N_1 = N \uplus (\tilde{N}_{\tilde{g}/\tilde{n}} \setminus \tilde{N}_{\tilde{g}/\tilde{\leftindex^0{n}}})$,
        \item $\leftindex^0{n}_1 = \leftindex^0{n}$,
        \item The labeling function
        \begin{align*}
            \mathcal{L}_1(n) &=
            \begin{cases}
              \mathcal{L}(n), & n \in N \\
              \tilde{\mathcal{L}}(n), & n \in \tilde{N}_{\tilde{g}/\tilde{n}} \setminus \tilde{N}_{\tilde{g}/\tilde{\leftindex^0{n}}}
            \end{cases}
        \end{align*}
        \item The successor function
        \begin{align*}
            \mathcal{S}_1(n)[i] &=
            \begin{cases}
              \mathcal{S}(n)[i], & n \in N \\
              \tilde{\mathcal{S}}(n)[i], & n, \tilde{\mathcal{S}}(n)[i] \in \tilde{N}_{\tilde{g}/\tilde{n}} \setminus \tilde{N}_{\tilde{g}/\tilde{\leftindex^0{n}}} \\
              \mathcal{H}(\tilde{\mathcal{S}}(n)[i]), & n \in \tilde{N}_{\tilde{g}/\tilde{n}} \setminus \tilde{N}_{\tilde{g}/\tilde{\leftindex^0{n}}}, \tilde{\mathcal{S}}(n)[i] \in \tilde{N}_{\tilde{g}/\tilde{\leftindex^0{n}}}
            \end{cases}
        \end{align*}
    \end{itemize}
    \item Construct $g_2 \triangleq \anglebrackets{N_2, \leftindex^0{n}_2, \mathcal{L}_2, \mathcal{S}_2}$ by substituting all references to $\mathcal{H}(\tilde{\leftindex^0{n}})$ with references to $\tilde{n}$ in $g_1$.
    \begin{itemize}
        \item $N_2 = N_1$,
        \item 
        \begin{align*}
            \leftindex^0{n}_2 &=
            \begin{cases}
                \tilde{n}, & \leftindex^0{n}_1 = \mathcal{H}(\tilde{\leftindex^0{n}}) \\
                \leftindex^0{n}_1, & \text{otherwise}
            \end{cases}
        \end{align*}
        \item $\mathcal{L}_2 = \mathcal{L}_1$,
        \item 
        \begin{align*}
            \mathcal{S}_2(n)[i] &=
            \begin{cases}
                \tilde{n}, & \mathcal{S}_1(n)[i] = \mathcal{H}(\tilde{\leftindex^0{n}}) \\
                \mathcal{S}_1(n)[i], & \text{otherwise}
            \end{cases}
        \end{align*}
    \end{itemize}
    \item Construct $g_3 = g_2/\leftindex^0{n}_2$ by deriving the subgraph of $g_2$ rooted at $\leftindex^0{n}_2$, as per Definition~\ref{def:nlg}.
\end{enumerate}
\end{definition}

\begin{example}[Graph Rewriting System]
\label{eg:grs}
An example of a graph rewriting system functioning the same as the TRS in Example~\ref{eg:trs} is depicted in Figure~\ref{fig:grs}, where $S$ consist of two NLGs $s_0 = \hat{+}(x, 0)$ and $s_1 = \hat{+}(x, \textit{Su}(y)) \oplus \textit{Su}(+(x, y))$, and ${\to} = \braces{(s_0, n^{0\prime}_0), (s_1, n^{1\prime}_0)}$.

\begin{figure}[htbp!]
\centering
\includegraphics[width=0.7\textwidth]{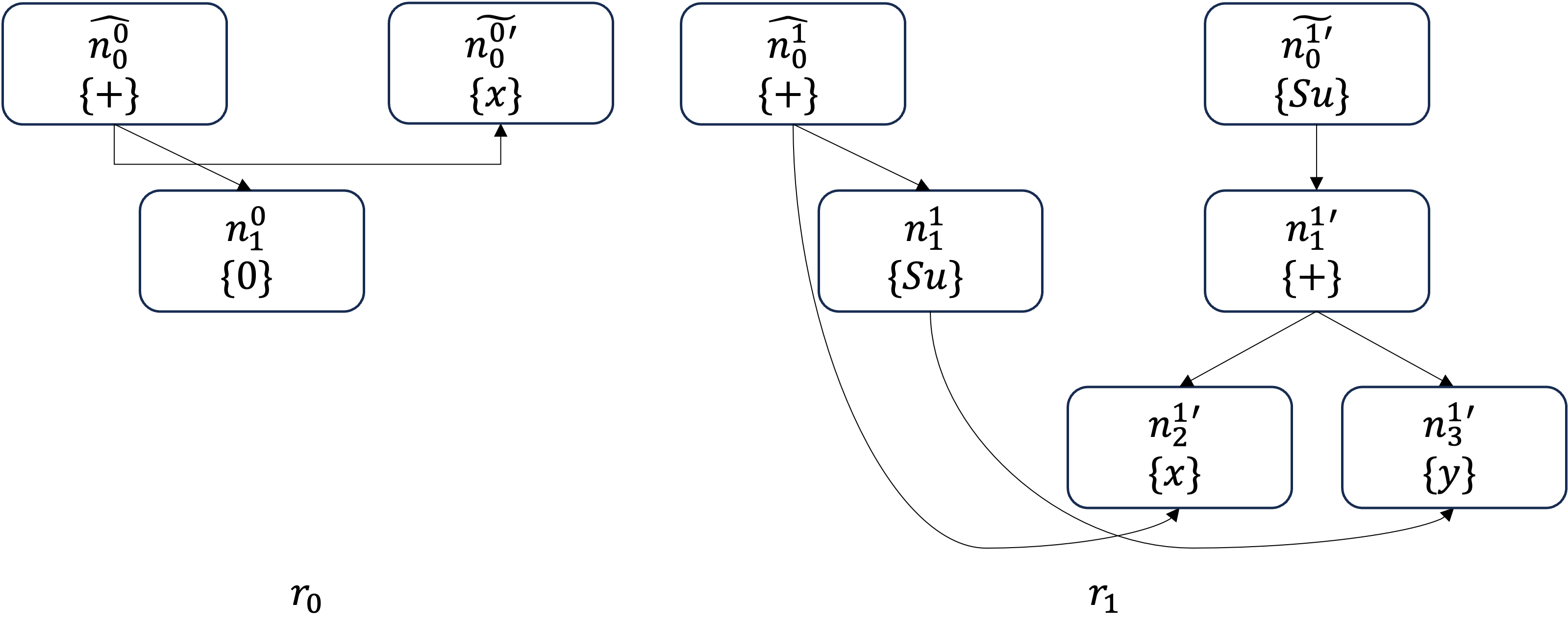}
\caption[Graph Rewriting System Example]{Graph rewriting system example.}
\label{fig:grs}
\end{figure}

Given an initial graph $g_0 = \hat{+}(\textit{Su}(0), \textit{Su}(0))$, the rewriting process sequentially applies transformation rules to $g_0$ if there is a homomorphism from the NLG of a rule to $g_0$. The process continues until no further homomorphisms are possible. For instance, by function $\textit{NLGRewrite}$ Definition~\ref{def:nlg_rewriting}, $r_1$ is first applied to $g_0$ and produces $g_1 = \hat{\textit{Su}}(+(\textit{Su}(0), 0))$ due to the homomorphism $s_1/n^1_0 \leadsto g_0$ as per Definition~\ref{def:nlg_homomorphism}. Then applying $r_0$ transforms $g_1$ to $g_2 = \hat{\textit{Su}}(\textit{Su}(0))$ due to $s_0/n^0_0 \leadsto g_1$. As no more homomorphisms exist, the process terminates with $g_2$ representing the natural number $2$.

\end{example}

Notably, graphs of GRSs can inherently represent shared and cyclic structures, where a node may have multiple predecessors. In contrast, TRSs, with their tree-like structures, cannot naturally represent a shared structure.

\subsubsection{Incomparability}
\label{sec:incomparability}

Transition systems and rewriting systems have their own advantages in abstraction. Transition systems are powerful in expressing stateful and temporal dynamics to define how a system operates step by step, while rewriting systems excel in expressing functional structure transformations to define what a system operates. 

In Example~\ref{eg:variable_assignment}, transition systems prove to be convenient in modeling stateful dynamics.

\begin{example}[Variable Assignment]
\label{eg:variable_assignment}
Value assignment is a common operation in imperative programming languages. For an operation where a variable $x$ is assigned the value of $x+1$, a transition system offers a convenient and intuitive representation through the transition relation $x \to x + 1$.

In contrast, the representation of this operation in a rewriting system introduces complexity, particularly in terms of encoding state context information to control termination. The complexity becomes increasingly notable with the escalation in the number of variables and the application of system compositions.

\end{example}

However, rewriting systems are convenient in modeling functional structure transformations, as shown in Example~\ref{eg:y_combinator}.

\begin{example}[Y Combinator]
\label{eg:y_combinator}
    The Y combinator is a higher-order function to implement recursion in languages without native support for recursive functions. In untyped lambda calculus, the Y combinator is typically represented as $\lambda f.(\lambda x. f(x~x))~(\lambda x. f(x~x))$.

    While formulating a transition system to model the Y combinator in a succinct way is challenging due to its potential to generate an infinite number of states, a TRS $\anglebrackets{S, {\to}}$ over $F = \braces{Y}$ and $V = \braces{f}$, as defined in Definition~\ref{def:trs}, can succinctly model the Y combinator, where $S = \braces{Y(f), f(Y(f))}, {\to} = \braces{(Y(f), f(Y(f))}$.
\end{example}

Therefore, transition systems are widely used in imperative programming (e.g., how a variable value changes from state to state in Java), while rewriting systems are widely used in functional programming (e.g., what a pure function does in Haskell).

A program can be modeled in different formal models depending on contexts, as discussed in Section~\ref{sec:incomparability}. Even with the same formal model, a program can be modeled in different ways. Some models are concrete with many details, while some models are rather abstract with a specific focus. For example, at the syntactic level, a program can be modeled by a parse tree (concrete model) or an abstract syntax tree (abstract model).

\subsection{Formal Logic}
\label{sec:formal_logic}

Formal logic is important in specifying system properties. Although formal models introduced in Section~\ref{sec:formal_model} can be used to specify properties, formal logic is more expressive and succinct. This section presents two representative formal logic: temporal logic and Hoare logic.

\subsubsection{Temporal Logic}
\label{sec:temporal_logic}

Temporal logic formulates temporal properties to reason about stateful dynamics regarding linear or branching time, particularly for transition systems.

\paragraph{Linear-Time Logic}

\begin{definition}[Linear Temporal Logic]
\label{def:ltl}
\textbf{Linear Temporal Logic} (LTL) is a linear-time temporal logic. The syntax of LTL formulae over set $A$ of atomic propositions is defined as
\[
\phi ::= \top \mid a \mid \phi_0 \land \phi_1 \mid \neg \phi \mid \bigcirc \phi \mid \phi_0 \mathsf{U} \phi_1,
\]
where $a \in A$.

Temporal modalities $\Diamond$ and $\Box$ are derived from $\mathsf{U}$ and defined as follows.
\begin{align*}
    \Diamond \phi &\triangleq \top \mathsf{U} \phi\\
    \Box \phi &\triangleq \neg \Diamond \neg \phi
\end{align*}

\end{definition}

\paragraph{Branching-Time Logic}

\begin{definition}[Computation Tree Logic]
\label{def:ctl}
    \textbf{Computation Tree Logic} (CTL) is a branching-time logic. The syntax of CTL formulae over set $A$ of atomic propositions is defined as
    \[
    \phi ::= \top \mid a \mid \phi_0 \land \phi_1 \mid \neg \phi \mid \exists \varphi \mid \forall \varphi
    \]
    where $a \in A$ and $\varphi$ is formed by
    \[
    \varphi ::= \bigcirc \phi \mid \phi_0 \mathsf{U} \phi_1
    \]

    Temporal modalities $\Diamond$ and $\Box$ are defined as follows.
    \begin{align*}
    \exists \Diamond \phi &\triangleq \exists(\top \mathsf{U} \phi)\\
    \forall \Diamond \phi &\triangleq \forall(\top \mathsf{U} \phi)\\
    \exists \Box \phi &\triangleq \neg \forall \Diamond \neg \phi\\
    \forall \Box \phi &\triangleq \neg \exists \Diamond \neg \phi\\
\end{align*}

The release operator $\mathsf{R}$ is defined as follows:
\begin{align*}
    \exists(\phi_0 \mathsf{R} \phi_1) = \neg \forall((\neg \phi_0) \mathsf{U} (\neg \phi_1))\\
    \forall(\phi_0 \mathsf{R} \phi_1) = \neg \exists((\neg \phi_0) \mathsf{U} (\neg \phi_1))\\
\end{align*}
\end{definition}

\begin{definition}[Universal Fragment of Computation Tree Logic]
\label{def:uctl}
    The \textbf{universal fragment} of CTL (UCTL) over set $A$ of atomic propositions is a fragment of CTL in Definition~\ref{def:ctl} without existential path quantifiers. The syntax of UCTL formulae over set $A$ of atomic propositions is defined as
    \[
    \phi ::= \top \mid \bot \mid a \mid \neg a \mid \phi_0 \land \phi_1 \mid \phi_0 \lor \phi_1 \mid \forall \varphi
    \]
    where $a \in A$ and $\varphi$ is formed by
    \[
    \varphi ::= \bigcirc \phi \mid \phi_0 \mathsf{U} \phi_1 \mid \phi_0 \mathsf{R} \phi_1
    \]
\end{definition}

\begin{definition}[Existential Fragment of Computation Tree Logic]
\label{def:ectl}
    The \textbf{existential fragment} of CTL (ECTL) over set $A$ of atomic propositions is a fragment of CTL in Definition~\ref{def:ctl} without universal path quantifiers. The syntax of ECTL formulae over set $A$ of atomic propositions is defined as
    \[
    \phi ::= \top \mid \bot \mid a \mid \neg a \mid \phi_0 \land \phi_1 \mid \phi_0 \lor \phi_1 \mid \exists \varphi
    \]
    where $a \in A$ and $\varphi$ is formed by
    \[
    \varphi ::= \bigcirc \phi \mid \phi_0 \mathsf{U} \phi_1 \mid \phi_0 \mathsf{R} \phi_1
    \]
\end{definition}

\subsubsection{Equational Logic}

Equational logic formulates functional properties to reason about structural transformations via equations, particularly for rewriting systems.

\begin{definition}[Equational Logic]
\label{def:el}
    \textbf{Equational logic} over set $F$ of function symbols and set $V$ of variables is a fragment of first-order logic (FOL) where formulae are formed by equations between structures (e.g., $\textit{Term}(F, V)$ in Definition~\ref{def:trs}, $\textit{NLG}(F, V)$ in Definition~\ref{def:grs}). An \textbf{equation}\index{Equation} is an ordered pair of structures denoted in the form of $s_0 \approx s_1$. The syllogisms are defined below.

    \begin{mathpar}
        \infer{s \approx s}{} \quad \textbf{Reflexivity}

        \infer{s_1 \approx s_0}{s_0 \approx s_1} \quad \textbf{Symmetry}
        
        \infer{s_0 \approx s_2}{s_0 \approx s_1, s_1 \approx s_2} \quad \textbf{Transitivity}
    \end{mathpar}
    \begin{mathpar}
        \infer{s[v := s_0] \approx s[v := s_1]}{s_0 \approx s_1} \quad \textbf{Leibniz}

        \infer{f(s_0,\dots,s_n) \approx f(s_0',\dots,s_n')}{s_0 \approx s_0', \dots, s_n \approx s_n'} \quad \textbf{Congruence}
    \end{mathpar}

    Here, $s, s_0, \dots, s_n \in \textit{Term}/\textit{NLG}(F, V), f \in F, v \in V$, where $n \geq 0$. The notation $s[v := s_0]$ denotes substituting $s_0$ for all occurrences of $v$ in $s$.
\end{definition}

\subsection{Formal Correctness}

\subsubsection{Formalization}

As presented in Section~\ref{sec:formal_model} and Section~\ref{sec:formal_logic}, we can use formal models to abstract programs, while using formal logic to formalize their properties. By default, this paper uses a \textit{model} to denote an abstract representation of a program in a formal model and uses a \textit{property} to denote a model property described in a formal logic interpretable in the context of the model. Formally, correctness is defined in Definition~\ref{def:correctness}.

\begin{definition}[Correctness]
\label{def:correctness}
\textbf{Correctness}\index{Correctness} is a satisfaction relation $\models$ between a model $\mathfrak{M}$ and a property $\phi$.

$\mathfrak{M}$ is correct with respect to $\phi$ if and only if $\mathfrak{M} \models \phi$ holds true.
\end{definition}

\textbf{Q1} illustrated in Section~\ref{sec:introduction} is to formulate a satisfaction relation defined in Definition~\ref{def:correctness} with a formal model and a formal logic.

\subsubsection{Correctness Operation}

From Definition~\ref{def:correctness}, two operations associated with correctness are derived.

\begin{definition}[Correctness Verification]
\label{def:correctness_verification}
A \textbf{correctness verification}, denoted as $\leftindex^V{\textit{Corr}}$, is a partial function that accepts a model $\mathfrak{M}$ and a set of properties $\Phi$. It returns $\top$ if $\forall \phi \in \Phi: \mathfrak{M} \vdash \phi$, and returns $\bot$ if $\exists \phi \in \Phi: \mathfrak{M}, \phi \vdash \bot$ upon termination.
\end{definition}

\begin{proposition}[Correctness Verification]
\label{pro:correctness_verification}
    If there exists a correctness verification $\leftindex^V{\textit{Corr}}$, such that $\leftindex^V{\textit{Corr}}(\mathfrak{M}, \Phi) = \top$, then $\mathfrak{M} \vdash \Phi$.

    \begin{proof}
        From Definition~\ref{def:correctness_verification}, $\leftindex^V{\textit{Corr}}(\mathfrak{M}, \Phi) = \top$ implies $\forall \phi \in \Phi: \mathfrak{M} \vdash \phi$. Therefore, $\mathfrak{M} \vdash \Phi$.
    \end{proof}
\end{proposition}

\begin{definition}[Soundness of Correctness Verification]
\label{def:soundness_correctness_verification}
A correctness verification is \textbf{sound} if $\mathfrak{M} \vdash \Phi \implies \mathfrak{M} \models \Phi$.
\end{definition}

\begin{definition}[Completeness of Correctness Verification]
\label{def:completeness_correctness_verification}
A correctness verification is \textbf{complete} if $\mathfrak{M} \models \Phi \implies \mathfrak{M} \vdash \Phi$.
\end{definition}

\begin{definition}[Correctness Implementation]
\label{def:correctness_implementation}
A \textbf{correctness implementation}, denoted as $\leftindex^I{\textit{Corr}}$, is a partial function that accepts a set of properties $\Phi$ and returns a model $\mathfrak{M}$ upon termination if there exists a correctness verification $\leftindex^V{\textit{Corr}}$, such that $\leftindex^V{\textit{Corr}}(\mathfrak{M}, \Phi) = \top$.
\end{definition}

Correspondingly, \textbf{Q2} illustrated in Section~\ref{sec:introduction} is to formulate a correctness verification defined in Definition~\ref{def:correctness_verification} and \textbf{Q3} is to formulate a correctness implementation defined in Definition~\ref{def:correctness_implementation}.

\subsubsection{Provable Correctness}

Correctness verification and implementation may not exist in some cases. For instance, given that a property $\phi$ specifies a model to decide whether a program with its input terminates, constructing $\mathfrak{M}$ and proving $\mathfrak{M} \models \phi$ is impractical due to the undecidability of the halting problem \cite{cook_complexity_2023}. Consequently, in practice, the focus often shifts to the concept of provable correctness, formally illustrated in Proposition~\ref{pro:provable_correctness}.

\begin{proposition}[Provable Correctness]
\label{pro:provable_correctness}
A set $\Phi$ of properties specifying a model $\mathfrak{M}$ is \textbf{provable}\index{Provable}, if there exists a sound and complete correctness verification $\leftindex^V{\textit{Corr}}$, such that $\leftindex^V{\textit{Corr}}(\mathfrak{M}, \Phi) \in \braces{\top, \bot}$.

\begin{proof}
    By Definition~\ref{def:correctness_verification}, \ref{def:soundness_correctness_verification}, and \ref{def:completeness_correctness_verification}, the computation of $\leftindex^V{\textit{Corr}}(\mathfrak{M}, \Phi)$ is the proof structure for $\Phi$, if $\leftindex^V{\textit{Corr}}(\mathfrak{M}, \Phi)$ terminates.
\end{proof}
\end{proposition}

Different from Proposition~\ref{pro:provable_correctness} that requires both soundness and completeness for correctness verification, provably correct models only require soundness, as in Proposition~\ref{pro:provably_correct_model}.

\begin{proposition}[Provably Correct Model]
\label{pro:provably_correct_model}
$\mathfrak{M}$ is a \textbf{provably correct model} with respect to a set $\Phi$ of properties, if there exists a sound correctness verification $\leftindex^V{\textit{Corr}}$, such that $\leftindex^V{\textit{Corr}}(\mathfrak{M}, \Phi) = \top$.

\begin{proof}
    As per Proposition~\ref{pro:correctness_verification}, $\leftindex^V{\textit{Corr}}(\mathfrak{M}, \Phi) = \top$ implies $\mathfrak{M} \vdash \Phi$. By Definition~\ref{def:soundness_correctness_verification}, if $\leftindex^V{\textit{Corr}}$ is sound, then $\mathfrak{M} \vdash \Phi \implies \mathfrak{M} \models \Phi$. Therefore, $\Phi$ holds for $\mathfrak{M}$, i.e., $\mathfrak{M}$ is provably correct.
\end{proof}
\end{proposition}
\section{Transition-Oriented Programming}
\label{sec:top}

\textbf{Transition-Oriented Programming} (TOP)\index{Transition-Oriented!Programming} is a programming paradigm focusing on articulating stateful dynamics and structural transformations of systems through formal models and logics. It leverages an expressive formal model (graph transition system) and formal logics to resolve the ambiguities present in requirement specifications. Besides, TOP integrates intrinsic correctness verification and bridges the gap between verified models and executables with side effects, thus intertwining correctness specification, verification, and implementation.

TOP combines two distinct approaches, bottom-up (modularization) and top-down (progressive specification), to facilitate correctness specification and verification through iterative and incremental processes. Besides, TOP enhances precise specification and efficient verification via context perception, including type refinement and transformation prioritization. Furthermore, TOP promotes automated proof techniques to streamline correctness verification, such as bounded model checking and iterative deepening theorem proving.

This section delves into the theoretical foundations of TOP and its practical implementation, particularly through the Seni language.

\subsection{Graph Transition System}
\label{sec:gts}

\subsubsection{Model}

\begin{definition}[Typed Node-Labeled Graph]
\label{def:tnlg}
    A \textbf{Typed Node-Labeled Graph} (TNLG)\index{Typed Node-Labeled Graph} over set $F$ of typed function symbols and set $V$ of typed variables, where $F \cap V = \emptyset$, is a quadruple $\anglebrackets{N, \leftindex^0{n}, \mathcal{L}, \mathcal{S}}$ paralleling that of a NLG in Definition~\ref{def:nlg} with restrictions applied to $\mathcal{S}$ as follows.
    \begin{itemize}
        \item $\forall n \in N, \mathcal{L}(n) \in F^0: |\mathcal{S}(n)| = 0$, where $F^0 \subseteq F$ is the set of nullary ($0$-ary) function symbols representing constants of \textbf{primitive types}\index{Primitive Type}, including Boolean, Integer, Float, and Character. The \textbf{node value}\index{Node Value} $\textit{Value}(n)$ of $n$ is a value of a primitive type.
        \item $\forall n \in N, \mathcal{L}(n) \in F^i, i > 0: |\mathcal{S}(n)| = i$, where $F^i \subseteq F$ is the set of $i$-ary function symbols containing $i$ typed parameters, such that $\mathcal{S}(n)[j], 0 < j \leq i$ represents the $j$-th typed parameter that is a typed function symbol or variable.
    \end{itemize}
    A \textbf{Closed TNLG} (CTNLG)\index{Closed Typed Node-Labeled Graph} and a \textbf{subgraph} of a TNLG are defined the same in Definition~\ref{def:nlg}.

\end{definition}

\begin{example}[Typed Node-Labeled Graph]
\label{eg:tnlg}

An example of a TNLG is depicted in Figure~\ref{fig:tnlg}, where each node is associated with a typed function symbol, characterizing it as a CTNLG. The binary operator $::$ binds the function symbol on the left side with the type on the right side. For any function symbol $f \in F \setminus F^0$, its type is represented by a sequence delineating the types of its parameters, enclosed within parentheses. For instance, node $\hat{n_0}$ corresponds to a binary function symbol $M$, of which the parameters are two unary function symbols, each taking an integer nullary function symbol as the parameter.

\begin{figure}[ht]
\centering
\includegraphics[width=0.5\textwidth]{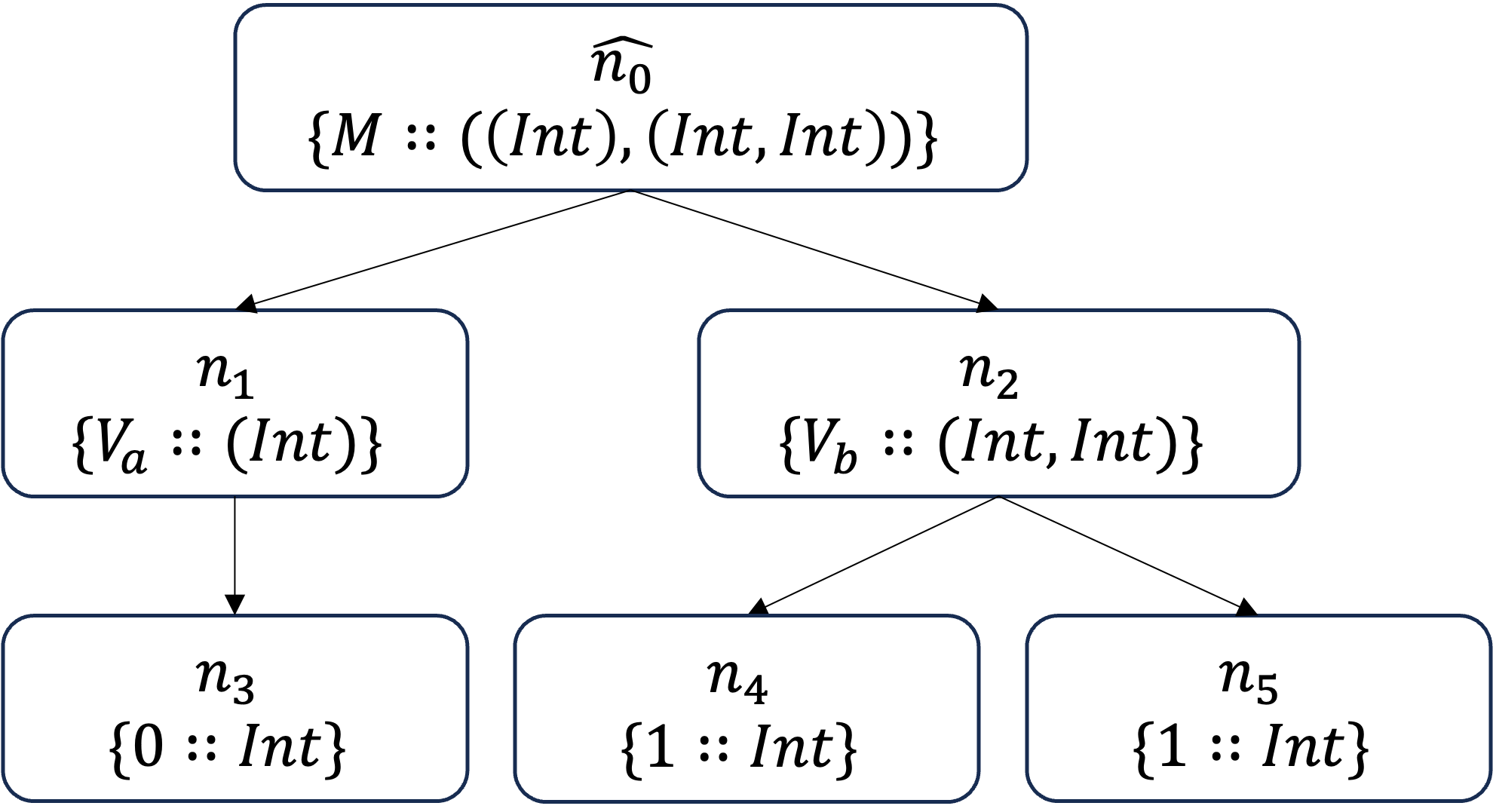}
\caption[Typed Node-Labeled Graph Example]{Typed node-labeled graph example.}
\label{fig:tnlg}
\end{figure}

\end{example}

\begin{definition}[Pattern Graph]
\label{def:pg}
    A \textbf{Pattern Graph} (PG) over set $F$ of typed function symbols and set $V$ of typed variables, where $F \cap V = \emptyset$, is a variant of TNLG as per Definition~\ref{def:tnlg}, defined as a quintuple
    \[
    \anglebrackets{N, \leftindex^0{n}, \mathcal{L}, \mathcal{S}, \mathcal{E}}
    \]
    \begin{itemize}
        \item $N, \leftindex^0{n}, \mathcal{L}, \mathcal{S}$ parallel that of a TNLG in Definition~\ref{def:tnlg}, and
        \item $\mathcal{E}: V \mapsto \textit{Condition}(V) \times \textit{Expression}(V)$ is a partial effect function that relates a condition and an expression to any variable $v$ that satisfies $\exists n, n' \in N, 0 < i \leq |\mathcal{S}(n')|: \mathcal{L}(n) = v \land n = \mathcal{S}(n')[i]$ with the $i$-th parameter of $\mathcal{L}(n')$ a primitive type.
    \end{itemize}

For $\mathcal{E}(v) = (c, e)$, $c$ is a propositional formula that applies value-level constraints to pattern matching, and $e$ is an expression that applies value change to $v$ after pattern matching, constructed using constants, variables and operators. $c$ is a \textbf{tautology}\index{Tautology} if $c = \top$. $e$ is an \textbf{idempotent expression}\index{Idempotent Expression} if $e = v$.
\end{definition}

\begin{example}[Pattern Graph]
\label{eg:pg}
Figure~\ref{fig:pg} illustrates an example of a PG that removes variables with duplicated values for garbage collection. The representation follows the conventions in Example~\ref{eg:tnlg} with additional details related to variable nodes with primitive types by $\mathcal{E}$. For a variable node associated with $v$ and $\mathcal{E}(v) = (c, e)$, the former square bracket represents $c$, and the latter square bracket represents $e$. For instance, $\mathcal{E}(x) = (\top, x + 1)$ for integer variable $x$ related to $n_1$.

\begin{figure}[ht]
\centering
\includegraphics[width=0.9\textwidth]{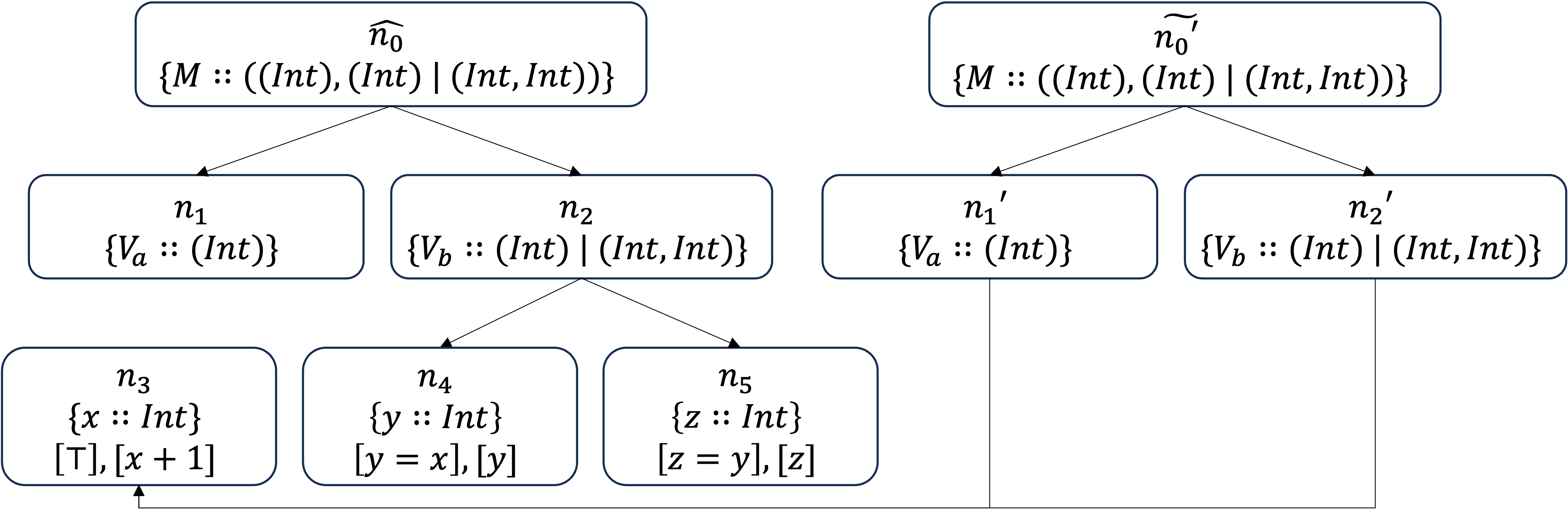}
\caption[Pattern Graph Example]{Pattern graph example.}
\label{fig:pg}
\end{figure}

\end{example}

\begin{definition}[Pattern Graph Homomorphism]
\label{def:pg_homomorphism}
    A \textbf{pattern graph homomorphism} $g_0 \rightsquigarrow g_1$ from a PG $g_0 = \anglebrackets{N_0, \leftindex^0{n}_0, \mathcal{L}_0, \mathcal{S}_0, \mathcal{E}_0}$ to a TNLG $g_1 = \anglebrackets{N_1, \leftindex^0{n}_1, \mathcal{L}_1, \mathcal{S}_1}$ over $F$ and $V$, where $F \cap V = \emptyset$, is a function $\mathcal{H}: N_0 \mapsto N_1$, such that $\mathcal{H}(\leftindex^0{n}_0) = \leftindex^0{n}_1$, and 
    \begin{itemize}
        \item for all $n \in N_0, \mathcal{L}_0(n) \in F$,
        \begin{itemize}
            \item $\mathcal{L}_1(\mathcal{H}(n)) = \mathcal{L}_0(n)$, and
            \item For each $i$-th element $n_i \triangleq \mathcal{S}_0(n)[i]$ in $\mathcal{S}_0(n)$, $\mathcal{S}_1(\mathcal{H}(n))[i] = \mathcal{H}(n_i)$, where $0 < i \leq |\mathcal{S}_0(n)|$;
        \end{itemize}
        \item for all $v \in V, \mathcal{E}_0(v) = (c, e)$, if there exists $n \in N_0, \mathcal{L}_0(n) = v$,
        \begin{itemize}
            \item $\textit{Value}_1 \models c$, i.e., substituting all variables in $c$ with their values from $\textit{Value}_1$ makes $c$ hold true.
        \end{itemize}
    \end{itemize}

\end{definition}

\begin{definition}[Pattern Graph Rewriting]
\label{def:pg_rewriting}
$\textit{PGRewrite}: G \times \tilde{G} \times \tilde{N} \mapsto G$ is a rewriting function, where $G \subseteq \textit{TNLG}(F, \emptyset)$, $\tilde{G} \subseteq \textit{PG}(F, V)$, $\forall \tilde{g} \in \tilde{G}, \tilde{g} = \anglebrackets{\tilde{N}, \tilde{\mathcal{L}}, \tilde{\leftindex^0{n}}, \tilde{\mathcal{S}}, \tilde{\mathcal{E}}}: (\tilde{g}, \tilde{n}) \in \tilde{G} \times \tilde{N} \implies \tilde{n} \in \tilde{N} \land \tilde{n} \neq \tilde{\leftindex^0{n}}$, and $\forall g \in G: \exists (\tilde{g}, \tilde{n}) \in \tilde{G} \times \tilde{N}: \tilde{g}/\tilde{\leftindex^0{n}} \rightsquigarrow g$.

Function $\textit{PGRewrite}$ follows the three steps defined in Definition~\ref{def:nlg_rewriting} with an additional step:
\begin{itemize}
    \item[4.] Construct $g_4 \triangleq \anglebrackets{N_4, \leftindex^0{n}_4, \mathcal{L}_4, \mathcal{S}_4}$ by applying effects defined for all $n \in N_3, \mathcal{L}_3(n) = v \in V, \mathcal{E}_0(v) = (c, e)$, such that $\textit{Value}_4(n) = \textit{Evaluate}(e, \textit{Value}_3)$, where $\textit{Evaluate}(e, \textit{Value}_3)$ evaluates $e$ to a value by substituting variables in $e$ with their values from $\textit{Value}_3$.
    
\end{itemize}
\end{definition}

\begin{definition}[Graph Transition System]
\label{def:gts}
    A \textbf{Graph Transition System} (GTS) over set $F$ of typed function symbols and set $V$ of variables, where $F \cap V = \emptyset$, is a quintuple
    \[
    \anglebrackets{P, \leftindex^0{P}, R, {\hookrightarrow}, \tilde{\leftindex^0{g}}}
    \]
    \begin{itemize}
        \item $P$ is a set of places,
        \item $\leftindex^0{P} \subseteq P$ is a set of initial places,
        \item $R \subseteq \tilde{G} \times \tilde{N}$ is a transformation relation, where $\tilde{G} \subseteq \textit{PG}(F, V)$ is a set of PGs over $F$ and $V$ as per Definition~\ref{def:pg}, and $\forall \tilde{g} \triangleq \anglebrackets{\tilde{N}, \tilde{\mathcal{L}}, \tilde{\leftindex^0{n}}, \tilde{\mathcal{S}}, \tilde{\mathcal{E}}} \in \tilde{G}: (p, \tilde{g}, \tilde{n}, p') \in {\hookrightarrow} \implies \tilde{n} \in \tilde{N} \land \tilde{n} \neq \tilde{\leftindex^0{n}}$,
        \item ${\hookrightarrow} \subseteq P \times R \times P$ is a pattern matching transition relation, and
        \item $\tilde{\leftindex^0{g}} \in \textit{PG}(F, V)$ is the initial pattern.
    \end{itemize}

\end{definition}

\begin{definition}[Labeled Kripke Structure Semantics of a Graph Transition System]
\label{def:lks_gts}
    Let $\mathfrak{G} \triangleq \anglebrackets{P, \leftindex^0{P}, R, {\hookrightarrow}, \tilde{\leftindex^0{g}}}$ be a GTS over $F$ and $V$. The corresponding LKS, denoted as $\textit{LKS}(\mathfrak{G})$, is the tuple $\anglebrackets{S, \leftindex^0{S}, L, {\to}, \mathcal{L}}$ over set $P \cup \textit{PG}(F, V)$ of atomic propositions,
    \begin{itemize}
        \item $S = P \times \textit{TNLG}(F, \emptyset)$,
        \item $\leftindex^0{S} = \braces{(p, g) \mid p \in \leftindex^0{P}, \tilde{\leftindex^0{g}} \rightsquigarrow g}$, where $\rightsquigarrow$ is defined in Definition~\ref{def:pg_homomorphism},
        \item $L = R \subseteq \tilde{G} \times \tilde{N}$,
        \item ${\to} \subseteq S \times L \times S$ is defined by the following rule:
        \[
        \frac{(p \xhookrightarrow{\tilde{g}, \tilde{n}} p') \land (\tilde{g}/\tilde{\leftindex^0{n}} \rightsquigarrow g)}
        {(p, g) \xrightarrow{\tilde{g}, \tilde{n}} (p', \textit{PGRewrite}(g, \tilde{g}, \tilde{n}))}
        \]
        where $\textit{PGRewrite}$ is defined in Definition~\ref{def:pg_rewriting}, and
        \item $\mathcal{L}$ is a labeling function such that $\mathcal{L}((p, g)) = \braces{p} \cup \braces{\tilde{g} \in \tilde{G} \mid \tilde{g}/\tilde{\leftindex^0{n}} \rightsquigarrow g}$.
    \end{itemize}

    A GTS is terminal if its LKS interpretation is terminal.
\end{definition}

\subsubsection{Property}

\paragraph{Temporal Property}
\label{sec:temporal_property}

\begin{definition}[Path]
\label{def:path}
    A \textbf{path} of a GTS interpreted over a LKS $\anglebrackets{S, \leftindex^0{S}, L, {\to}, \mathcal{L}}$ according to Definition~\ref{def:lks_gts} is either
    \begin{itemize}
        \item a finite state sequence $[s_0, s_1, \dots, s_n]$ over $S$, such that $s_0 \in \leftindex^0{S}$, $s_n$ is a terminal state, and $\forall 0 < i \leq n: s_i \in \mathcal{S}(s_{i-1})$, or
        \item an infinite state sequence $[s_0, s_1, \dots]$ over $S$, such that $\forall i > 0: s_i \in \mathcal{S}(s_{i-1})$.
    \end{itemize}
    $\textit{Path}(\mathfrak{G})$ is the set of all paths.
\end{definition}

\begin{definition}[Trace]
\label{def:trace}
    A \textbf{trace}\index{Trace} of a GTS over a LKS $\textit{LKS}(\mathfrak{G}) \triangleq \anglebrackets{S, \leftindex^0{S}, L, {\to}, \mathcal{L}}$ is defined as
    \begin{itemize}
        \item $[\mathcal{L}(s_0), \mathcal{L}(s_1), \dots, \mathcal{L}(s_n)]$ for a finite path $[s_0, s_1, \dots, s_n]$, and
        \item $[\mathcal{L}(s_0), \mathcal{L}(s_1), \dots]$ for an infinite path $[s_0, s_1, \dots]$.
    \end{itemize}

    $\textit{Trace}(\mathfrak{G})$, the traces of $\mathfrak{G}$, is a set of traces defined for $\textit{Path}(\mathfrak{G})$.
\end{definition}

\begin{definition}[Semantics of LTL over GTS]
Given a GTS $\mathfrak{G} \triangleq \anglebrackets{P, \leftindex^0{P}, R, {\hookrightarrow}, \tilde{\leftindex^0{g}}}$ over $F$ and $V$, and an LTL formula $\phi$ formed in Definition~\ref{def:ltl} over $P \cup \textit{TNLG}(F, V)$, $\mathfrak{G} \models \phi$ if and only if $\textit{Trace}(\mathfrak{G}) \subseteq W(\phi)$.

Here, $W(\phi) = \braces{w \in A^{\omega} \mid w \models \phi}$, where $A = \wp(P \cup \textit{TNLG}(F, V))$, $A^{\omega}$ is an $\omega$-language for $\textit{LKS}(\mathfrak{G})$. For each word $w \in  A^{\omega}$, $w$ is an infinite sequence $[A_0, A_1, \dots]$. The satisfaction relation ${\models} \subseteq A^{\omega} \times \textit{Language}(\textit{LTL})$ is the smallest relation defined as follows.
\begin{align*}
    w &\models \top\\
    w &\models a &&\iff a \in A_0\\
    w &\models \phi_0 \land \phi_1 &&\iff w \models \phi_0 \land w \models \phi_1\\
    w &\models \neg \phi &&\iff w \not\models \phi\\
    w &\models \bigcirc \phi &&\iff w[A_1, \dots] \models \phi\\
    w &\models \phi_0 \amalg \phi_1 &&\iff (\exists j \geq 0: w[A_j, \dots] \models \phi_1) \land (\forall 0 \leq i < j: w[A_i, \dots] \models \phi_0)\\
    w &\models \Diamond \phi &&\iff \exists i \geq 0: w[A_i, \dots] \models \phi\\
    w &\models \Box \phi &&\iff \forall i \geq 0: w[A_i, \dots] \models \phi
\end{align*}

\end{definition}

\begin{definition}[Semantics of CTL over GTS]
    Given a GTS $\mathfrak{G}$ over $F$ and $V$, and a CTL formula $\phi$ formed in Definition~\ref{def:ctl} over $P \cup \textit{TNLG}(F, V)$, $\mathfrak{G} \models \phi$ if and only if $\forall s \in \leftindex^{0}S: s \models \phi$ for $\textit{LKS}(\mathfrak{G}) \triangleq \braces{S, \leftindex^0{S}, L, {\to}, \mathcal{L}}$. The satisfaction relation ${\models} \subseteq S \times \textit{Language}(\textit{CTL})$ is the smallest relation defined as follows.
    \begin{align*}
        s &\models \top\\
        s &\models a &&\iff a \in \mathcal{L}(s)\\
        s &\models \neg \phi &&\iff s \not\models \phi\\
        s &\models \phi_0 \land \phi_1 &&\iff s \models \phi_0 \land s \models \phi_1\\
        s &\models \exists \varphi &&\iff \exists \textit{path} \in \textit{Path}(s): \textit{path} \models \varphi\\
        s &\models \forall \varphi &&\iff \forall \textit{path} \in \textit{Path}(s): \textit{path} \models \varphi\\
        \textit{path} &\models \bigcirc \phi &&\iff \textit{path}[1] \models \phi\\
        \textit{path} &\models \phi_0 \amalg \phi_1 &&\iff \exists j \geq 0: \textit{path}[j] \models \phi_1 \land (\forall 0 \leq k < j: \textit{path}[k] \models \phi_0)\\
        \textit{path} &\models \Diamond \phi &&\iff \exists j \geq 0: s_j \models \phi\\
        \textit{path} &\models \Box \phi &&\iff \forall j \geq 0: s_j \models \phi
    \end{align*}
\end{definition}

\paragraph{Functional Property}
\label{sec:functional_property}

\begin{definition}[Structure Path]
\label{def:structure_path}
    A \textbf{structure path} of a GTS is a finite CTNLG sequence $[g_0, g_1, \dots, g_n]$ over $\textit{TNLG}(F, \emptyset)$, such that
    \begin{itemize}
        \item $[(p_0, g_0), (p_1, g_1), \dots, (p_n, g_n)]$ is a path variant of $\textit{Path}(\mathfrak{G})$ without the constraint $(p_0, g_0) \in \leftindex^0{S}$ , and
        \item $\forall 0 < i \leq n: (p_i, g_i) \in \mathcal{S}((p_{i-1}, g_{i-1}))$.
    \end{itemize}
\end{definition}

\begin{definition}[Instantiation of Typed Node-Labeled Graph]
    An \textbf{instantiation}\index{Instantiation} of a TNLG $g$ is a function $\textit{Instance}(g, \mathcal{V})$ that takes $g$ and a total variable evaluation function $\mathcal{V}$ well-defined for $\textit{Variable}(g)$ and produces a CTNLG by assigning concrete values to all typed variables present in $g$ according to $\mathcal{V}$.
\end{definition}

\begin{definition}[Semantics of EL over GTS]
\label{def:el_semantics}
    Given a GTS $\mathfrak{G} = \anglebrackets{P, \leftindex^0{P}, R, {\hookrightarrow}, \tilde{\leftindex^0{g}}}$ and an EL formula $\leftindex^L{g} \approx \leftindex^R{g}$ over $F$ and $V$, where $\leftindex^L{g}, \leftindex^R{g} \in \textit{TNLG}(F, V)$. $\mathfrak{G} \models \leftindex^L{g} \approx \leftindex^R{g}$ if and only if 
    \begin{itemize}
        \item for any $\leftindex^L{\mathcal{V}}$, there exists $\leftindex^R{\mathcal{V}}$ and a CTNLG $g$, such that $[\textit{Instance}(\leftindex^L{g}, \leftindex^L{\mathcal{V}}), \dots, g]$ and $[\textit{Instance}(\leftindex^R{g}, \leftindex^R{\mathcal{V}}), \dots, g]$ are structure paths of $\mathfrak{G}$, and
        \item for any $\leftindex^R{\mathcal{V}}$, there exists $\leftindex^L{\mathcal{V}}$ and a CTNLG $g$, such that $[\textit{Instance}(\leftindex^R{g}, \leftindex^R{\mathcal{V}}), \dots, g]$ and $[\textit{Instance}(\leftindex^L{g}, \leftindex^L{\mathcal{V}}), \dots, g]$ are structure paths of $\mathfrak{G}$,
    \end{itemize}
    where $\forall v \in \textit{Variable}(\leftindex^L{g}) \cap \textit{Variable}(\leftindex^R{g}): \leftindex^L{\mathcal{V}}(v) = \leftindex^R{\mathcal{V}}(v)$.
\end{definition}

\begin{definition}[Symbolic Instantiation of Typed Node-Labeled Graph]
\label{def:symbolic_instantiation}
    A \textbf{symbolic instantiation} of a TNLG $g$ is a function $\textit{Symbolic}(g)$ that produces a CTNLG by substituting $\textit{Variable}(g)$ with nullary function symbols in $g$.
\end{definition}

\begin{definition}[Derivation Tree]
\label{def:derivation_tree}
    A \textbf{derivation tree} of a TNLG $g$ regarding a GTS $\anglebrackets{P, \leftindex^0{P}, R, {\hookrightarrow}, \tilde{\leftindex^0{g}}}$ over $F$ and $V$ is a tree $\anglebrackets{N, \leftindex^0{n}, \mathcal{S}}$
    \begin{itemize}
        \item $N \subseteq \textit{TNLG}(F, V)$ is a set of nodes,
        \item $\leftindex^0{n} = \textit{Symbolic}(g), \leftindex^0{n} \in N$, and
        \item $\mathcal{S}(n) = \braces{\textit{PGRewrite}(n, \tilde{g}, \tilde{n}) \mid (\tilde{g}, \tilde{n}) \in R, \tilde{g}/\tilde{\leftindex^0{n}} \rightsquigarrow n}$.
    \end{itemize}
\end{definition}

\begin{theorem}[EL Satisfiability]
\label{the:el_satisfiability}
$\leftindex^L{g} \approx \leftindex^R{g}$ holds for $\mathfrak{G}$ if $\textit{DerivationTree}(\leftindex^L{g})$ and $\textit{DerivationTree}(\leftindex^R{g})$ with respect to $\mathfrak{G}$ have at least one common node.
\begin{proof}
    Let $g$ be a common node present in both $\textit{DerivationTree}(\leftindex^L{g})$ and $\textit{DerivationTree}(\leftindex^R{g})$. According to Definition~\ref{def:derivation_tree}, there exist structure paths $[\textit{Symbolic}(\leftindex^L{g}), \dots, \textit{Symbolic}(g)]$ and $[\textit{Symbolic}(\leftindex^R{g}), \dots, \textit{Symbolic}(g)]$. 
    
    Correspondingly, for any $\leftindex^L{\mathcal{V}}$ that substitutes symbolic values in $\textit{Symbolic}(\leftindex^L{g})$, $\textit{Symbolic}(g)$, and shared symbolic values of $\leftindex^L{g}$ and $\leftindex^R{g}$, the structure paths remain unaffected. Therefore, for each such $\leftindex^L{\mathcal{V}}$, there exists a corresponding $\textit{Symbolic}(\leftindex^R{g})$ to ensure that $\forall v \in \textit{Variable}(\leftindex^L{g}) \cap \textit{Variable}(\leftindex^R{g}): \leftindex^R{\mathcal{V}}(v) = \leftindex^L{\mathcal{V}(v)}$. Analogously, this relationship is bidirectional, as each $\leftindex^R{\mathcal{V}}$ implies the existence of a corresponding $\leftindex^L{\mathcal{V}}$.
    
    By Definition~\ref{def:el_semantics}, this completes the proof.
\end{proof}
\end{theorem}

\subsubsection{Implementation}
\label{sec:gts_implementation}

GTSs interact with external environments through side effects encapsulated by their transitions. A \textbf{side effect} refers to any operation that alters the state of external environments, a typical example of which is performing Input/Output (I/O) operations.

For a GTS containing a transformation relation $R$, each $r \in R$ preserves its functional purity while encapsulating two side effects: prior effect and later effect. A \textbf{prior effect} transpires before $r$, such as consuming a signal from the external environment. Conversely, a \textbf{later effect} manifests after $r$, such as displaying a message on a console via a command transmission to the external environment.

\begin{theorem}[Provably Correct Implementation]
\label{the:provably_correct_implementation}
    A provably correct GTS with side effects is a provably correct implementation with respect to a set of temporal and functional properties.

    \begin{proof}
        Let $\mathfrak{S}$ denote a provably correct GTS $\mathfrak{S}$ associated with a set $\Phi$ of temporal and functional properties. A sound correctness verification that mechanizes $\mathfrak{S} \vdash \Phi$ exists to ensure $\mathfrak{S} \models \Phi$, as per Definition~\ref{def:gts} and Proposition~\ref{pro:provably_correct_model}. 
        
        By Definition~\ref{def:correctness_implementation}, the process of formulating the specification ($\mathfrak{S}$ and $\Phi$) is a correctness implementation. Additionally, side effects alter the external environment but do not affect the functional purity of the transformation relation of $\mathfrak{S}$. Therefore, the implementation, $\mathfrak{S}$ with side effects, preserves $\Phi$ and is provably correct.
    \end{proof}
\end{theorem}

\subsection{The Seni Language}
\label{sec:seni}

The Seni language is a transition-oriented programming language that facilitates the development of provably correct systems by practicalizing the TOP features elaborated upon in subsequent Section~\ref{sec:modularization}, Section~\ref{sec:progressive_specification}, Section~\ref{sec:context_perception}, and Section~\ref{sec:proof_automation}.

The etymology of "Seni" is derived from the romanization of the Japanese term "\begin{CJK}{UTF8}{ipxm}遷移\end{CJK}," meaning "transition". This nomenclature profoundly reflects the fundamental purpose of the Seni language. Specifically, Seni effectively mechanizes GTSs as illustrated in Section~\ref{sec:gts} to capture both stateful dynamics and structural transformations that are formulated as temporal and functional properties to be verified by its intrinsic correctness verification.

\subsubsection{Core Syntax}

The core syntax of the Seni language focuses on constructing a PGTS as per Definition~\ref{def:pgts}, of which the Extended Backus–Naur Form (EBNF) is shown in Definition~\ref{def:seni_syntax}.

\begin{definition}[Core Syntax of Seni]
\label{def:seni_syntax}
    \begin{align*}
    \textit{Unit} &= \braces{\textit{TypeDecl}}, \braces{\textit{SysDecl}} \\
    \textit{TypeDecl} &= \textbf{type}, \textit{TypeID}, \text{"="}, \textit{ConsDecl}, \braces{\text{"|"}, \textit{ConsDecl}} \\
    \textit{SysDecl} &= [\textbf{default}], \textbf{system}, \textit{SysID}, \textit{SysParam}, \textit{SysBody} \\
    \textit{SysBody} &= \text{"\{"}, \braces{\textit{SymDecl}}, \braces{\textit{VarStmt}}, \braces{\textit{InstStmt}}, \textit{InitStmt}, \braces{\textit{RuleDecl}}, \text{"\}"} \\
    \textit{InstStmt} &= \textit{InstID} \mid \textbf{as} \mid \textit{SysID} \textit{SysArg} \\
    \textit{SymDecl} &= \textit{Symbol}, \text{"::"}, \textit{Type} \mid \textit{SymVarDecl} \\
    \textit{VarStmt} &= \textit{Variable}, \text{"::"}, \textit{Type} \\
    \textit{SymVarDecl} &= \textbf{var}, \textit{Symbol}, \text{"::"}, \textit{Type} \\
    \textit{InitStmt} &= \textit{InitParam}, \text{"="}, \textit{GraphStmt} \mid \textit{RelationStmt} \\
    \textit{RuleDecl} &= \textit{RuleID}, \text{"="}, \textit{Rule} \\
    \textit{PropDecl} &= \textit{LTLDecl} \mid \textit{CTLDecl} \mid \textit{ELDecl} \\
\end{align*}
\end{definition}

The Seni language extends the primitive type with string, list, stack, and queue as built-in types. Particularly, the graph type is handy for representing an inferable complicated type.

\subsubsection{Execution by Rewriting}

The Seni language represents TNLGs and PGs in a natural way, as shown in Example~\ref{eg:seni_garbage_collection}.

\begin{example}[Garbage Collection]
\label{eg:seni_garbage_collection}
A Seni program implementing the garbage collection in Example~\ref{eg:pg} is shown in Listing~\ref{lst:seni_garbage_collection}.

\begin{lstlisting}[caption={Garbage collection by Seni.}, language=Seni, mathescape=true, numbers=left, label={lst:seni_garbage_collection},basicstyle=\footnotesize]
default system Main {

    init() = GarbageCollector(1, 1)

}

system GarbageCollector {

    M :: ((Int), (Int) | (Int, Int))
    V_a :: (Int)
    V_b :: (Int) | (Int, Int)
    x, y, z :: Int

    init(a :: Int, b :: Int) = M(
        V_a(a),
        V_b(b, 1)
    )

    r = ^M(
            V_a(x),
            V_b(
                y[y = x],
                z[z = y]
            )
        )
        ->
        ~M(
            V_a(x),
            V_b(
                x[x + 1]
            )
        )

    before r {
        print("Collecting")
    }

    after r {
        print("Collected")
    }
}

\end{lstlisting}

The default system \textit{Main} initializes a \textit{GarbageCollector} by assigning its arguments at line 3. Lines 9 to 11 declare three typed function symbols $M$, $V_a$, and $V_b$ representing a memory model with two program variables. Subsequently, line 12 declares a typed variable used to define PGs. The code block spanning lines 14 to 17 creates the CTNLG of the initial state. Following this, the block from lines 19 to 32 declares the rule $r$ by constructing a PG and identifying $\hat{M}$ as the root and $\tilde{M}$ a post-transformation root by tokens \textasciicircum and \textasciitilde. These tokens can be omitted if the first function symbols are roots. For the branch rooted at $\hat{M}$, propositional formulae in square brackets are conditions to guard the execution of the rule. For the branch rooted at $\tilde{M}$, expressions in square brackets are effects to be applied after applying the rule. Additionally, lines 34 to 40 define two I/O functions associated with rule $r$.
\end{example}

The execution of Seni programs relies on a virtual machine that implements $\textit{PGRewrite}$ as defined in Definition~\ref{def:pg_rewriting}. A Seni virtual machine continuously rewrites the initial CTNLG defined in a Seni program according to the defined PGs until no pattern rewriting rule can be applied. Meanwhile, it calls I/O functions associated with the PGs each time a PG is applied.

\subsubsection{Temporal Property Verification}

The Seni language offers an intuitive way to formulate temporal properties in temporal logics, including LTL (Definition~\ref{def:ltl}) and CTL (Definition~\ref{def:ctl}), as shown in Example~\ref{eg:seni_mutual_exclusion}.

\begin{example}[Mutual Exclusion]
\label{eg:seni_mutual_exclusion}
A Seni program implementing the semaphore-based mutual exclusion is shown in Listing~\ref{lst:seni_mutual_exclusion}.

\begin{lstlisting}[caption={Mutual Exclusion by Seni.}, language=Seni,mathescape=true, numbers=left, label={lst:seni_mutual_exclusion},basicstyle=\footnotesize]
system Process with (s :: Int) {

    var id :: String
    var loc :: Int

    init(_id :: String) = (
        id(_id),
        loc(0)
    )

    toWait = [loc = 0] -> {loc: 1}
    enterCrit = [loc = 1 & s > 0] -> {loc: 2, s: s - 1}
    exitCrit = [loc = 2] -> {loc: 0, s: s + 1}

    expose prop crit = [loc = 2]

    ltl G F crit
}

\end{lstlisting}

Line 3 uses the keyword \textit{var} to declare a unary function symbol $\textit{id}$ with a unique anonymous integer variable as the parameter. Notably, the rules describing PGs are simplified since there is no structure transformation. For instance, line 11 states that the rule $\textit{toWait}$ will be applied if there is a homomorphism from $\textit{loc}$ to the CTNLG of the current state with the variable condition $\textit{loc} = 0$ and the effect changing the value of the node bound to $\textit{loc}$ to $1$ will be applied after the transition. Additionally, line 17 declares an LTL property using the proposition declared in line 15. The exposed proposition is also accessible by other modules.

\end{example}

The verification of temporal properties is carried out by default before the execution. A Seni virtual machine embeds bounded model checking techniques to automate verification processes. The details will be presented in Section~\ref{sec:bmc}.

\subsubsection{Functional Property Verification}

Functional properties are naturally encoded in equational logic (Definition~\ref{def:el}) with the Seni language, as shown in Example~\ref{eg:seni_addition}.

\begin{example}[Binary Addition Function]
\label{eg:seni_addition}
    A Seni program implementing the binary addition function in Example~\ref{eg:trs} is shown in Listing~\ref{lst:seni_add}.

\begin{lstlisting}[caption={Binary addition function by Seni.}, language=Seni,mathescape=true, numbers=left, label={lst:seni_add}, basicstyle=\footnotesize]
type Nat = Zero | Su(Nat)

default system Main {
    init() = BinaryAddition(Su(Zero), Su(Zero))
}

system BinaryAddition {

    Plus :: (Nat, Nat)
    x, y :: Nat

    init(left :: Nat, right :: Nat) = Plus(left, right)

    r0 = Plus(x, Zero) -> x
    r1 = Plus(x, Su(y)) -> Su(Plus(x, y))

    el Plus(Su(x), Su(Zero)) = Su(Su(x))

}

\end{lstlisting}

The keyword \textit{type} in line 1 creates a user-defined recursive type \textit{Nat} presenting natural numbers. The initial statement in line 12 creates a CTNLG in a succinct representation. Since there are only structural transformations, the rules can be simplified, as shown in lines 14 and 15. Additionally, line 17 declares a functional property by keyword \textit{el}.
\end{example}

The same as temporal properties, a Seni virtual machine verifies functional properties before the execution by automated theorem proving techniques that will be discussed in Section~\ref{sec:idtp}.

\subsection{Modularization}
\label{sec:modularization}

\textbf{Modularization}\index{Modularization} is a bottom-up approach to compose a large model by a set of modules.

\subsubsection{Module}
\label{sec:module}

A module is a self-contained model for scalability and reusability. It is architecturally engineered to be seamlessly integrated into larger, more complex models, where it interacts with other modules to implement global functionalities. In this manner, correctness verification and implementation of complex models are scalable by decomposing global correctness into manageable segments, represented by the local correctness of individual modules. This decomposition not only simplifies the verification and implementation process but also enhances the manageability and maintainability of complex models.

Besides, a module can function as an abstract model encapsulating generalized functionalities. This abstraction enables the instantiation of a module into multiple concrete models, each tailored to specific requirements through the provision of relevant arguments. Therefore, formulating modules mitigates the need for redundant efforts in developing similar models from scratch, thereby enhancing reusability. As shown in Example~\ref{eg:seni_mutual_exclusion}, two \textit{Process} systems encapsulating primary functionalities are instantiated by a higher-level system \textit{Main} with different arguments. Formally, an abstract model is a parameterized graph transition system defined in Definition~\ref{def:pgts}.

\begin{definition}[Parameterized Graph Transition System]
\label{def:pgts}
    A \textbf{Parameterized Graph Transition System} (PGTS) is a GTS $\bar{\mathfrak{G}}$ over $F$ and $V$ with typed parameters binding with $\bar{V} \subseteq V$, such that for all $(\leftindex^0{p}, \leftindex^0{g}) \in \leftindex^0{S}$ of $\textit{LKS}(\bar{\mathfrak{G}})$, $\leftindex^0{g} \in \textit{TNLG}(F, \bar{V})$.

    A PGTS is \textbf{initialized}\index{Initialization} if $\leftindex^0{g}$ is instantiated.
\end{definition}

Notably, the correctness of an abstract model can be verifiable through a set of correctness verification processes. These processes are used to rigorously verify the model against all conceivable parameter values, ensuring its correctness for all potential concrete models.

However, the correctness verification is usually computationally intensive and may not terminate within a reasonable timeframe, especially when the parameter space is vast and multidimensional. A common strategy involves constraining the ranges of parameters. By limiting the scope of parameter values to a smaller manageable subset, the correctness verification has the potential to be efficient. This strategy reduces the computational burden by focusing on a narrowed yet representative concrete models to be generated.

\subsubsection{Concurrency}
\label{sec:concurrency}

Multiple modules can compose a concurrent model by asynchronous or synchronous composition.

\begin{definition}[Asynchronous Concurrency of Graph Transition Systems]
\label{def:async_gts}
Given two GTSs $\mathfrak{G}_0 \triangleq \anglebrackets{P_0, \leftindex^0{P}_0, R_0, {\hookrightarrow}_0, \tilde{\leftindex^0{g}}_0}$ over $F_0$ and $V_0$, and $\mathfrak{G}_1 \triangleq \anglebrackets{P_1, \leftindex^0{P}_1, R_1, {\hookrightarrow}_1, \tilde{\leftindex^0{g}}_1}$ over $F_1$ and $V_1$, the \textbf{asynchronous concurrency}\index{Asynchronous Concurrency} of $\mathfrak{G}_0$ and $\mathfrak{G}_1$, denoted as $\mathfrak{G}_0 \interleave \mathfrak{G}_1$, is defined over $F_0 \cup F_1$ and $V_0 \cup V_1$:
\[
\mathfrak{G}_0 \interleave \mathfrak{G}_1 \triangleq \anglebrackets{P_0 \times P_1, \leftindex^0{P}_0 \times \leftindex^0{P}_1, R_0 \uplus R_1, {\hookrightarrow}, \tilde{\leftindex^0{g}}_0 \oplus \tilde{\leftindex^0{g}}_1}
\]
where ${\hookrightarrow}$ is defined by the rules:
    \begin{mathpar}
    \frac{p_0 \xhookrightarrow{r}_0 p_0'}{\anglebrackets{p_0, p_1} \xhookrightarrow{r} \anglebrackets{p_0', p_1}}
    
    \frac{p_1 \xhookrightarrow{r}_1 p_1'}{\anglebrackets{p_0, p_1} \xhookrightarrow{r} \anglebrackets{p_0, p_1'}}
    \end{mathpar}
\end{definition}

\begin{example}[Asynchronous Concurrency]
The \textit{Process} module coded in Listing~\ref{lst:seni_mutual_exclusion} of Example~\ref{eg:seni_mutual_exclusion} can be composed as an asynchronous module, as shown in Listing~\ref{lst:seni_async}.

\begin{lstlisting}[caption={Asynchronous Concurrency by Seni.}, language=Seni,mathescape=true, numbers=left, label={lst:seni_async},basicstyle=\footnotesize]
default system Main {

    var s :: Int

    p1 as Process("p1") with (s)
    p2 as Process("p2") with (s)

    init() = async [p1, p2] {s: 1}

    ltl G (not p1.crit or not p2.crit)

}
\end{lstlisting}

Line 5 instantiates a \textit{Process} with a string and passes a shared TNLG into the instance. Line 8 initializes an asynchronous model composed of two \textit{Process} instances sharing a TNLG $\textit{s}(1)$. Line 10 declares an LTL property using the exposed proposition \textit{crit} of the \textit{Process} module.
\end{example}

\begin{definition}[Synchronous Concurrency of Graph Transition Systems]
\label{def:sync_gts}
Given two GTSs $\mathfrak{G}_0 \triangleq \anglebrackets{P_0, \leftindex^0{P}_0, R_0, {\hookrightarrow}_0, \tilde{\leftindex^0{g}}_0}$ over $F_0$ and $V_0$, and $\mathfrak{G}_1 \triangleq \anglebrackets{P_1, \leftindex^0{P}_1, R_1, {\hookrightarrow}_1, \tilde{\leftindex^0{g}}_1}$ over $F_1$ and $V_1$, the \textbf{synchronous concurrency}\index{Synchronous Concurrency} of $\mathfrak{G}_0$ and $\mathfrak{G}_1$, denoted as $\mathfrak{G}_0 \parallel \mathfrak{G}_1$, is defined over $F_0 \cup F_1$ and $V_0 \cup V_1$:
\[
\mathfrak{G}_0 \parallel \mathfrak{G}_1 \triangleq \anglebrackets{P_0 \times P_1, \leftindex^0{P}_0 \times \leftindex^0{P}_1, R_0 \cup R_1, {\hookrightarrow}, \tilde{\leftindex^0{g}}_0 \oplus \tilde{\leftindex^0{g}}_1}
\]
where ${\hookrightarrow}$ is defined by the rules:
\begin{itemize}
    \item For $r \notin R_0 \cap R_1$:
    \begin{mathpar}
        \frac{p_0 \xhookrightarrow{r}_0 p_0'}
        {\anglebrackets{p_0, p_1} \xhookrightarrow{r} \anglebrackets{p_0', p_1}}
        
        \frac{p_1 \xhookrightarrow{r}_1 p_1'}
        {\anglebrackets{p_0, p_1} \xhookrightarrow{r} \anglebrackets{p_0, p_1'}}
    \end{mathpar}
    \item For $r \in R_0 \cap R_1$:
    \begin{align*}
        \frac{p_0 \xhookrightarrow{r}_0 p_0' \land p_1 \xhookrightarrow{r}_1 p_1'}
        {\anglebrackets{p_0, p_1} \xhookrightarrow{r} \anglebrackets{p_0', p_1'}}
    \end{align*}
\end{itemize}

\end{definition}

\subsection{Progressive Specification}
\label{sec:progressive_specification}

\textbf{Progressive specification}\index{Progressive Specification} is a top-down approach to refine a model through a set of iterations.

\subsubsection{Iteration}

An iteration takes a model associated with a set of properties as the input and produces a refined model with more details. An input model is typically an approximation that encapsulates a smaller set of functionalities than its output model.

During an iteration, the initial step executes a correctness verification against the input model with respect to its predefined properties. This verification assesses whether the model satisfies the established provable correctness. If the verification returns $\bot$, indicating the input model violates one of the predefined properties, the model is subjected to a rectification phase. This phase is dedicated to resolving any deficiencies until the model satisfies the established provable correctness.

Subsequent to the correctness verification, the iteration progresses to a refinement stage. In this stage, the primary objective is to further align the model with the demands by making small increments for the model. This is achieved through a dual approach: firstly, by concretizing and establishing additional properties that better resonate with the intended functionalities; and secondly, by extending the structure and transitions of the model to fulfill the functionalities. The iteration concludes with another round of correctness verification applied to the refined model with respect to all specified properties.

\subsubsection{Correctness Preservation}

Each iteration benefits from the bisimulation and simulation techniques \cite{alur_alternating_1998,pasareanu_concrete_2005,baier_principles_2008,ding_formalism-driven_2022}. These techniques enable the preservation of the established provable correctness without necessitating the correctness verification of old properties. This not only improves the verification efficiency but also ensures that the refined model maintains consistency with the input model while evolving to meet new requirements. Formally, bisimulation and simulation are defined in Definition~\ref{def:bisimulation} and Definition~\ref{def:simulation}, respectively.

\begin{definition}[Bisimulation]
\label{def:bisimulation}
    Given two GTSs $\mathfrak{G}_0$ and $\mathfrak{G}_1$ over $F$ and $V$, $\mathfrak{G}_0$ and $\mathfrak{G}_1$ are \textbf{bisimulation-equivalent}\index{Bisimulation Equivalence}, denoted $\mathfrak{G}_0 \sim \mathfrak{G}_1$, if there exists a \textbf{bisimulation}\index{Bisimulation} ${\stackrel{s}{\sim}} \subseteq S_0 \times S_1$ regarding $\textit{LKS}(\mathfrak{G}_0) \triangleq \anglebrackets{S_0, \leftindex^0{S}_0, L_0, {\to}_0, \mathcal{L}_0}$ and $\textit{LKS}(\mathfrak{G}_1) \triangleq \anglebrackets{S_1, \leftindex^0{S}_1, L_1, {\to}_1, \mathcal{L}_1}$  such that
    
    \begin{itemize}
        \item $\forall s_0 \in \leftindex^0{S}_0: \exists s_1 \in \leftindex^0{S}_1: (s_0, s_1) \in {\stackrel{s}{\sim}}$ and $\forall s_1 \in \leftindex^0{S}_1: \exists s_0 \in \leftindex^0{S}_0: (s_0, s_1) \in {\stackrel{s}{\sim}}$, and
        \item for all $(s_0, s_1) \in {\stackrel{s}{\sim}}$,
        \begin{itemize}
            \item $\mathcal{L}_0(s_0) = \mathcal{L}_1(s_1)$,
            \item if $(s_0, s_0') \in {\to}_0$, then there exists $(s_1, s_1') \in {\to}_1$ such that $(s_0', s_1') \in {\stackrel{s}{\sim}}$, and
            \item if $(s_1, s_1') \in {\to}_1$, then there exists $(s_0, s_0') \in {\to}_0$ such that $(s_0', s_1') \in {\stackrel{s}{\sim}}$.
        \end{itemize}
    \end{itemize}
\end{definition}

\begin{example}[Bisimulation]
\label{eg:bisimulation}
The \textit{toWait} rule of the \textit{Process} system in Example~\ref{eg:seni_mutual_exclusion} can be augmented by an additional rule \textit{toWaitLog} associating with an I/O operation writing the waiting information to a file, as coded in Listing~\ref{lst:seni_bisimulation} and visualized in Figure~\ref{fig:bisimular}.

\begin{lstlisting}[caption={Bisimulating the Mutual Exclusion Program.}, language=Seni,mathescape=true, numbers=left, label={lst:seni_bisimulation},basicstyle=\footnotesize]
system Process with (s :: Int) {

    var id :: String
    var loc :: Int

    init(_id :: String) = (
        id(_id),
        loc(0)
    )

    toWait = [loc = 0] -> {loc: 1}
    toWaitLog = [loc = 0] -> {loc: 1}
    enterCrit = [loc = 1 & s > 0] -> {loc: 2, s: s - 1}
    exitCrit = [loc = 2] -> {loc: 0, s: s + 1}

    after toWaitLog {
        write("system.log").append(id + " is waiting")
    }

    expose prop crit = [loc = 2]
}

\end{lstlisting}

\begin{figure}[htbp!]
\centering
\includegraphics[width=0.13\textwidth]{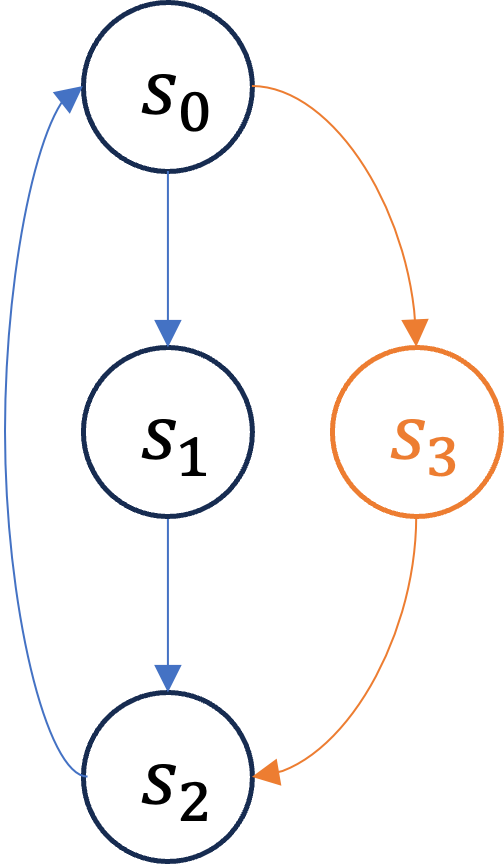}
\caption[Kripke Structure of Bisimular GTSs]{Kripke structure of bisimular GTSs. The blue transitions and states $\braces{s_0, s_1, s_2}$ are interpreted from the GTS in Example~\ref{eg:seni_mutual_exclusion}, while the orange transitions and state $s_3$ are interpreted from its bisimular GTS in Example~\ref{eg:bisimulation}.}
\label{fig:bisimular}
\end{figure}

While the introduction of the \textit{toWaitLog} rule introduces an additional state into the underlying LKS, the resultant GTS is bisimulation-equivalent to the original one. This equivalence indicates that, despite the new state, the stateful dynamics of the system are fundamentally unaltered.
\end{example}

\begin{definition}[Simulation]
\label{def:simulation}
    Given two GTSs $\mathfrak{G}_0$ and $\mathfrak{G}_1$ over $F$ and $V$, $\mathfrak{G}_0$ simulates $\mathfrak{G}_1$, denoted $\mathfrak{G}_0 \succeq \mathfrak{G}_1$, if there exists a \textbf{simulation}\index{Simulation} ${\succeq^s} \subseteq S_0 \times S_1$ regarding $\textit{LKS}(\mathfrak{G}_0) \triangleq \anglebrackets{S_0, \leftindex^0{S}_0, L_0, {\to}_0, \mathcal{L}_0}$ and $\textit{LKS}(\mathfrak{G}_1) \triangleq \anglebrackets{S_1, \leftindex^0{S}_1, L_1, {\to}_1, \mathcal{L}_1}$ such that
    
    \begin{itemize}
        \item $\forall s_1 \in \leftindex^0{S}_1: \exists s_0 \in \leftindex^0{S}_0: (s_0, s_1) \in {\succeq^s}$, and
        \item for all $(s_0, s_1) \in {\succeq^s}$,
        \begin{itemize}
            \item $\mathcal{L}_0(s_0) = \mathcal{L}_1(s_1)$, and
            \item if $(s_1, s_1') \in {\to}_1$, then there exists $(s_0, s_0') \in {\to}_0$ such that $(s_0', s_1') \in {\succeq^s}$.
        \end{itemize}
    \end{itemize}

    $\mathfrak{G}_0$ and $\mathfrak{G}_1$ are \textbf{simulation-equivalent}\index{Simulation Equivalence}, denoted as $\mathfrak{G}_0 \simeq \mathfrak{G}_1$, if $\mathfrak{G}_0 \succeq \mathfrak{G}_1$ and $\mathfrak{G}_1 \succeq \mathfrak{G}_0$.
\end{definition}

\begin{example}[Simulation]
\label{eg:simulation}
An unfair \textit{Process} system in Example~\ref{eg:seni_mutual_exclusion} contains an additional rule \textit{continueCrit = [loc = 2] -> {loc: 1}}, which simulates a refined GTS coded in Listing~\ref{lst:seni_simulation}.

\begin{figure}[htbp!]
\centering
\includegraphics[width=0.15\textwidth]{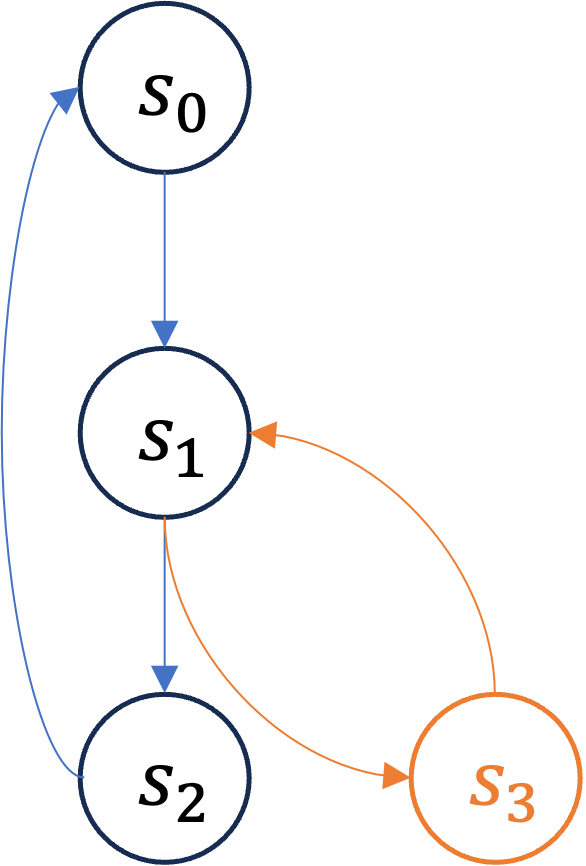}
\caption[Kripke Structure of Simular GTSs]{Kripke structure of simular GTSs. The blue transitions and states $\braces{s_0, s_1, s_2}$ are interpreted from the GTS of the unfair \textit{Process}, while the orange transitions and state $s_3$ are interpreted from its simular GTS in Example~\ref{eg:simulation}.}
\label{fig:simular}
\end{figure}

\begin{lstlisting}[caption={Simulating the Mutual Exclusion Program.}, language=Seni,mathescape=true, numbers=left, label={lst:seni_simulation},basicstyle=\footnotesize, float=*]
system Process with (s :: Int) {

    var id :: String
    var loc :: Int

    init(_id :: String) = (
        id(_id),
        loc(0)
    )

    toWait = [loc = 0] -> {loc: 1}
    enterCrit = [loc = 1 & s > 0] -> CRIT{loc: 2, s: s - 1}
    enterContinuousCrit = [loc = 1 & s > 0] -> CONT_CRIT{loc: 2, s: s - 1}
    continueCrit = CONT_CRIT[loc = 2] -> {loc: 1}
    exitCrit = CRIT[loc = 2] -> {loc: 0, s: s + 1}

    after toWaitLog {
        write("system.log").append(id + " is waiting")
    }

    expose prop crit = [loc = 2]
}

\end{lstlisting}

Lines 12 and 13 explicitly specify the places \textit{CRIT} and \textit{CONT\_CRIT}. This specification is instrumental in managing the stateful transitions with fine granularity. 
\end{example}

An iteration can prove to preserve the correctness of the last iteration by bisimulation and simulation.

\begin{theorem}[Correctness Preservation]
\label{the:correctness_preservation}
For non-terminal $\mathfrak{G}_0$ and $\mathfrak{G}_1$ over $F$ and $V$,
\begin{itemize}
    \item[(a)] if $\mathfrak{G}_0 \sim \mathfrak{G}_1$, then $\forall \phi \in \textit{Language}(\textit{LTL}) \cup \textit{Language}(\textit{CTL}): \mathfrak{G}_0 \models \phi \iff \mathfrak{G}_1 \models \phi$,
    \item[(b)] if $\mathfrak{G}_0 \simeq \mathfrak{G}_1$, then $\forall \phi \in \textit{Language}(\textit{LTL}) \cup \textit{Language}(\textit{UCTL}) \cup \textit{Language}(\textit{ECTL}): \mathfrak{G}_0 \models \phi \iff \mathfrak{G}_1 \models \phi$,
    \item[(c)] if $\mathfrak{G}_0 \succeq \mathfrak{G}_1$, then $\forall \phi \in \textit{Language}(\textit{LTL}) \cup \textit{Language}(\textit{UCTL}): \mathfrak{G}_0 \models \phi \implies \mathfrak{G}_1 \models \phi$.
\end{itemize}
    
\begin{proof}[Proof Outline]
    The proof of (a) follows that bisimulation in Definition~\ref{def:bisimulation} is finer than CTL*, a formal logic that subsumes LTL and CTL, by structural induction over LTL and CTL formulae. The proofs of (b) and (c) follow similar induction over the structure of logic formulae.
\end{proof}
\end{theorem}

\subsection{Context Perception}
\label{sec:context_perception}

\textbf{Context perception} enables GTSs to behave according to contexts, which improves the efficiency of verification. A context contains constraint information about typed function symbols and pattern transition relations. 

\begin{definition}[Context of Graph Transition Systems]
\label{def:context_gts}
    A context of a GTS $\anglebrackets{P, \leftindex^0{P}, R, {\hookrightarrow}, \tilde{\leftindex^0{g}}}$ over $F$ and $V$ is a pair
    \[
        \anglebrackets{\textit{Refine}, \textit{Prior}}
    \]
    \begin{itemize}
        \item $\textit{Refine}: F \cup V \mapsto \textit{Language}(FOL)$ is a type refinement function that relates FOL formulae to typed function symbols and variables, and
        \item $\textit{Prior}: R \times R \mapsto \mathbb{N}$ is a transformation prioritization function such that $\textit{Prior}(r, r')$ is the priority differential between $r$ and $r'$, i.e., $r$ is accorded a priority that is higher by $\textit{Prior}(r, r')$ units than that of $r'$.
    \end{itemize}
\end{definition}

\subsubsection{Type Refinement}

Each typed function symbol and variable can have its type refined by associating it with a predicate in a context. A predicate is a FOL formula assumed to hold for any element of the refined type. These refinement types provide precise type information to reason about systems and reduce the state space of property verification.

\begin{example}[Type Refinement]
\label{eg:type_refinement}
In the mutual exclusion program described in Example~\ref{eg:seni_mutual_exclusion}, the function symbol \textit{loc} is defined to accept an anonymous integer variable as its parameter. Notably, within the \textit{Process} system, the values bound to \textit{loc} are limited to ${0, 1, 2}$. This observation indicates that the generic integer type originally associated with \textit{loc} and its parameter is overly broad. To refine this, predicate $\braces{v \in \textit{Int} \mid 0 \leq v < 3}$ can be used to effectively narrow the range of permissible values. This refinement enhances the precision of the type system in the \textit{Process} system. Consequently, the \textit{Main} system is updated to reflect this type refinement, as detailed in the revised program with a context in Listing~\ref{lst:seni_context_type}.

\begin{lstlisting}[caption={Mutual Exclusion Program with a Type Refinement Context.}, language=Seni,mathescape=true, numbers=left, label={lst:seni_context_type},basicstyle=\footnotesize]
context TypeRefinement for Process {
    refine var loc :: {v :: Int | v >= 0 and v < 3}
}

default system Main {

    var s :: Int

    p1 as Process("p1") with (s) under TypeRefinement
    p2 as Process("p2") with (s) under TypeRefinement

    init() = async [p1, p2] {s: 1}

    ltl G (not p1.crit or not p2.crit)

}
\end{lstlisting}

Notably, the context in Listing~\ref{lst:seni_context_type} refines both the type of \textit{loc} and its bound variable. The refinement can also operate on a function-symbol or variable level. For example, the statement \textit{refine loc :: ({v :: Int | v >=0 and v < 3})} identifies $\textit{Refine}(\textit{loc}) = (\braces{v \in \textit{Int} \mid 0 \leq v < 3})$ within a context.
\end{example}

As illustrated in Example~\ref{eg:type_refinement}, the integer type associated with \textit{loc} and its parameter is refined in the context to explicitly state the expected behavior and reduce the state space of the verification of the formulated temporal properties. 

\subsubsection{Transformation Prioritization}

In GTSs, when multiple patterns are concurrently matched, the selection of PGs for transformation is nondeterministic. Consequently, the Seni language assigns identical priority levels to all PGs by default. Nonetheless, PGs can be prioritized in contexts by assigning integers to each concurrently matched rule or by formulating inequations. Example~\ref{eg:transformation_prioritization} illustrates the method of formulating an inequation to prioritize a rule.

\begin{example}[Transformation Prioritization]
\label{eg:transformation_prioritization}
    For the bisimulation-equivalent GTS in Example~\ref{def:bisimulation}, the priority of rules \textit{toWait} and \textit{toWaitLog} is assigned in a context programmed in Listing~\ref{lst:seni_context_transformation}.

\begin{lstlisting}[caption={Transformation Prioritization Context.}, language=Seni,mathescape=true, numbers=left, label={lst:seni_context_transformation},basicstyle=\footnotesize]
context TransformationPrioritization for Process {
    prior toWait > toWaitLog
}

default system Main {

    var s :: Int

    p1 as Process("p1") with (s) under TypeRefinement
    p2 as Process("p2") with (s) under TypeRefinement

    init() = async [p1, p2] {s: 1}

    ltl G (not p1.crit or not p2.crit)

}
\end{lstlisting}

The inequation statement in line 2 serves as a concise notation to establish that the priority of \textit{toWait} supersedes that of \textit{toWaitLog}, irrespective of the actual magnitude of their priority differential. This formulation ensures that \textit{toWait} is consistently applied prior to \textit{toWaitLog} within the specified context.
\end{example}

Notably, the inequation method is particularly pertinent in scenarios where the objective is to completely eradicate nondeterminism, which effectively dictates a deterministic order of rule application.

The method of integer assignment operates on a probabilistic basis, where the likelihood of a rule application is determined by the ratio of its assigned integer value to the aggregate sum of all assigned integers to the concurrent rules. This method enables a nuanced control over the rule application to preserve certain nondeterminism.

Transformation prioritization can improve the efficiency of verification processes by directing them toward an expedient convergence. Besides, it facilitates fine-grained control over the execution.

\subsection{Proof Automation}
\label{sec:proof_automation}

TOP promotes \textbf{proof automation}\index{Proof Automation} techniques to facilitate property verification by default. For temporal property verification, bounded model checking techniques are introduced. For functional property verification, automated theorem proving techniques are introduced.

\subsubsection{Bounded Model Checking}
\label{sec:bmc}

\textbf{Bounded Model Checking} (BMC) is a model checking technique that leverages Satisfiability Modulo Theories (SMT) solvers to efficiently check temporal properties as illustrated in Section~\ref{sec:temporal_property}.

\begin{figure}[htbp!]
\centering
\includegraphics[width=0.3\textwidth]{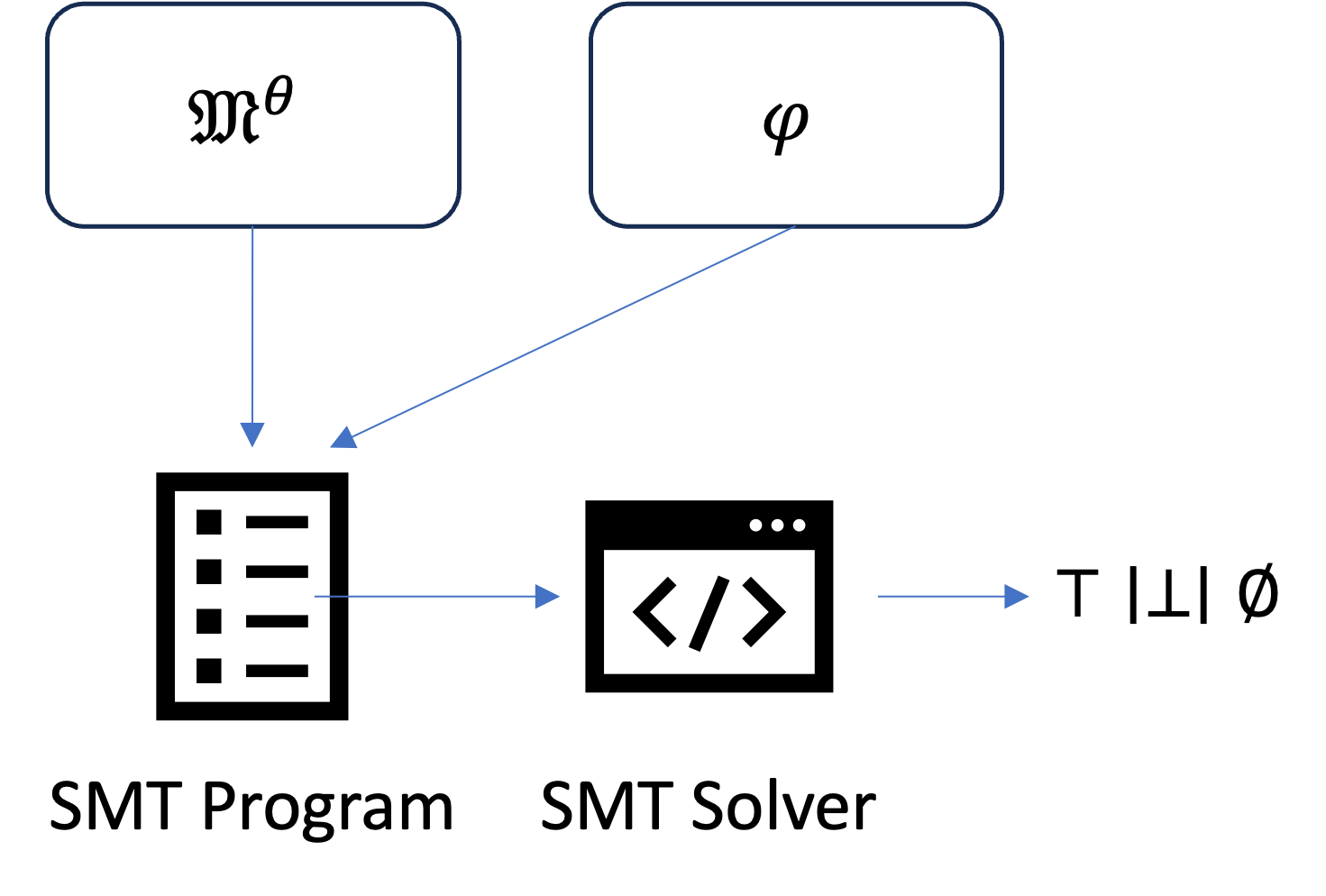}
\caption[Bounded Model Checking]{Overview of the bounded model checking for temporal properties.}
\label{fig:bmc}
\end{figure}

BMC transforms the temporal property verification problem into a satisfiability problem, as outlined in Figure~\ref{fig:bmc}. System structures and behaviors are encoded for a fixed number of steps, and a formula encoding a temporal property is satisfiable if and only if a counterexample to the property being verified exists within those steps. The SMT solver is then used to check the satisfiability of this formula and report a counterexample if a violation is found.

Compared to traditional model checking techniques that attempt to explore the entire state space and can suffer from the state explosion problem \cite{valmari_state_1996}, BMC focuses on a fixed number of steps for a quicker exploration in systems where violations manifest in the early stages of execution. However, BMC does not guarantee the absence of violations beyond the specified bound, which necessitates careful consideration of the bound's adequacy in practical applications.

In TOP, GTSs and temporal properties formulated in LTL and CTL are encoded into SMT programs by Algorithm~\ref{alg:bmc_gts}.

\begin{algorithm}
\caption{Bounded Model Checking of GTSs with Temporal Properties.}
\label{alg:bmc_gts}
\begin{algorithmic}[1]
\Require $\mathfrak{G}, \phi \in \textit{Language}(LTL) \cup \textit{Language}(CTL), \theta \in \mathbb{N}^+$
\Ensure $\top$ or $\textit{path} \in \textit{Path}(\mathfrak{G})$
\State $\mathfrak{K} \gets \textit{LKS}(\mathfrak{G})$ \Comment{By Definition~\ref{def:lks_gts}}

\ForEach{$k \in \braces{0, \dots, \theta}$}
    \State $f_M^k \gets \textit{UnfoldToSMT}(\mathfrak{K}, k)$ \Comment{Unfold $\mathfrak{K}$ for $k$ steps}
    \State $f_\phi^k \gets \textit{TranslateToSMT}(\neg \phi, k)$ \Comment{Translate $\neg \phi$ with respect to the bound $k$}
    \State $(\textit{sat}, \textit{path}) \gets \textit{SMTSolver}(f_M^k \land f_\phi^k)$
    \If{$\textit{sat} = \top$}
        \State \Return $\textit{path}$
    \EndIf
\EndFor
\State \Return $\top$
\end{algorithmic}
\end{algorithm}

Notably, the termination of Algorithm~\ref{alg:bmc_gts} with $\top$ implies that $\mathfrak{G} \models \phi$ holds within bound $\theta$. If $\theta$ attains or exceeds the completeness threshold \cite{biere_bounded_2009} of $\mathfrak{G}$ regarding $\phi$, then $\mathfrak{G} \models \phi$ can be conclusively stated.

\subsubsection{Iterative Deepening Theorem Proving}
\label{sec:idtp}

\textbf{Iterative Deepening Theorem Proving} (IDTP) is an automated theorem proving technique that capitalizes on the Iterative Deepening Depth-First Search (IDDFS) algorithm to efficiently verify functional properties as illustrated in Section~\ref{sec:functional_property}.

\begin{figure}[htbp!]
\centering
\includegraphics[width=0.44\textwidth]{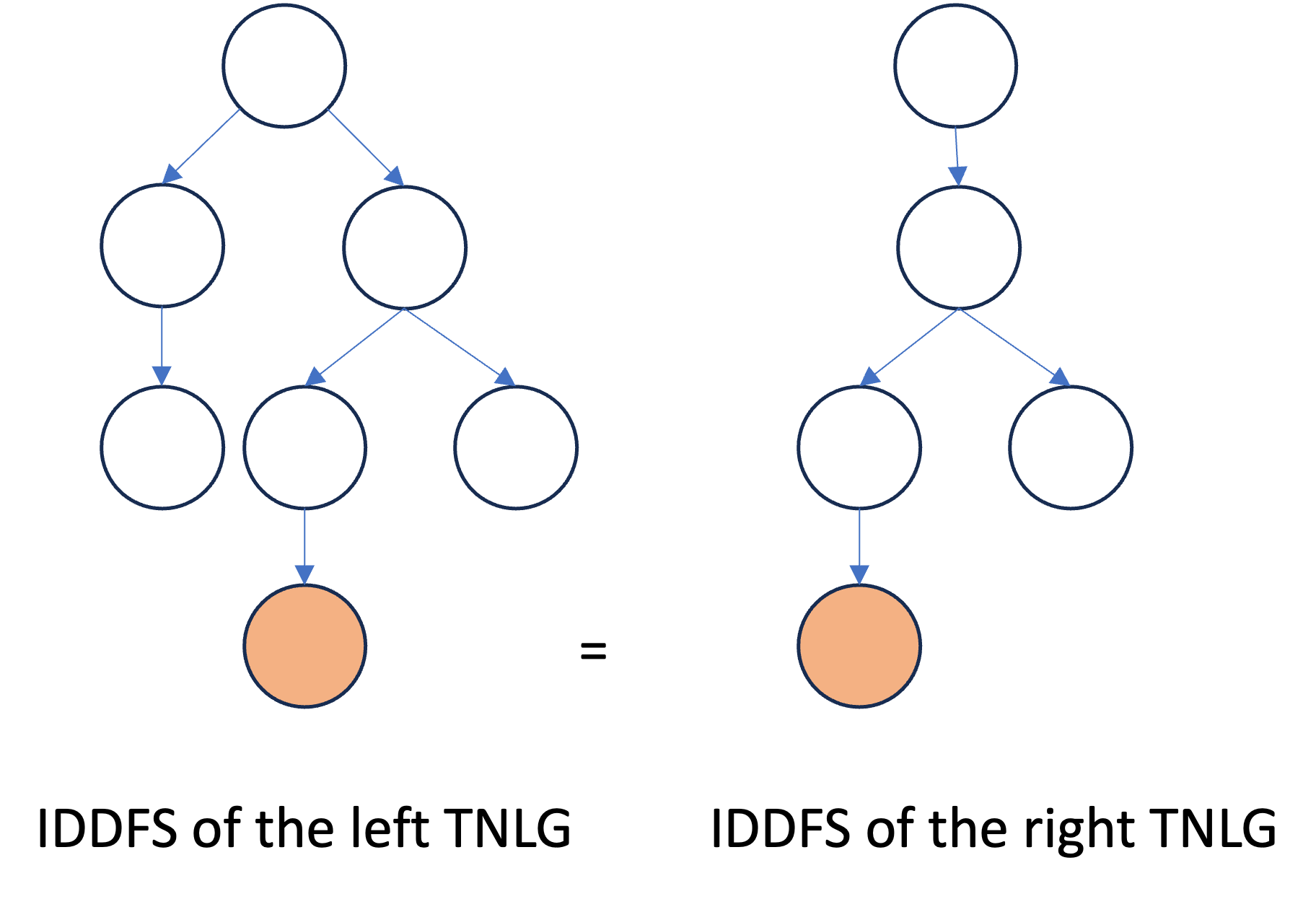}
\caption[Iterative Deepening Theorem Proving]{Overview of the iterative deepening theorem proving for functional properties.}
\label{fig:idtp}
\end{figure}

In accordance with the semantics of EL in Definition~\ref{def:el_semantics}, IDTP concurrently applies transformation rules to TNLGs on both sides of an EL formula to identify a shared TNLG, which establishes the logical equivalence, as outlined in Figure~\ref{fig:idtp}. IDTP amalgamates the space-efficiency of Depth-First Search (DFS) with the completeness of Breadth-First Search (BFS) to enable a gradual exploration of the search space.

The core of IDTP relies on an algorithm for the Depth-Limited Search (DLS) of symbolic instances, as introduced in Algorithm~\ref{alg:dls_gts}. Based on the DLS algorithm, IDTP is illustrated in Algorithm~\ref{alg:idtp_gts}. In essence, Algorithm~\ref{alg:idtp_gts} is an implementation of Theorem~\ref{the:el_satisfiability}.

\begin{algorithm}
\caption{Depth-Limited Search of Symbolic Instances.}
\label{alg:dls_gts}
\begin{algorithmic}[1]
\Require $g \in \textit{TNLG}(F, V)$, depth $d \in \mathbb{N}$, and set $T \subseteq \textit{TNLG}(F, V)$ of target symbolic instances to search.
\Ensure $\emptyset$ if a target instance is found, or set $D$ of derived symbolic instances.
\If{$d = 0 \land g \in T$}
    \State \Return $\emptyset$
\EndIf

\If{$d > 0$}
    \State $D = \textit{Derive}(g, R)$ \Comment{$\textit{Derive}$ implements $\mathcal{S}$ in Definition~\ref{def:derivation_tree}.}
    \ForEach{$g' \in D'$}
        \State $D' \gets \textit{DLS}(g', d - 1, T)$
        \If{$D' = \emptyset$}
            \State \Return $\emptyset$
        \Else
            \State $D \gets D \cup D'$
        \EndIf
    \EndFor
\EndIf
\State \Return $D$

\end{algorithmic}
\end{algorithm}

\begin{algorithm}
\caption{Iterative Deepening Theorem Proving of Functional Properties.}
\label{alg:idtp_gts}
\begin{algorithmic}[1]
\Require transformation relation $R$ of a GTS, EL formula $(\leftindex^L{g}, \leftindex^R{g}) \in \textit{Language}(EL)$, and depth bound $\theta \in \mathbb{N}^+$.
\Ensure $\top$ if the EL formula holds. $\bot$ otherwise.
\State $(D_l, D_r) \gets (\textit{Symbolic}(\leftindex^L{g}), \textit{Symbolic}(\leftindex^R{g}))$
\ForEach{$d \in \braces{0, \dots, \theta}$}
    \State $D_l' \gets \textit{DLS}(\leftindex^L{g}, d, D_r)$ \Comment{$\textit{DLS}$ is illustrated in Algorithm~\ref{alg:dls_gts}.}
    \If{$D_l' = \emptyset$}
        \State \Return $\top$
    \Else
        \State $D_l \gets D_l \cup D_l'$
    \EndIf
    \State $D_r' \gets \textit{DLS}(\leftindex^R{g}, d, D_l)$
    \If{$D_r' = \emptyset$}
        \State \Return $\top$
    \Else
        \State $D_r \gets D_r \cup D_r'$
    \EndIf
\EndFor
\State \Return $\bot$

\end{algorithmic}
\end{algorithm}

\subsubsection{External Proof Assistance}
\label{sec:external_proof_assistance}

TOP is formulated to integrate a diverse array of proof techniques for verifying both temporal and functional properties. Nonetheless, the evolution of a practical language necessitates ongoing developmental efforts to assimilate these techniques effectively. To address this challenge, TOP is engineered to support compatibility with external proof assistants. The current version of Seni provides fully automated proof mechanisms as illustrated in Section~\ref{sec:bmc} and Section~\ref{sec:idtp}. Consequently, external proof tools can be used when these built-in automated proofs are either ineffective or fail to conclude within a reasonable time. 

Consider, for example, a snippet of a Seni program in Listing~\ref{lst:seni_admitted_property}. The coded functional property \textit{Rev(Rev l) = l} asserts that applying the reversal function to any list twice will yield the original list. Due to the absence of mathematical induction capabilities in the current Seni, the automated verification process is unable to reach a conclusion promptly. In such cases, an external proof assistant, such as Lean \cite{moura_lean_2015,moura_lean_2021}, can be used with the lemma \textit{Rev(Append(l0, l1)) = Append(Rev(l1), Rev(l0))} to efficiently give the proof. Correspondingly, this externally verified property is marked by the \textit{admitted} keyword.

\begin{lstlisting}[caption={Admitted Property.}, language=Seni,mathescape=true, numbers=left, label={lst:seni_admitted_property},basicstyle=\footnotesize]
type IntList = Nil | Cons(Int, IntList)
system Main {

    Append :: (IntList, IntList)
    Rev :: (IntList)

    h :: Int
    l, l0, l1 :: IntList

    r0 = Append(Nil, l) -> l
    r1 = Append(Cons(h, l0), l1) -> Cons(h, Append(l0, l1))
    r2 = Rev(Nil) -> Nil
    r3 = Rev(Cons(h, l)) -> Append(Rev(l), Cons(h, Nil))

    admitted el Rev(Rev l) = l

}
\end{lstlisting}

\section{Applicability}

\subsection{Transition-Oriented Development of Distributed Protocols}
\label{sec:tod}

A distributed protocol constitutes a comprehensive set of rules agreed upon by nodes to synchronize their behaviors and maintain state consistency. These protocols adeptly navigate various challenges endemic to distributed computing, such as event ordering \cite{lamport_time_2019}, data consistency \cite{lamport_part-time_2019}, fault tolerance \cite{castro_practical_1999}, and concurrent control \cite{bernstein_concurrency_1987}.

Distributed Ledger Technology (DLT) epitomizes the application of distributed protocols to build decentralized systems. Initially conceived as the underlying protocol for Bitcoin, a decentralized payment system \cite{nakamoto_bitcoin_2019}, DLT has gained widespread adoption. Its applicability extends across diverse domains, addressing critical security and privacy concerns, such as the Internet of Things (IoT) \cite{dai_blockchain_2019}, data persistence \cite{ding_dagbase_2020,ding_derepo_2020}, and security infrastructures \cite{ding_bloccess_2023}. Among the various distributed protocols for DLT, consensus mechanisms hold a place of central importance. They serve as the protocol enabling the nodes in a distributed system to agree on a consistent data value, especially in the absence of central authorities. Examples of these mechanisms include proof-based mechanisms (e.g., Proof of Work (PoW) \cite{dwork_pricing_1992}) for permissionless DLT, Practical Byzantine Fault Tolerance (PBFT) \cite{castro_practical_1999}, and Raft \cite{ongaro_search_2014} for permissioned DLT.

The allure of DLT primarily stems from its distinctive features, such as immutability, robust fault tolerance, non-repudiation, transparency, traceability, and auditability. In a world increasingly reliant on data as a crucial asset, DLT offers enhanced protection against the vulnerabilities inherent in centralized systems by distributing control and restoring ownership to the data proprietors. Nonetheless, the development of a reliable decentralized system that upholds these characteristics is far from straightforward or cost-free. Even experienced architects and developers encounter formidable challenges in this endeavor \cite{fu_evmfuzzer_2019}. A case in point is Geth, the predominant implementation of the Ethereum Virtual Machine (EVM), which is not impervious to vulnerabilities that can induce consensus errors. Notably, a recent vulnerability\footnote{\url{https://github.com/ethereum/go-ethereum/security/advisories/GHSA-9856-9gg9-qcmq}} led to a chain split following the London hard fork, demonstrating that even minor flaws can precipitate substantial security concerns \cite{yang_finding_2021}. Meanwhile, these vulnerabilities are hard to identify, yet threatening the core distributed protocols \cite{yang_finding_2021}, such as the consensus mechanism.

Current methods for verifying decentralized systems typically fall into two categories: implementation-level verification and model-based verification. The former scrutinizes the execution details, addressing runtime errors (e.g., null pointer, division by zero, buffer overflow), function exceptions (e.g., undefined behaviors, unexpected algorithm output), and concurrency issues (e.g., deadlocks, race conditions). Although crucial to verify implementations and even worthwhile analyzing bytecodes, these methods are limited in their ability to detect underlying design flaws. Nevertheless, the implementation-level verification is a unilateral strategy and hard to unravel the design flaw. Conversely, the latter involves verifying the extracted models from implementations to identify design faults. However, extracting abstract models from complex implementation is nontrivial. Extracted models may be overly simplified and lack meaningful properties, or too complicated for effective verification \cite{valmari_state_1996}. Additionally, ensuring that the extracted model aligns with the intended demands poses a significant conformance issue, where any deviation renders the verified properties unreliable. Moreover, a standardized development process is still missing \cite{destefanis_smart_2018}, especially regarding the development of provably correct decentralized systems.

To mitigate the aforementioned challenges effectively, \textbf{Transition-Oriented Development} (TOD), an iterative and incremental development process, is introduced by leveraging TOP in the development of distributed protocols, with a focus on decentralized systems. TOD represents a sophisticated evolution of Formalism-Driven Development (FDD), a concept initially proposed in \cite{ding_formalism-driven_2022} and further refined in \cite{ding_formalism-driven_2022-1}. The essence of TOD lies in its pragmatic application of TOP within the development lifecycle, guiding the process toward the generation of precise specifications and provably correct implementations.

\subsubsection{Overview}

TOD is an iterative and incremental development process that promotes formal methods throughout its lifecycle. It is devised to take advantage of TOP to eliminate design ambiguity, prove model properties, compile correctness implementation with mainstream programming languages, and ensure conformance among design, model, and implementation.

In fact, the philosophy of iterative and incremental development process has been widely practiced in agile development \cite{shore_art_2007}. Nevertheless, both iteration and increment are not formally defined in agile processes. Typically, iteration means enhancing systems progressively, whereas increment refers to the piecewise delivery of the system. However, it is hard to give a well-defined explanation about what an iteration or increment produces and relations between two iterations and relations between an iteration and an increment. In TOD, iteration and increment are defined based on Graph Transition System (GTS) theory, including modeling, refinement, and verification. With these well-defined theories, iterations and increments can be rigorously managed and used to produce verifiable deliveries.

\begin{figure}[htbp!]
\centering
\includegraphics[width=0.39\textwidth]{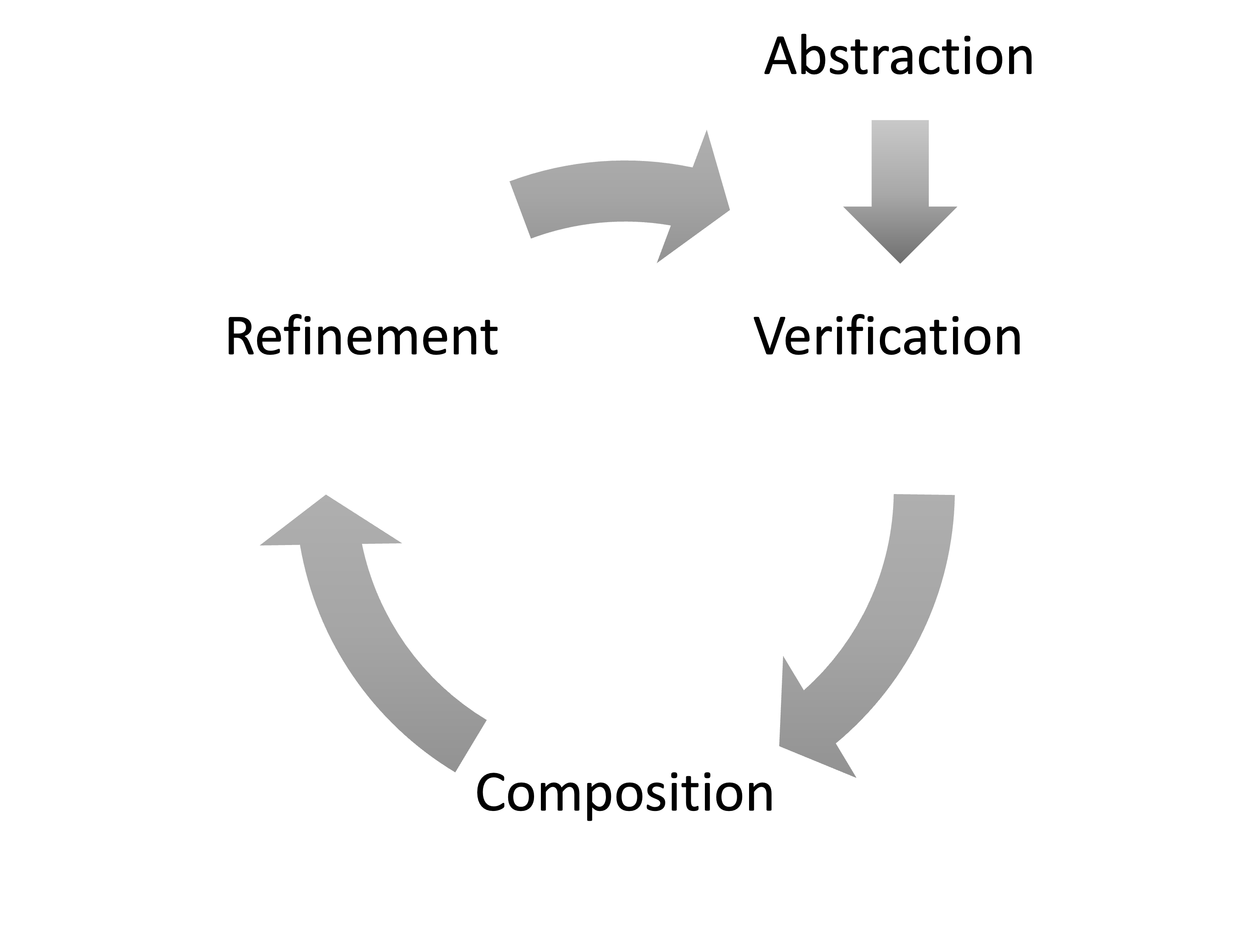}
\caption[Overview of Transition-Oriented Development]{Overview of transition-oriented development.}
\label{fig:tod_overview}
\end{figure}

In TOD, an iteration formulates a model, proves model properties, implements the model, verifies the model implementation, and integrates or delivers the milestone. An increment organizes subsystems together as a higher-level system. Concretely, an iteration contains four stages: abstraction, verification, composition, and refinement, which is shown in Figure~\ref{fig:tod_overview}. All stages are supported by the Seni language introduced in Section~\ref{sec:seni}. \textit{Abstraction Stage} produces a GTS as a module. In \textit{Verification Stage}, the created GTS is verified by either built-in correctness verification mechanisms or external tools. \textit{Composition Stage} integrates multiple verified modules into higher-level systems or delivered. \textit{Refinement Stage} accepts a module from the last iteration and produces a more detailed GTS while preserving and extending properties and can also launch a new iteration.

\subsubsection{Abstraction Stage}
\label{sec:abstraction_stage}
In \textit{Abstraction Stage}, the goal is to produce a GTS, a rigorous design, from the ground up. Compared to the system graph, originally formulated in FDD, the GTS is more expressive, especially regarding its functional property specifiability.

\paragraph{System Graph Parameter}
Variables of system graphs can be represented as a structure where a unary function symbol binds a variable in GTSs. As introduced in Example~\ref{eg:seni_mutual_exclusion}, the Seni language uses the keyword \textit{var} to conveniently declare such a structure. Besides, channels of system graphs can be established by shared variables and side effects of GTSs, as introduced in Section~\ref{sec:concurrency} and Section~\ref{sec:gts_implementation}.

\paragraph{State Declarator}
State declarators of system graphs enable a succinct representation of stateful dynamics because only the most critical features of states identified by interesting state variables need to be explicitly specified, while the evaluation of other state variables is inferred from the preceding state. For GTSs, the concept of places functions in a similar way and is also equipped with the capability of state inference.

\begin{lemma}[State Inference]
\label{lem:state_inference}
Consider a labeled Kripke structure $\anglebrackets{S, \leftindex^0{S}, L, {\to}, \mathcal{L}}$ interpreted from a GTS, as per Definition~\ref{def:lks_gts}. Let $\textit{Value}$ and $\textit{Value}'$ be node value evaluation functions of $s$ and $s'$, where $s, s' \in S, (s, s') \in {\to}$ interpreted according to a pattern graph containing an effect function $\mathcal{E}$ associated with variable nodes $\leftindex^{\mathcal{E}}{N} = \braces{n \in N \mid \mathcal{L}(n) \in \textit{Domain}(\mathcal{E})}$. The following formula holds.

\[
\forall n \in \textit{Domain}(\textit{Value}): \textit{Value}'(n)= \begin{cases} 
      \textit{Evaluate}(\mathcal{E}(\mathcal{L}(n))_e, \textit{Value}) & n \in \leftindex^{\mathcal{E}}{N} \\
      \textit{Value}(n) & n \notin \leftindex^{\mathcal{E}}{N}.
    \end{cases}
\]

\begin{proof}
    The proof follows Definition~\ref{def:pg_rewriting} and Definition \ref{def:lks_gts}.
\end{proof}
\end{lemma}

In Example~\ref{eg:seni_mutual_exclusion}, the rule \textit{toWait} of \textit{Process} only identifies the change of the node value of the child of \textit{loc} without explicitly identifying the child of \textit{id}. By Lemma~\ref{lem:state_inference}, the interpreted LKS has the child of \textit{id} remaining unchanged.

\subsubsection{Verification Stage}
\label{sec:verification_stage}
\textit{Verification Stage} produces a verifiable model based on the input GTS associated with the specified admissible behaviors as properties. The properties are formally verified by internal verification algorithms by default and can also be verified by external verification tools. 

\paragraph{Internal Verification}

Inherited from TOP, TOD intertwines correctness specification and verification. Consequently, TOD provides internal verification mechanisms. For instance, the Seni language mechanizes a bounded model checking and an iterative deepening theorem proving technique, as introduced in Section~\ref{sec:proof_automation}.

Compared to FDD that only promotes the usage of external verification tools, these internal verification mechanisms enhance the applicability of TOD by reducing the dependence on extra tools for verification, thereby minimizing the risk of conformance issues and the requirement of the domain knowledge of formal verification tools.

\paragraph{External Verification}

A practical TOD tool takes time to evolve into a sophisticated tool with versatile correctness verification. To bridge the gap, TOD enables compatibility with external verification tools.

Similar to the idea proposed in FDD \cite{ding_formalism-driven_2022,ding_formalism-driven_2022-1}, the modeling language of TOD generates verification-oriented programs for temporal property verification, such as SMT \cite{de_moura_z3_2008} programs, Promela \cite{holzmann_model_1997} programs, and NuSMV \cite{cimatti_nusmv_2002} programs. As illustrated in Section~\ref{sec:abstraction_stage}, the Seni language that is capable of describing GTSs is expressive enough to program system graphs. Therefore, it is trivial to implement the generation algorithm by slightly modifying the operational semantics of the verification program generation of FDD.

Besides, TOD supports the verification of functional properties via proof assistants, as introduced in Section~\ref{sec:external_proof_assistance}. These external proof tools, such as F* \cite{swamy_dependent_2016} and Lean4 \cite{moura_lean_2021}, can be used when internal verification mechanisms are either ineffective or fail to conclude within a reasonable time.

\subsubsection{Composition Stage}
\label{sec:composition_stage}
\textit{Composition Stage} serves the overall bottom-up approach that composes multiple GTSs into a higher-level GTS, which is an incremental process. There are three types of compositions in this stage: concatenation, embedding, and concurrency.

\paragraph{Concatenation}

The concatenation of GTSs is to merge their corresponding elements, which is detailed in Algorithm~\ref{alg:concatenating_gts}. With the support of the concatenation composition, complicated business logic can be separated into smaller modules.

\begin{algorithm}
\caption{Concatenating $\mathfrak{G}_0$ and $\mathfrak{G}_1$.}
\label{alg:concatenating_gts}
\begin{algorithmic}[1]
\Require $\mathfrak{G}_i = \anglebrackets{P_i, \leftindex^0{P}_i, R_i, {\hookrightarrow}_i, \tilde{\leftindex^0{g}}_i}$ over $F_i$ and $V_i$, where $i \in \braces{0, 1}$
\Ensure $\anglebrackets{P, \leftindex^0{P}, R, {\hookrightarrow}, \tilde{\leftindex^0{g}}}$ or exceptions

\ForEach{$i, j \in \braces{0, 1}$}
    \If{$F_i \cup V_j \neq \emptyset$}
        \State \Return \textbf{exception} \textit{Uncontactable}
    \EndIf
\EndFor
\State \Return $\anglebrackets{P_0 \cup P_1, \leftindex^0{P}_0 \cup \leftindex^0{P}_1, R_0 \cup R_1, {\hookrightarrow}_0 \cup {\hookrightarrow}_1, \tilde{\leftindex^0{g}}_0 \oplus \tilde{\leftindex^0{g}}_1}$
\end{algorithmic}
\end{algorithm}

\begin{example}[Concatenation]
    Given three GTSs \textit{GreaterThan} and \textit{LessThan} in Listing~\ref{lst:seni_greater_than} and Listing~\ref{lst:seni_less_than}, respectively. A concatenated GTS \textit{Inequality} is in Listing~\ref{lst:seni_concatenation}. A higher-level GTS can instantiate a \textit{GreaterThan} module by \textit{Inequality.GreaterThan(Su(Zero), Su(Zero))}.
\begin{lstlisting}[caption={Module for $>$.}, language=Seni,mathescape=true, numbers=left, label={lst:seni_greater_than}, basicstyle=\footnotesize, float=*]
system GreaterThan {

    GT :: (Nat, Nat)
    x, y :: Nat

    init(left :: Nat, right :: Nat) = GT(left, right)

    r0 = GT(Su(x), Zero) -> True
    r1 = GT(Zero, Su(x)) -> False
    r2 = GT(Su(x), Su(y)) -> GT(x, y)

}
\end{lstlisting}

\begin{lstlisting}[caption={Module for $<$.}, language=Seni,mathescape=true, numbers=left, label={lst:seni_less_than}, basicstyle=\footnotesize, float=*]
system LessThan {

    LT :: (Nat, Nat)
    x, y :: Nat

    init(left :: Nat, right :: Nat) = LT(left, right)

    r0 = LT(Su(x), Zero) -> False
    r1 = LT(Zero, Su(x)) -> True
    r2 = LT(Su(x), Su(y)) -> LT(x, y)

}
\end{lstlisting}

\begin{lstlisting}[caption={Concatenated module for inequality.}, language=Seni,mathescape=true, numbers=left, label={lst:seni_concatenation}, basicstyle=\footnotesize, float=*]
system Inequality {
    GreaterThan, LessThan
}
\end{lstlisting}

\end{example}

\paragraph{Embedding}

Embedding a GTS into another GTS is commonly used to formulate nested structures, as detailed in Algorithm~\ref{alg:embedding_gts}.

\begin{algorithm}
\caption{Embedding $\mathfrak{G}_0$ into a node of $\tilde{\leftindex^1{g}}$ of $\mathfrak{G}_1$.}
\label{alg:embedding_gts}
\begin{algorithmic}[1]
\Require $\mathfrak{G}_i = \anglebrackets{P_i, \leftindex^0{P}_i, R_i, {\hookrightarrow}_i, \tilde{\leftindex^0{g}}_i}$ over $F_i$ and $V_i$, where $i \in \braces{0, 1}, (\tilde{\leftindex^1{g}}, \_) \in R_1$, and $\theta$
\Ensure $\anglebrackets{P, \leftindex^0{P}, R, {\hookrightarrow}, \tilde{\leftindex^0{g}}}$ or exceptions

\ForEach{$i, j \in \braces{0, 1}$}
    \If{$F_i \cup V_j \neq \emptyset$}
        \State \Return \textbf{exception} \textit{Unembeddable}
    \EndIf
\EndFor

\State $\mathfrak{K}_1 \gets \textit{LKS}(\mathfrak{G}_1)$ \Comment{By Definition~\ref{def:lks_gts}}
\State $G \gets \textit{Unfold}(\mathfrak{K}_1, \theta)$ \Comment{Unfold $\mathfrak{K}_1$ for $\theta$ steps to get a set of terminal CTNLGs.}
\If{$|G| = 0$}
    \State \Return \textbf{exception} \textit{Nonterminal Embedding}
\ElsIf{$|G| > 1$}
    \State \Return \textbf{exception} \textit{Nondeterministic Embedding}
\EndIf

\State $R \gets R_1$
\State ${\hookrightarrow} \gets {\hookrightarrow}_1$

\ForEach{$(g, n) \in R$}
    \If{$g = \tilde{\leftindex^1{g}}$}
        \State $g' \gets \textit{UpdateNode(g, G[0])}$ \Comment{Update the pattern graph that embeds $\mathfrak{G}_0$.}
        \State $R \gets (R \setminus \braces{(g, n)}) \cup \braces{(g', n)}$
        \ForEach{$(p, (g, n), p') \in {\hookrightarrow}$}
            \State ${\hookrightarrow} \gets ({\hookrightarrow} \setminus \braces{(p, (g, n), p')}) \cup \braces{(p, (g, n), p')}$
        \EndFor
    \EndIf
\EndFor

\State \Return $\anglebrackets{P_1, \leftindex^0{P}_1, R, {\hookrightarrow}, \tilde{\leftindex^0{g}}_1}$
\end{algorithmic}
\end{algorithm}

Notably, the invocation relation between two GTSs is established by the embedding composition, as exemplified in Example~\ref{eg:embedding_gts}.

\begin{example}[Embedding]
\label{eg:embedding_gts}
The \textit{Process} in Example~\ref{eg:seni_mutual_exclusion} can embed the \textit{BinaryAddition} in Example~\ref{eg:seni_addition} together with functional modules \textit{EqualTo}, \textit{GreaterThan}, and \textit{BinarySubtraction}, as shown in Listing~\ref{lst:seni_embedding}.

\begin{lstlisting}[caption={Process with embedded modules.}, language=Seni,mathescape=true, numbers=left, label={lst:seni_embedding}, basicstyle=\footnotesize, float=*]
system Process with (s :: Nat) {

    var id :: String
    var loc :: Nat

    init(_id :: String) = (
        id(_id),
        loc(Zero)
    )

    toWait = [EqualTo(loc, Zero)] -> {loc: Su(Zero)}
    enterCrit = [EqualTo(loc, Su(Zero)) & GreaterThan(s, Zero)] -> {loc: Su(Su(Zero)), s: BinarySubtraction(s, Su(Zero))}
    exitCrit = [EqualTo(loc, Su(Su(Zero)))] -> {loc: Zero, s: BinaryAddition(s, Su(Zero))}

}
\end{lstlisting}

\end{example}

\paragraph{Concurrency}

Concurrency composition is introduced in Section~\ref{sec:concurrency}, including asynchronous and synchronous concurrency. The algorithms are trivial to implement according to Definition~\ref{def:async_gts} and Definition~\ref{def:sync_gts}.

\subsubsection{Refinement Stage}
\label{sec:refinement_stage}
\textit{Refinement Stage} accepts a GTS from the last iteration and produces a more detailed GTS while preserving and extending properties. This stage also produces executables and functions as a milestone for delivery.

\paragraph{Property-Preserving Refinement}
TOD utilizes both bisimulation and simulation techniques introduced in Section~\ref{sec:progressive_specification} to support the refining process. The original purposes of these techniques are generally to optimize the verification process and improve verification efficiency by compacting a model while preserving its properties. However, \textit{Refinement Stage} inverses the original purpose to extend a small model into a big one while preserving its properties by Theorem~\ref{the:correctness_preservation}.

\paragraph{Executable Generation}

TOD intertwines correctness specification, verification, and implementation by default. Therefore, a GTS is executable with side effects according to Theorem~\ref{the:provably_correct_implementation}, which is supported by the Seni language. However, language consistency plays an important role in maintaining a large-scale system developed in other programming languages, such as Java, C++, and Go. Correspondingly, generating programs from GTSs for other programming languages can be necessary to eliminate the consistency.

It is straightforward to construct a GTS in functional programming languages, as its related definitions and theorems are mechanizable as pure functions. Meanwhile, the impure side effects are isolated by the monad. However, mechanizing a GTS in object-oriented programming languages can be challenging, regarding the correctness preservation and performance. It is an open challenge for the TOD tool.

\subsection{Transition-Oriented Security Analysis of Security Infrastructures}
\label{sec:tosa}

Security infrastructures, such as public key infrastructures \cite{brands_rethinking_2000}, access control frameworks \cite{ribot_theory_2003}, and identity management systems \cite{recordon_openid_2006}, are integral to the security paradigm in information and communications technologies (ICTs) to ensure confidentiality, integrity, and availability of the protected resources.

Nonetheless, implementing a trustworthy security infrastructure is challenging, especially within centralized architectural designs where a central authority governs a critical component or mechanism. In centralized architectures, the entire infrastructure is vulnerable to security breaches if a critical component is compromised. This vulnerability manifests as a Single Point of Failure (SPoF), where the malfunction or penetration of a critical component can lead to the breakdown of the entire infrastructure. In response to these vulnerabilities, recent advancements in decentralized security infrastructures have emerged, offering solutions to the limitations inherent in centralized systems. Notable developments include decentralized access control \cite{ding_bloccess_2023} and decentralized identity management \cite{ding_self-sovereign_2022}. These decentralized systems distribute the control and management of security mechanisms, thereby mitigating the risks associated with SPoFs and enhancing the overall robustness of the infrastructure.

However, decentralized security architectures are not immune to threats, including unauthorized access, impersonation attacks, Man-in-the-Middle (MitM) attacks, and Distributed Denial of Service (DDoS) attacks. Consequently, conducting a formal security analysis of these frameworks is essential for making architectural decisions. To bridge the gap in formal security analysis of security infrastructures, \textbf{Transition-Oriented Security Analysis} (TOSA) is introduced to analyze security properties against threat models, exemplified by SecureSSI \cite{ding_model-driven_2023}, a security analysis framework for Self-Sovereign Identity (SSI) systems.

\begin{figure}[htbp!]
\centering
\includegraphics[width=0.39\textwidth]{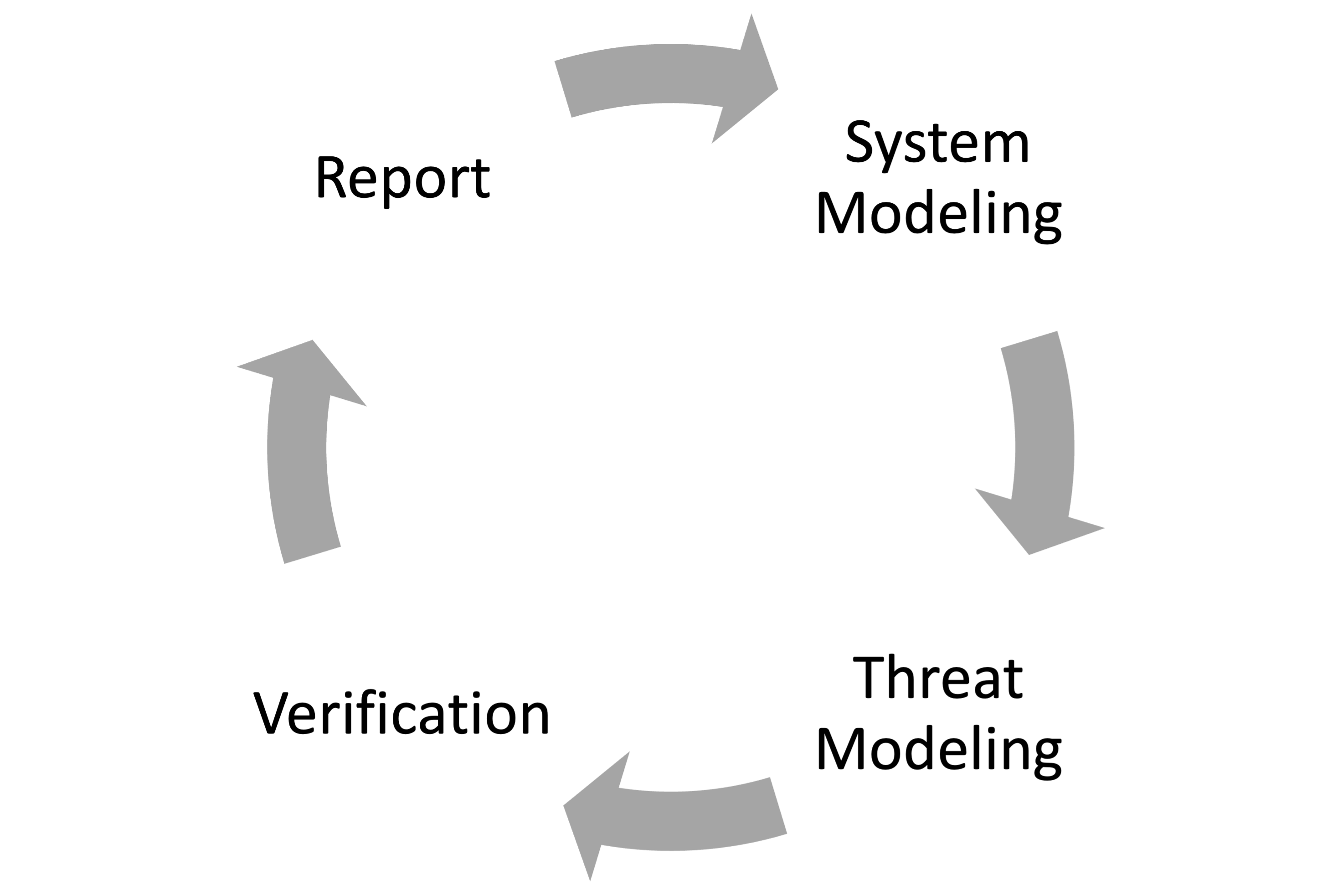}
\caption[Overview of Transition-Oriented Security Analysis]{Overview of transition-oriented security analysis.}
\label{fig:tosa_overview}
\end{figure}

TOSA is a security analysis framework that unifies system and threat modeling through GTSs. The specified systems and threats are composed into a concurrent system, as introduced in Section~\ref{sec:concurrency}. Besides, TOSA promotes the intrinsic correctness verification for specifications, thereby obviating the need for additional formal verification efforts. Additionally, the temporal property verification via the bounded model checking technique introduced in Section~\ref{sec:bmc} can report counterexamples for reproducing attack vectors. The workflow is visualized in Figure~\ref{fig:tosa_overview}.

In contrast to TOD introduced in Section~\ref{sec:tod}, TOSA aims to analyze systems rather than produce deliverable implementations. Besides, TOSA verifies security properties regarding formulated threat models, as opposed to the intrinsic properties inherent to system designs. This focus implies that the properties addressed within TOSA invariably entail the interaction of multiple, relatively independent GTSs.

\subsection{Transition-Oriented Program Analysis}
\label{sec:topa}

Decentralized applications (DApps) are implemented by smart contracts that are programs deployed and automatically executed on blockchains like Ethereum \cite{wood_ethereum_2014} and Hyperledger Fabric \cite{androulaki_hyperledger_2018}. Decentralized finance (DeFi), a domain-specific DApps, is an emerging paradigm in finance, decoupling financial instruments from traditional centralized intermediaries. Over the past five years, DeFi has provided many types of financial services, such as decentralized exchanges (DEXs) \cite{xu_sok_2023}, derivative trading \cite{eskandari_feasibility_2017}, and lending \cite{bartoletti_sok_2021}. Although the DeFi benefits are remarkable (e.g., high availability, transparency, tamper-proofing) \cite{zheng_blockchain_2018}, developing secure DeFi services is still challenging \cite{zou_smart_2019}.

Table~\ref{tab:defi_vulnerabilities} presents an investigation of 80 (with damages ranging from 2,400 to 600 million dollars) real-world DeFi incidents occurring between November 2017 and December 2022, significantly endangering market health. The investigation further categorizes them into six types based on root causes and ordered by severity, computed by the average loss in US dollars per incident. Notably, this investigation dismisses the incidents caused by outdated vulnerabilities like insecure integer arithmetic \cite{torres_osiris_2018,gao_easyflow_2019,rodler_evmpatch_2021} given that recent smart contract compilers (e.g., Solidity v0.8+), libraries (e.g., OpenZeppelin), and virtual machines (e.g., go-ethereum) will prevent these vulnerabilities. Existing methods related to smart contract security, published in top-tier venues and journals from 2018 to 2022, have proved to be adequate for finding generic vulnerabilities with common patterns \cite{singh_blockchain_2020,durieux_empirical_2020}, such as insufficient validation, arbitrary external call, and typical reentrancy vulnerabilities based on symbolic execution \cite{mossberg_manticore_2019,krupp_teether_2018,permenev_verx_2020}, model checking \cite{wang_detecting_2019,frank_ethbmc_2020}, semantic pattern analysis \cite{hildenbrandt_kevm_2018,tsankov_securify_2018}, and fuzzing \cite{jiang_contractfuzzer_2018,nguyen_sfuzz_2020,choi_smartian_2021}.

\begin{table}
\caption[Typical DeFi Incidents]{Typical DeFi incidents.}
\centering
\label{tab:defi_vulnerabilities}
\resizebox{\textwidth}{!}{\begin{tabular}{l l c c l}
\toprule
\textbf{Root Cause} & \textbf{Description} & \textbf{Total} & \textbf{Loss} & \textbf{Incident Examples (Year, Affected Value)}\\
\midrule
    BF & Business Logic Flaw & 23 & \$1.4B & Eleven Finance (2021, \$4.5M), Parity (2017, \$280M) \\
    RE & Reentrancy & 9 & \$147M & Rari Capital (2022, \$80M), Grim Finance (2021, \$30M) \\
    PM & Price Oracle Manipulation & 20 & \$92M & Lodestar Finance (2022, \$4M) , Harvest Finance (2020, \$33.8M) \\
    IV & Insufficient Validation & 10 & \$12M & OlympusDAO (2022, \$292K), Anyswap (2022, \$1.4M) \\
    AF & Access Control Flaw & 13 & \$14M & TempleDAO (2022, \$2.3M), Reaper Farm (2022, \$1.7M) \\
    UE & Unexpected External Call & 5 & \$3M & Rubic (2022, \$1.5M), Rabby Wallet (2022, \$200K) \\
\bottomrule
\end{tabular}}
\end{table}

Unfortunately, the top three severe root causes responsible for the most substantial financial losses pose a formidable challenge for existing methods. Consider business logic flaws, the most severe vulnerability type, are exploited based on understanding the conformity issues between requirement specifications and implementations. On account of their particularity, highly automated methods \cite{krupp_teether_2018,wang_detecting_2019,nguyen_sfuzz_2020} are not applicable, while other methods \cite{tsankov_securify_2018,mossberg_manticore_2019,choi_smartian_2021} are not scalable to these patternless vulnerabilities. Besides, vulnerabilities stemming from potential price oracle manipulations bring attackers exceptionally profitable arbitrage opportunities with low attack costs via flash loans. However, neither methods based on control flow \cite{krupp_teether_2018,wang_detecting_2019} and semantic pattern analysis \cite{hildenbrandt_kevm_2018,tsankov_securify_2018} nor symbolic execution \cite{mossberg_manticore_2019,so_smartest_2021} and fuzzing \cite{jiang_contractfuzzer_2018,nguyen_sfuzz_2020} can efficiently find these vulnerabilities. Additionally, reentrancy variants, exemplified by the Rari Capital hack in 2022, where no typical recursive reentry via a callback function occurred, have emerged as new threats since this specific pattern proves challenging for existing reentrancy detectors \cite{mossberg_manticore_2019,wang_detecting_2019,choi_smartian_2021}.

Motivated by finding DeFi security issues in the early stage, \textbf{Transition-Oriented Program Analysis} (TOPA) is introduced to leverage TOP in finding DeFi vulnerabilities, exemplified by Context-Sensitive Concolic Verification (CSCV) \cite{ding_hunting_2024}. Different from TOSA discussed in Section~\ref{sec:tosa}, TOPA analyzes concrete programs rather than abstract protocols. Besides, TOPA extends its analytical purview to encompass both security and safety considerations. Additionally, the analysis of TOPA incorporates both temporal and functional properties.

\begin{figure}[htbp!]
\centering
\includegraphics[width=0.39\textwidth]{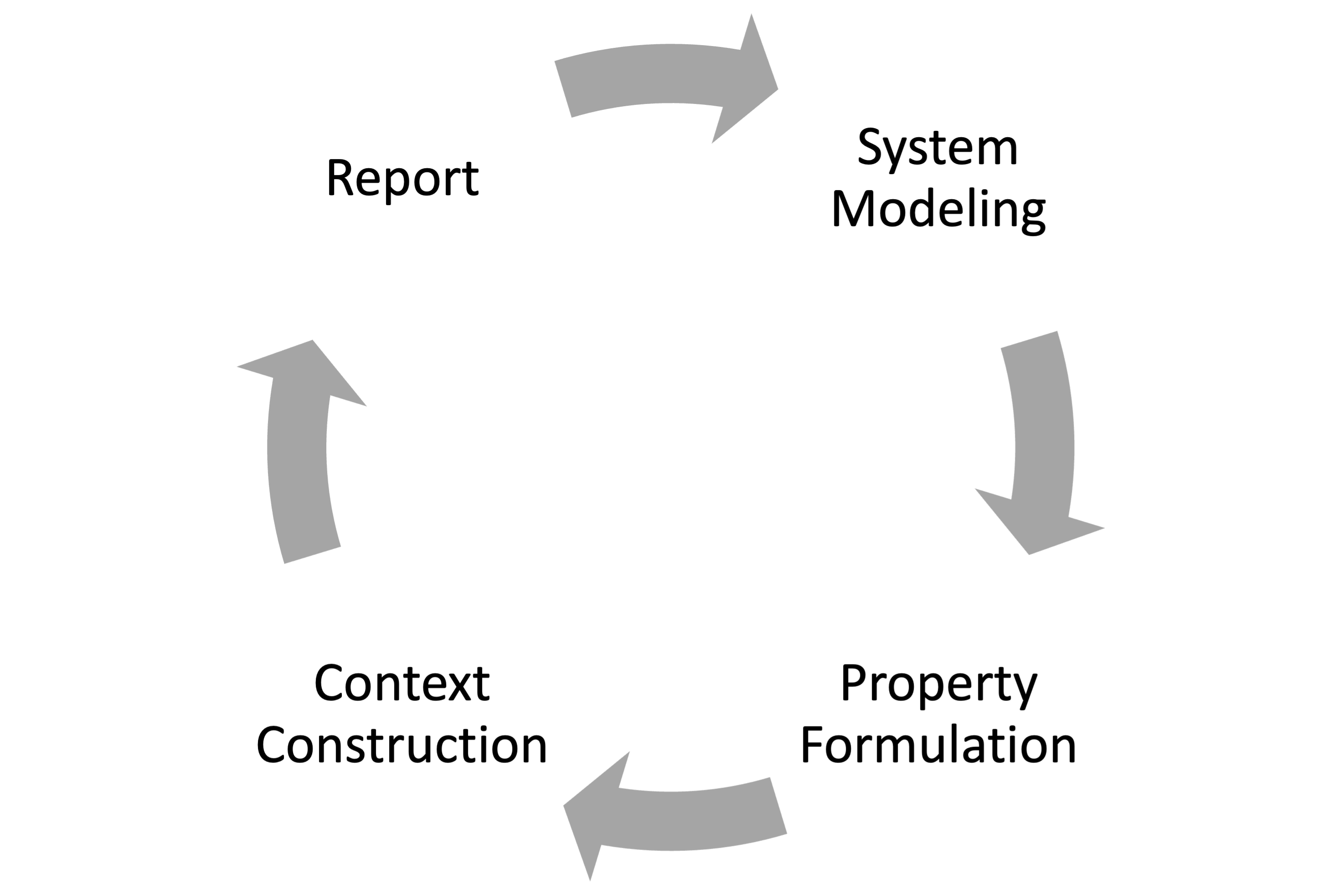}
\caption[Overview of Transition-Oriented Program Analysis]{Overview of transition-oriented program analysis.}
\label{fig:topa_overview}
\end{figure}

The workflow of TOPA is illustrated in Figure~\ref{fig:topa_overview}. The initial step involves system modeling, where programs are abstracted into formal models, and properties specifying normal operational behaviors are formulated. Concurrently, TOPA promotes contexts, extended from the context perception feature of TOP in Section~\ref{sec:context_perception}, to optimize correctness verification. If a property violation is identified, TOPA reports the threat action path along with any malicious assignments, thereby facilitating the identification and mitigation of potential vulnerabilities.

\section{Related Work}

Formal methods have been advancing software engineering to enable the development of safe and secure systems. As discussed in Section~\ref{sec:introduction} and Section~\ref{sec:correctness}, formal methods can be roughly categorized into: formal specification, formal verification, and formal implementation.

\subsection{Formal Specification}

Formal languages introduced in Section~\ref{sec:formal_model} and Section~\ref{sec:formal_logic} of Section~\ref{sec:correctness} are of paramount importance for formal specification. However, there is a marked preference for formal languages that offer higher expressiveness in practice.

To illustrate, process algebras are widely employed for reasoning about concurrent systems and communication protocols. Notable examples include Communication Sequential Processes (CSP) \cite{hoare_communicating_1978}, Calculus of Communicating Systems (CCS) \cite{milner_calculus_1980}, Language Of Temporal Ordering Specification (LOTOS) \cite{bolognesi_introduction_1987}, and $\pi$-calculus \cite{milner_calculus_1992}. Concurrently, program logics such as Hoare logic \cite{hoare_axiomatic_1969}, lambda calculus \cite{barendregt_lambda_1984}, and Z notation \cite{spivey_z_1992} are instrumental in the reasoning about program behavior and structure.

With the advancement of modern system development, there has been a notable shift towards formal languages that align with prevalent programming paradigms, such as object-oriented and functional programming. Languages such as Alloy \cite{jackson_alloy_2002}, Dafny \cite{leino_dafny_2010}, Lean \cite{moura_lean_2021}, and F* \cite{swamy_dependent_2016} exemplify this trend, which proves to be more comprehensible to practitioners.

\subsection{Formal Verification}

Formal verification tools that mechanize correctness verification defined in Definition~\ref{def:correctness_verification} can be broadly classified into two categories: model checking and theorem proving.

Model checking techniques \cite{clarke_model_1994,baier_principles_2008,biere_bounded_2009,clarke_jr_model_2018}, known for their high degree of automation, are often challenged by the state explosion problem \cite{clarke_model_2012}. This issue arises from the exponential growth in the number of states that need to be examined as system complexity increases. To mitigate this, symbolic model checkers like NuSMV \cite{cimatti_nusmv_2002} employ binary decision diagrams, which offer a more efficient representation of stateful dynamics. Additionally, bounded model checkers like CBMC \cite{kroening_cbmcc_2014} adopt a different strategy by unrolling the transition system for a predetermined number of steps, thereby circumventing the need to explore the entire state space.

In contrast, theorem proving techniques \cite{gordon_introduction_1993,fitting_first-order_2012,bibel_automated_2013,loveland_automated_2016} boast greater generality and expressiveness but typically necessitate human intervention, commonly leveraging proof assistants to facilitate correctness verification through human-machine collaboration. Prominent proof assistants enable a more interactive and comprehensive way of correctness verification, including Isabelle \cite{nipkow_isabellehol_2002}, F* \cite{swamy_dependent_2016}, and Lean4 \cite{moura_lean_2021}.

\subsection{Formal Implementation}

Formal implementation tools serve as a conduit between formal models and executables, which mechanizes correctness implementation defined in Definition~\ref{def:correctness_implementation}.

Model-Driven Engineering (MDE) \cite{kent_model_2002,schmidt_model-driven_2006,da_silva_model-driven_2015} prioritizes system development through the management of formal models. In MDE, formal models transcend their traditional role as mere specifications, becoming the fundamental basis from which executables are derived. For instance, there is a focus on the code generation from formal models, exemplified by Executable UML \cite{mellor_executable_2002} and the Eclipse Modeling Framework \cite{steinberg_emf_2008}. Additionally, there is a research community dedicated to improving the interoperability of formal models through model transformation techniques, as seen in DUALLy \cite{malavolta_providing_2009}. Complementing these approaches, formalism-driven development, as discussed in \cite{ding_formalism-driven_2022, ding_formalism-driven_2022-1}, introduces a verification-focused aspect by generating verification-oriented executables.

\section{Conclusion}
\label{sec:conclusion}

This paper systematically discussed system correctness through three aspects: specification, verification, and implementation. First, it formalized correctness with formal models and logics by unifying the representations of transition systems and rewriting systems and their typical variants as the fundamentals for a unified theoretical framework. Meanwhile, formal correctness elucidates the essence of \textbf{Q1}, \textbf{Q2}, and \textbf{Q3} introduced in Section~\ref{sec:introduction}.

Second, this paper presented TOP, a programming paradigm that facilitates the development of provably correct systems. The central tenet of TOP is to seamlessly intertwine specification, verification, and implementation within a unified theoretical framework. This ensures that a provably correct specification is a provably correct implementation, thus providing a theoretical solution to \textbf{Q2} and \textbf{Q3}. The framework is centered around GTSs, a formalism capturing both stateful dynamics and structural transformations. Besides, TOP promotes intrinsic verification support to prove correctness satisfiability regarding the formulated GTSs and their temporal and functional properties. This verification process is grounded in model-checking and theorem-proving methods, yet is specifically refined for GTSs. Crucially, TOP eliminates the gap between formal models and program implementations to enable executable formal models with side effects. 

Third, this paper presented the Seni language, a TOP language with fully-fledged features. Seni supports intuitively constructing GTSs with side effects and formulating temporal and functional properties. Besides, it combines a bottom-up (modularization) and a top-down (progressive specification) approach to enhance specifiability and verifiability. Modularization focuses on improving scalability and reusability by enabling model parameterization and composition, while progressive specification enables an iterative and incremental process for model and property specifications via model refinement techniques. Additionally, Seni introduces context perception to facilitate precise specification and efficient verification. Furthermore, the current version of Seni also intrinsically supports correctness verification at compile time through fully automated techniques, including bounded model checking and iterative deepening theorem proving. Concurrently, Seni demonstrates a practical solution to \textbf{Q1}, \textbf{Q2}, and \textbf{Q3}.

Finally, this paper illustrated three practical scenarios with TOP. It initially delineated transition-oriented development, a development process designed to develop provably correct distributed protocols. Subsequently, the paper introduced transition-oriented security analysis for security infrastructures by TOP, exemplified by SecureSSI formally assessing self-sovereign identity systems against threat models that address security concerns. Lastly, this paper presented transition-oriented program analysis through CSCV in identifying vulnerabilities of decentralized applications, particularly targeting DeFi systems.

\subsection{TOP Vision}
\label{sec:vision}

In addition to the three application scenarios illustrated in Section~\ref{sec:tod}, Section~\ref{sec:tosa}, and Section~\ref{sec:topa}, TOP holds the potential for the development, verification, and optimization of compiling tools, explainable artificial intelligence, and hardware.

\subsubsection{Compiling Tool Development}

TOP provides a comprehensive and coherent method for programming language design and implementation. At the syntactic level, TOP affords a robust formalism, enabling the precise definition of languages by facilitating the formal specification of both sequential and functional models of computation, which is pivotal in establishing a consistent and unambiguous language structure. Moving to the semantic level, TOP enables an effective and intuitive representation of the operational semantics \cite{plotkin_structural_1981} of languages through GTSs, which offers a provable mapping between the language semantics and their execution mechanisms. Furthermore, TOP promotes the formulation of both temporal and functional properties that are critical in specifying the behavior of compiling tools and ensures their correctness through formal verification.

Additionally, existing compiling tools can be encoded into GTSs associated with properties for formal analysis to identify vulnerabilities and explore potential optimization via the proof automation techniques, including BMC and IDTP introduced in Section~\ref{sec:proof_automation}.

\subsubsection{Explainable Artificial Intelligence}

TOP can facilitate the development of explainable AI systems. Based on formal logic, TOP enables the precise specification of machine learning models, thereby articulating the reasoning behind their decisions. Meanwhile, the expected model properties of training and prediction processes can be encoded into GTSs that precisely describe model behaviors. By the intrinsic verification of TOP, these properties are formally verified before the real processes to ensure model safety in practical applications.

Besides, encoding neural networks into GTSs under TOP provides a clear window into the operations of both training and predicting phases. This transparency is critical in understanding the relationship between input features and their corresponding outputs, which is particularly beneficial in counterfactual explanations that illustrate how variations in certain inputs could lead to different outcomes.

Particularly, TOP offers a structured method of unraveling the complex pattern within graph-structured data since Graph Neural Networks (GNN)\cite{wu_comprehensive_2020}, known for their effectiveness in handling graph data, can greatly benefit from the logical reasoning provided by TOP.

\subsubsection{Hardware Design}

TOP has the potential to facilitate hardware design, especially in the context of High-Level Synthesis (HLS) \cite{gajski_highlevel_2012}. GTSs are expressive to encode circuit designs and their corresponding expected properties. A noteworthy parallel exists between the communication protocols designed for circuits, such as handshake mechanisms \cite{edwards_compositional_2019}. In the same manner, the modularization feature of TOP, as discussed in Section~\ref{sec:modularization}, along with the progressive specification approach detailed in Section~\ref{sec:progressive_specification}, endows TOP with the capability to specify large-scale circuit designs effectively. Furthermore, the formal verification of TOP plays a crucial role in optimizing circuit designs and ensuring their safety. Additionally, the potential of a TOP language, such as Seni introduced in Section~\ref{sec:seni}, extends to generating Register-Transfer Level (RTL) designs that can be subsequently synthesized to the gate level through the application of a logic synthesis tool.

\subsection{Future Direction}

As a nascent programming paradigm, TOP is on a trajectory toward establishing an ecosystem that facilitates the development of provably correct systems across diverse domains. Looking ahead, the Seni language, as a practical embodiment of TOP, plays a pivotal role in advancing the theoretical framework that underpins TOP.

Concretely, Seni is set to augment syntactic sugar to be more expressive in modeling complex structures and behaviors. This enhancement will include the integration of functional styles reminiscent of programming languages like Haskell, specifically for formulating transformation relations of GTSs. Besides, Seni aims to incorporate a suite of ready-to-use modules designed to streamline the development of real-world applications, such as network modules at the syntactic level, operating system modules, and enhanced I/O modules.

Besides, Seni is positioned to broaden its repertoire of proof techniques and optimize compatibility with existing tools. The roadmap includes the introduction of the mathematical induction capability, complemented by semi-automated theorem proving techniques. The objective is to enhance Seni's verification prowess and further refine the theoretical framework to incorporate usable proof techniques.

Moreover, Seni is strategically set to leverage interactions with language processing methods like Large Language Models (LLMs) \cite{brown_language_2020} to aid in the comprehension of model and property specifications, which is particularly pertinent when dealing with complex models and large-size formulas. Meanwhile, introducing machine learning approaches presents a promising avenue to assist developers who may not possess extensive mathematical expertise. Based on these approaches, Seni aims to facilitate a more intuitive and user-friendly way of understanding and working with the concepts of TOP and lowers the barriers to entry for developers with diverse backgrounds.

As Seni evolves, its applicability is expected to become wider. Seni will continue to be applied in the studied domains, such as distributed protocols, security infrastructures, and decentralized applications. Simultaneously, Seni intends to branch out into other promising fields, as outlined in Section~\ref{sec:vision}, such as the development of compiling tools, the exploration of explainable artificial intelligence, and advancements in hardware design. This strategic expansion is not merely an extension of Seni's applicability but a testament to the growing influence of TOP as a formidable programming paradigm to develop provably correct systems.

\bibliographystyle{unsrtnat}
\bibliography{references}  






\end{document}